\documentclass[11pt,letterpaper]{article}

\usepackage[margin=1in]{geometry}

\usepackage{
amsmath,
amssymb,
amsthm,
mathtools,
mathrsfs
}

\usepackage{algorithm}
\usepackage{algpseudocode}

\usepackage{booktabs}
\usepackage{graphicx}
\usepackage{xspace}
\usepackage{enumitem}
\usepackage{mdframed}
\usepackage{array}

\usepackage[dvipsnames]{xcolor}
\usepackage{tikz}
\usetikzlibrary{arrows.meta}
\usepackage[colorlinks=true,linkcolor=MidnightBlue,citecolor=Plum,urlcolor=MidnightBlue]{hyperref}
\usepackage{aliascnt}
\usepackage[nameinlink,noabbrev,capitalise]{cleveref}

\usepackage{etoolbox}
\makeatletter
\pretocmd{\@startsection}{\@nobreakfalse}{}{}
\renewcommand\paragraph{\@startsection{paragraph}{4}{\z@}%
  {1.5ex \@plus .5ex \@minus .2ex}{-1em}%
  {\normalfont\normalsize\bfseries}}
\makeatother

\newtheorem{theorem}{Theorem}[section]
\newaliascnt{lemma}{theorem}
\newtheorem{lemma}[lemma]{Lemma}
\aliascntresetthe{lemma}
\newaliascnt{corollary}{theorem}
\newtheorem{corollary}[corollary]{Corollary}
\aliascntresetthe{corollary}
\newaliascnt{proposition}{theorem}
\newtheorem{proposition}[proposition]{Proposition}
\aliascntresetthe{proposition}
\newaliascnt{claim}{theorem}

\aliascntresetthe{claim}
\newaliascnt{assumption}{theorem}
\newtheorem{assumption}[assumption]{Assumption}
\aliascntresetthe{assumption}
\theoremstyle{definition}
\newaliascnt{definition}{theorem}
\newtheorem{definition}[definition]{Definition}
\aliascntresetthe{definition}
\newaliascnt{notation}{theorem}

\aliascntresetthe{notation}
\newaliascnt{remark}{theorem}
\newtheorem{remark}[remark]{Remark}
\aliascntresetthe{remark}
\crefname{assumption}{Assumption}{Assumptions}
\Crefname{assumption}{Assumption}{Assumptions}

\newcommand{\F}{\mathbb F_q}
\newcommand{\Ftwo}{\mathbb F_2}
\newcommand{\Mat}{\mathbb F_q^{n\times n}}
\newcommand{\GL}{\operatorname{GL}_n(\mathbb F_q)}
\newcommand{\GLc}[1]{\operatorname{GL}_{#1}(\mathbb F_q)}
\newcommand{\Id}{\mathrm{Id}}
\newcommand{\rank}{\operatorname{rank}}
\newcommand{\Tr}{\operatorname{tr}}

\newcommand{\wt}{\operatorname{wt}}

\newcommand{\enc}{\mathsf{enc}}
\newcommand{\T}{\top}

\newcommand{\E}{\mathbb E}

\newcommand{\LINEQ}{\mathsf{LINEQ}}
\newcommand{\MM}{\mathsf{MM}}
\newcommand{\COORD}{\mathsf{COORD}}

\newcommand{\FINDRANK}{\mathsf{FINDRANK}}
\newcommand{\TRI}{\oplus\mathsf{TRI}}
\newcommand{\agr}{\operatorname{agr}}

\newcommand{\NP}{\mathsf{NP}}
\newcommand{\poly}{\operatorname{poly}}
\newcommand{\polylog}{\operatorname{polylog}}
\newcommand{\Otilde}{\widetilde O}

\newcommand{\eps}{\varepsilon}
\newcommand{\Ocal}{\mathcal O}
\newcommand{\Acal}{\mathcal A}
\newcommand{\Bcal}{\mathcal B}
\newcommand{\Pcal}{\mathcal P}
\newcommand{\Ecal}{\mathcal E}
\newcommand{\Dcal}{\mathcal D}
\newcommand{\Scal}{\mathcal S}

\title{Hardness Self-Amplification in Linear Algebra\\
and Fine-Grained Cryptography}
\author{%
Geoffroy Couteau\thanks{Universit\'e Paris Cit\'e, CNRS, IRIF.
  \href{mailto:couteau@irif.fr}{\texttt{couteau@irif.fr}}}
\and
Nikolas Melissaris\thanks{Universit\'e Paris Cit\'e, CNRS, IRIF.
  \href{mailto:nikolas@irif.fr}{\texttt{nikolas@irif.fr}}}
\and
Thejas Radhika Sajith\thanks{Chennai Mathematical Institute.
  \href{mailto:thejasrs.mcs2026@cmi.ac.in}{\texttt{thejasrs.mcs2026@cmi.ac.in}}}}
\date{}

\begin{document}

\pagenumbering{roman}
\maketitle

\begin{abstract}
\normalsize
We study the possibility of basing fine-grained cryptography on the hardness of fundamental
problems in linear algebra. Our main technical contribution is a hardness
self-amplification method based on random low-rank updates: we show how to turn an algorithm that solves an inverse-polynomial fraction of random linear systems into a solver which succeeds for every right-hand side of almost every matrix. Our reduction increases the runtime only linearly in the inverse of the original success probability. In particular, this yields fine-grained one-way functions from the almost worst-case hardness of linear systems over large enough fields.

Using the same approach, we also obtain tighter worst-case to average-case reductions for matrix multiplication and online matrix-vector multiplication, as well as variants of our reductions that recover exact solutions from approximate algorithms, whose expected coordinate agreement is only $1/q+\varepsilon$, with inverse-quadratic dependence on $\varepsilon$ up to logarithmic factors, over every finite field.  Finally, we also use it to connect the existence of fine-grained key agreement to the problem of computing the rank of a matrix: over a suitable sequence of growing characteristic-two fields, the almost-everywhere average-case hardness of rank computation (for any of a large class of distributions) implies infinitely-often fine-grained key agreement.

\end{abstract}
\clearpage
\setcounter{tocdepth}{2}
{\setlength{\parskip}{0pt}\tableofcontents}
\clearpage
\pagenumbering{arabic}

\section{Introduction}
\label{sec:intro}

In this work, we study the possibility of basing fine-grained cryptography
on the hardness of fundamental problems in linear algebra. Our guiding
goal is to obtain assumptions whose failure would imply useful
algorithmic progress: either a cryptographic primitive is secure, or a
central computational problem admits a faster algorithm. Motivated by a simple connection between the average-case hardness of solving linear systems and the existence of fine-grained one-way functions (FGOWFs), we focus on the problem of
solving linear systems and related tasks, and ask how these problems connect to fine-grained cryptography.

Our main technical contribution connects FGOWFs to the \emph{almost} worst-case hardness of solving linear systems. At its heart is a new hardness self-amplification procedure based on random walks
with low-rank updates. For linear systems, our approach turns a weak average-case
solver into a solver for every right-hand side of almost every matrix,
up to an exponentially small exceptional fraction, and yields fine-grained one-way functions from the nonexistence of such almost-everywhere solvers. Our method also applies beyond this cryptographic
motivation: it gives tighter worst-case to average-case reductions for
matrix multiplication and related problems, as well as tighter ``worst-case exact'' to ``average-case approximate'' reductions. Our
reductions work over every finite field, with overhead inverse-linear in the success probability in the exact setting and inverse-quadratic in the agreement advantage for the approximate setting. Before stating our results, we briefly discuss our original motivation for studying them, through the lens of fine-grained cryptography.

\paragraph{Fine-grained cryptography.}
Fine-grained cryptography studies primitives whose security is required
only against adversaries running in moderate polynomial time $N^c$,
where $c>1$ and $N$ is the honest parties' running time. The standard
motivation is to obtain cryptography from weaker assumptions compared to coarse-grained primitives. The seminal example is Merkle's
key agreement~\cite{Mer78}: it achieves security against
adversaries running in time $\ll N^2$, using only a random oracle (a variant of Merkle's protocol can further be based on exponentially secure one-way functions~\cite{BGI08}), an assumption not known to imply standard public-key cryptography.

What makes an assumption \emph{good} for a fine-grained primitive? Cryptographers have a natural preference for worst-case over average-case assumptions, or for
unstructured over algebraic assumptions (as in Merkle's protocol), since they appear to conceptually \emph{weaken} the security requirements. But this does not always settle the question: for example, it is unclear whether exponentially secure
one-way functions are a ``better'' assumption than polynomially secure
public-key encryption. Instead, a good guiding principle follows from the program of excluding Pessiland laid out by Impagliazzo~\cite{Imp95}: we should aim for assumptions whose falsification
would yield useful advances in algorithms. Such results yield a desirable win-win
situation: either some form of cryptographic hardness holds, or an
important computational problem admits a faster algorithm.

Brzuska and Couteau~\cite{BC22,BC25} pursued this goal for strong forms
of average-case hardness in $\NP$: they asked whether the nonexistence
of FGOWFs would imply algorithms
slightly faster than brute-force for all $\NP$ problems on average,
when amortizing over multiple instances. Unfortunately, their main result is negative, ruling out black-box reductions achieving this win-win result. Therefore, we pursue a different route: rather than aiming for speedups for general $\NP$ problems (which is perhaps too much to hope for with current techniques given~\cite{BC22}), we seek to find useful win-win results from concrete problems for which faster algorithms would have far-reaching consequences. In this work, we focus on such problems in the realm of linear algebra.

\paragraph{Solving linear systems.}
\label{par:intro-mm-exponents}
Solving a linear system of $\Theta(n)$ equations in $n$ variables takes
$n^{\omega+o(1)}$ operations with the fastest known methods, where
$\omega$ is the matrix multiplication exponent, with $\omega<2.372$~\cite{ADVXXZ25} (a recent unpublished result by
OpenAI claims $\omega\le9/4$ over $\mathbb C$, and a subsequent preprint
claims the same bound over every field~\cite{OpenAIMM26,ABZ26}).
In light of our current knowledge, a nearly quadratic algorithm for solving
linear systems would be a major breakthrough. Our investigation starts from a simple observation: if solving random
linear systems with non-vanishing probability requires time
$n^{2+\eta}$, for some $\eta>0$, then $f(M,x)=(M,Mx)$ is a FGOWF over sufficiently large finite fields. Evaluation takes
$\Theta(n^2)$ operations, so the exponent gap between evaluation and
inversion is $1+\eta/2$. However, this falls short of providing a useful win-win: solving linear systems is ubiquitous in computer science, but solving \emph{uniformly random} linear systems (with non-vanishing probability) appears much less consequential. We therefore
ask how much more can be obtained from such an algorithm, and what forms of one-wayness follow from worst-case hardness. Additionally, we investigate whether related linear
algebra problems can provide stronger cryptographic consequences such as fine-grained \emph{public-key} cryptography.

\subsection{Our results on fine-grained cryptography}
\label{sec:results}

\paragraph{Almost FGOWFs from worst-case hardness.}
We start with a few simple observations to derive relaxed forms of one-wayness from the worst-case hardness of solving linear systems. Given a linear system $(M,y)$, randomizing the right-hand side via $y \mapsto y + M\cdot x'$ for a random $x'$ takes only quadratic time: the expensive part is randomizing \emph{the matrix $M$}. We leverage this observation by focusing on the permutation $f_M(x) = M\cdot x$ for a fixed matrix $M$. First, $x\mapsto Mx$, with $M$ as a public parameter, is a family of \emph{auxiliary-input} FGOWFs~\cite{OW93}: for each adversary there exists at least one $M$ on which inversion is hard. Second, Levin's universal search method~\cite{Lev73} lets us turn this auxiliary-input FGOWF into a single sequence of public matrices $(M_n)$ for which $x\mapsto M_nx$ is one-way against all \emph{uniform} adversaries within the stated time bound. Hence, this yields a family of uniform FGOWFs (with a polynomial gap between evaluation and inversion) in a model where all parties are given access to a public advice string.

\paragraph{FGOWFs from almost worst-case hardness.}
To obtain ordinary FGOWFs, we develop a procedure for amplifying
the hardness of solving linear systems. Our approach forms the main technical contribution of our work. Our linear-system amplification result (\cref{prop:rank-k}) shows that FGOWFs follow even if only a very small (in fact, exponentially small) fraction of matrices remain slightly hard. To provide a concrete parameter setting,
fix any $\eta>0$ and $0<a<0.32$\footnote{The dual matrix multiplication exponent exceeds $0.32$~\cite{VXXZ24}; a recent preprint claims a lower bound of $0.5$~\cite{ABZ26}.}. Suppose that, for every uniform time-$n^{2+\eta}$ algorithm and every sufficiently large $n$, for more than a $2^{-n^a}$ fraction of invertible matrices $M$, there
exists at least one right-hand side $y$ on which the algorithm fails with probability
greater than $1/3$. We refer to this as \emph{almost worst-case hardness}.
Under this assumption, our theorem implies that $f(M,x)=(M,Mx)$ is a FGOWF over sufficiently
large fields of polynomial size. Thus an algorithm may solve every
system associated with almost every matrix, and the remaining hardness
still suffices for cryptography. The precise tradeoff between running
time, inversion probability, and field size is given in
\cref{sec:amplification-owf}. Our reduction from approximate solvers
(\cref{sec:approximate-lineq}) further shows that recovering even a small
constant fraction of the coordinates of $x$ is hard
(\cref{cor:approximate-owf}).

The two reductions above (almost-FGOWFs from worst-case hardness, and FGOWFs from almost worst-case hardness) leave open the goal of proving an unconditional fine-grained worst-case to average-case reduction for solving linear systems. We view it as the main open problem of our work. Such a reduction could give the stronger win-win that either FGOWFs exist or arbitrary linear systems can be solved in nearly quadratic time. Unfortunately, this goal appears to be out of reach of the techniques recently developed in unconditional reductions for matrix multiplication and linear problems, as solving linear systems lacks the ``low-degree'' structure of these problems\footnote{A recent work~\cite{VZ25} overcomes this by using \emph{trapdoor matrices}, a cryptographic primitive which implies in particular (standard) one-way functions (and is currently known from a very limited set of assumptions).}.

\paragraph{Key agreement and matrix rank.} We complement our study by asking whether hardness in linear algebra can support fine-grained public-key cryptography. We focus on the following natural two-message protocol: Alice sends $(M,a^\T M)$ for uniform $M,a$, Bob sends $Mb$ for a uniform $b$, and both parties agree on $K=a^\T Mb$. For invertible $M$, recovering $K$ can be shown equivalent (up to quadratic additional work) to recovering one coordinate of the solution to a random linear system. Unfortunately, we do not know how to relate this task to the average-case hardness of \emph{solving} linear systems. Instead, we provide preliminary results towards obtaining a useful win-win by relating this task to the problem of computing the rank of a matrix. Matrix rank computation is another fundamental problem in linear algebra. Here, we focus on its average-case hardness under a natural family of ``hard'' distributions (the uniform distribution is easy for matrix rank: random matrices are close to full rank): we consider any distribution that first samples a rank, and then samples a uniform matrix of that rank. We prove the following: over a suitable sequence of growing fields of characteristic two, almost-everywhere average-case hardness of rank computation implies infinitely-often fine-grained key agreement. This yields the following win-win: either (infinitely often) fine-grained key agreement exists or the rank-finding problem can be (infinitely often) solved efficiently on average even for such ``hard'' distributions.

\subsection{Hardness self-amplification}
\label{sec:hardness-self-amp}

Our FGOWF from almost worst-case hardness is a \emph{hardness self-amplification} procedure: it turns any time-$T$ algorithm solving random invertible linear systems with probability $\alpha\ge n^{-O(1)}$ into a time-$O\!\left((T+n^{2+o(1)})/\alpha\right)$ algorithm which, for all but a $2^{-n^{0.32}}$ fraction of the invertible matrices $M$, solves $Mx=y$ with probability at least $2/3$ for every right-hand side $y$. We show that our approach to hardness self-amplification enjoys applications beyond our original cryptographic motivation.

\paragraph{Matrix multiplication, OMv, and approximate algorithms.} We extend our approach to other important problems: matrix multiplication and online matrix-vector multiplication (OMv). For these problems, it follows from classical self-correction results~\cite{BLR93,Lip91} that hardness self-amplification can be turned into worst-case hardness. We also further consider the setting of \emph{approximate} algorithms that return only partially correct answers, which has received a significant amount of attention in recent years. In particular, starting from an approximate solver with $\eps$-advantage over a random guess, we obtain an exact solver with an overhead polynomial in $1/\eps$ (in fact, the overhead is quadratic). We summarize our new bounds in \cref{tab:intro-summary}. Our new approach also yields hardness self-amplification for triangle parity: it turns an algorithm computing the parity of the number of triangles in $G(n,1/2)$ with probability $1/2+\eps$ into one with high constant success probability, using $O(1/\eps^2)$ calls and $O((T+n^2)/\eps^2)$ time. Combining this with~\cite{BBB19} gives a worst-case reduction, whose queries have $3n$ vertices (see \cref{thm:triangles,cor:triangles-wc}).

\begin{table}[htbp]
\centering
\begin{tabular}{@{}lcc@{}}
\toprule
Problem & Exact success $\alpha$
        & Expected agreement $1/q+\eps$ \\
\midrule
Linear systems
 & $O((T+n^{2+o(1)})/\alpha)$
 & $O((T+n^2)\eps^{-2}\log(n/\eps)+n^{2+o(1)})$ \\
Matrix multiplication
 & $O((T+n^2)/\alpha)$
 & $\Otilde((T+n^2)/\eps^2)$ \\
OMv preprocessing
 & $O((P+n^2)/\alpha)$
 & $\Otilde((P+n^2)/\eps^2)$ \\
OMv query
 & $O((T+n)/\alpha)$
 & $\Otilde((T+n)/\eps^2)$ \\
\bottomrule
\end{tabular}
\caption{Exact recovery from average-case algorithms over every finite field, with success probability at least $2/3$. Here $T$ is the oracle or query time and $P$ is preprocessing time. Matrix multiplication and OMv have worst-case guarantees. Linear systems have guarantees for every right-hand side outside a $2^{-n^{0.32}}$ fraction of matrices, assuming $\alpha\ge n^{-O(1)}$ for exact answers. The notation $\Otilde$ hides
factors polylogarithmic in $n,q,1/\eps$.}
\label{tab:intro-summary}
\end{table}

\paragraph{Comparison with previous reductions.}
For exact matrix multiplication, our $O(1/\alpha)$ overhead removes the factor $\log^2(1/\alpha)\log\log(1/\alpha)$ from~\cite{HS23}. For OMv, we reduce the $\alpha^{-1}\polylog(1/\alpha)\cdot O(\log n)$ oracle queries of \cite{HS23,HS25a} to $O(1/\alpha)$. In the approximate agreement setting, previous unconditional uniform reductions for
matrix multiplication over prime fields had exponential~\cite{HS25a,SS25} or doubly
exponential~\cite{HS25b} dependence on $1/\eps$. Our quadratic
dependence, up to logarithmic factors, allows polynomially small
$\eps$ over every finite field and answers the question of Shinkar
and Singh~\cite[Remark~2]{SS25} of obtaining a reduction in time
$\poly(1/\eps)\cdot\Otilde(T)$ when $T\ge n^2$. For OMv, we obtain
$O(\eps^{-2}\log(nq))$ oracle queries, replacing the
$2^{\poly(q,1/\eps)}\cdot O(\log n)$ bound of~\cite{HS25a}.
Polynomial dependence on $1/\eps$ was already known over large
fields~\cite{HS25a}. Vaikuntanathan and Zamir~\cite{VZ25} obtain
reductions over prime fields under the assumption that trapdoor
matrices exist. Our reductions are unconditional. Detailed comparisons appear in \cref{tab:mm,tab:mm-approx}.

\subsection{Related work}
\label{sec:related}

Past work on fine-grained cryptographic primitives includes one-way
functions~\cite{BC22,BC25}, proofs of work~\cite{BRSV18}, and public-key
encryption~\cite{LLW19}. Other applications include multiparty
non-interactive key exchange~\cite{ACMS23,BCS24} and anonymous
transfer~\cite{ACM22}. Following Merkle's seminal work, fine-grained key
agreement has been studied through query-complexity bounds~\cite{BM09},
communication lower bounds~\cite{HMORY19}, distributed variants~\cite{DH21},
and perfect completeness~\cite{Maz25}. Alman, Huang, and Yeo~\cite{AHY25}
connect standard one-way functions to average-case conjectures for
$k$-SUM and Zero-$k$-Clique. Security against bounded depth attackers has
also been studied~\cite{DVV16,CG18,EWT19,WP22a,WP22b,WSP24}.

Worst-case to average-case reductions and hardness self-amplification
for matrix multiplication, OMv, and related problems were developed
in~\cite{AGGS22,HS22,HS23}. Error-correcting reductions over finite fields
were studied in~\cite{GSS24,HS25a,HS25b,SS25,VZ25}. Hirahara and
Shimizu~\cite{HS26int} extend this direction to binary matrices with integer
products, using expander walks in an encoding for approximate list decoding.
Reductions for fixed linear maps appear in~\cite{HS26}, and quantum
reductions in~\cite{AGGSS24}. Related work also treats subgraph
counting~\cite{GR18,BBB19,DLW20} and amplification beyond Boolean
functions~\cite{SY26}. Our additive walk is a lazy walk on the classical
bilinear forms graph~\cite{Del78,BF26}, in which two matrices are adjacent
when their difference has rank one.

\Needspace{8\baselineskip}
\section{Technical overview}
\label{sec:overview}

\vspace{-.4cm}

\subsection{Hardness self-amplification via random walks}
\label{sec:overview-walk}

At the heart of our approach is a conceptually simple strategy: to amplify an average-case solver, we perform a random walk on the graph whose edges connect matrices whose difference is low-rank, updating the linear system at each step. We show that this walk is efficient (it takes $O(n^2)$ work to update the system while preserving its solution), and that it (almost always) quickly hits matrices on which the solver succeeds. In the following, we fix an algorithm $\Ocal$ which solves $Mx=y$ with probability $\alpha$ on a uniform invertible matrix $M\in\GL$ and a uniform vector $y\in\F^n$. For this exposition, we assume for simplicity that $q\ge3$ and consider only updates of rank at most one.

\paragraph{Starting point.} Without loss of generality, the solution $x=M^{-1}y$ can be arbitrary: for a uniform
$z\gets\F^n$, the system $(M,y+Mz)$ has the uniform solution $x+z$. We can therefore query $\Ocal$ on this system and subtract $z$ from the answer, using $O(n^2)$ additional operations. The difficulty lies in randomizing $M$: a natural approach would be to send $(M,y)$ to $(RM,Ry)$ for
a uniform invertible matrix $R$, making $RM$ uniform (over invertible matrices) and leaving the solution unchanged. However, computing $RM$ requires a matrix
multiplication, which is too slow for our reduction. Instead, we seek a rerandomization with two features: first, given the fixed solver $\Ocal$, there should be an efficient way to apply the rerandomization while preserving the ability
to recover the solution. Second, starting from \emph{most} matrices, a short sequence of these rerandomization steps should reach matrices on which $\Ocal$ succeeds.

\paragraph{Moving by low-rank updates.}
Using random low-rank shifts is a natural idea: computing
$N=M+uv^\T$ for random $u,v\in\F^n$ takes $O(n^2)$ operations. Furthermore, when $M$ and $N$ are invertible, the Sherman--Morrison formula
recovers $M^{-1}y$ from $N^{-1}y$ and $N^{-1}u$ with only $O(n)$ additional operations: a solver for $N$ gives a solver for $M$ using two calls. We use a slight refinement of this idea which preserves the solution directly: sample
$u,v\gets\F^n$, set $R=I_n+uv^\T$, and define the candidate update to be $(M,y)\longmapsto(RM,Ry)
 =\bigl(M+u(v^\T M),\;y+u(v^\T y)\bigr)$. Computing the update takes $O(n^2)$ operations. Using the identity $\det(I_n+uv^\T)=1+v^\T u$, we can check whether $R$ is invertible with a single inner product. If $R$ is
singular, we stay in place; otherwise, we move to $(RM,Ry)$, which has
the same solution. For a fixed invertible $M$, the vector $M^\T v$ is
uniform and independent of $u$, so $u(v^\T M)$ has the same distribution
as $uv^\T$, hence our walk is almost the same as the additive rank-one walk, up to staying in place whenever a proposal leaves $\GL$.

\begin{samepage}
\paragraph{Finding successful matrices.}
Why should such restricted low-rank moves find the successful matrices quickly? A simple way to understand this is to consider the following simplified scenario: suppose that success depends only on a \emph{linear measurement} of $M$, of the form
\[
 \ell_C(M)=\operatorname{tr}(C^\T M)=\sum_{i,j}C_{ij}M_{ij},
 \qquad C\ne0.
\]
\end{samepage}
Now, imagine grouping all matrices into $q$ layers, according to the value
of $\ell_C(M)$. Ignore for now the invertibility constraint and consider updating the matrix by adding a random rank-1 matrix. The change in the measurement is $\ell_C(M+uv^\T)-\ell_C(M)=v^\T C^\T u$. Fix $u$. If $C^\T u\ne0$, then the inner product with a uniform $v$ is
uniform in $\F$: the step therefore lands in each of the $q$ layers with equal probability, regardless of the starting layer. The only exception is when $C^\T u=0$, in which case the measurement is unchanged, but this event has probability $q^{-\rank C}\le1/q$. Thus, as seen by any
fixed nonzero linear test, a step completely refreshes the measurement except with probability at most $1/q$. The same statement holds for any success rule depending only on this measurement.

\paragraph{Beyond linear tests.}
Of course, the behavior of $\Ocal$ may depend on $M$ in a much more
complicated way. This is where Fourier analysis comes to the rescue: it allows us to express \emph{any} such
dependence as a sum of simple functions of linear tests. For a prime field, encode a test value $a$ by the phase $\psi(a)=e^{2\pi i a/q}$,
a point on the unit circle (these $q$ points have average zero).
Over other finite fields, we use a
nontrivial additive character, which has the same cancellation property. Then, every function $f:\Mat\to\mathbb C$ has a unique expansion
\[
 f(M)=\sum_{C\in\Mat}\widehat f(C)\cdot
       \underbrace{\psi\bigl(\ell_C(M)\bigr)}_{\chi_C(M)},
 \qquad \text{with }
 \widehat f(C)=\E_M\bigl[f(M)\cdot\overline{\chi_C(M)}\bigr].
\]
In our setting, we take $f(M)\in[0,1]$ to be the success probability of the solver on $M$. For intuition, imagine a vertex for each matrix, colored white
when $f(M)=0$, black when $f(M)=1$, and gray for intermediate values: the Fourier expansion expresses this arbitrary assignment of values as a linear combination of simple functions $\chi_C$. Each $\chi_C$ depends on just one scalar measurement, $\ell_C(M)$: all matrices with the same measurement receive the same phase on the unit circle.
Although these components are complex-valued, their sum recovers
the real values $f(M)$. The constant component
$\widehat f(0)=\E[f]$ assigns the same shade to every vertex,
corresponding to the average success probability. Now, let $(Pf)(M)=\E_{u,v}[f(M+uv^\T)]$ be the expected value of $f$ after walking from $M$ by one step.
In this picture, applying $P$ replaces the shade of each vertex by
the expected shade after one step of the walk from that vertex.
For every integer $t\ge0$, let $P^t f$ be the expected value of $f$
after $t$ steps. The calculation above gives
\[
 (P\chi_C)(M)=q^{-\rank C}\cdot\chi_C(M),
 \qquad
 (P^t f)(M)=\widehat f(0)+
 \sum_{C\ne0}\widehat f(C)\cdot q^{-t\cdot\rank C}\cdot\chi_C(M).
\]
Indeed, when $C^\T u\ne0$, the phases average to zero, and otherwise
the phase is unchanged. Thus, averaging over a step preserves the
constant part of every pattern and shrinks each other component by at
least a factor $q$. These components are orthogonal, so their squared
contributions add without cross terms. In particular, $\E_M\bigl[|(P^t f)(M)-\E[f]|^2\bigr]
 \le q^{-2t}\cdot\E_M\bigl[|f(M)-\E[f]|^2\bigr]$. This is the useful consequence of the Fourier decomposition: repeated
averaging rapidly flattens \emph{every} pattern, including an arbitrary
pattern of success probabilities. In our actual walk, a candidate step is rejected with probability $e=(1-q^{-n})/q\le1/q$ (when $R$ is singular), independently of the current invertible matrix, but this does not change the high-level bottom line: replacing these proposals by steps that stay
in place preserves the uniform stationary distribution on $\GL$.
Furthermore, the contraction factor increases from at most $1/q$ to at most $1/q+e\le2/q$. Consequently, the same bound on the
average squared deviation holds for the invertible walk, with
$q^{-2t}$ replaced by $(2/3)^{2t}$ when $q\ge3$. This is the averaging
guarantee provided by a constant \emph{spectral gap}.

\paragraph{Hitting successful queries.}
\label{par:overview-hitting}
We now run the above walk and query $\Ocal$ at every visited matrix, updating the system $(M,y)$ to $(RM,Ry)$ whenever the candidate step $R$ is invertible. Write
\[
 h(N)=\Pr_{c\gets\F^n,\,\Ocal}[\Ocal(N,c)=N^{-1}c],
 \qquad \mu=\E_{N\gets\GL}[h(N)]\ge\alpha.
\]
The spectral gap guarantee yields a standard hitting bound (a Chernoff-like concentration bound around the expected behavior that we described above), which we state
for success probabilities $h(N)\in[0,1]$ in \cref{lem:hitting}.
With fresh randomness at every query, it gives, for every integer $t\ge1$,
\[
 \E_{M_1\gets\GL}
 \Pr[\text{all $t$ queries fail}\mid M_1]
 \le \exp\bigl(-\alpha\cdot(t-1)/3\bigr).
\]
Thus, averaged over a uniform starting matrix, the failure probability
decreases exponentially with the number of queries. In turn, this immediately gives the desired statement for
most starting matrices: let $p_t(M)$ denote the failure probability
starting from $M$. Taking
$t=O(\alpha^{-1}\log(1/(\beta\delta)))$ makes $\E_M[p_t(M)]\le\beta\delta$,
and therefore, by a straightforward Markov inequality, $\Pr_{M\gets\GL}[p_t(M)>\delta]\le\beta$. Rerandomizing the solution to the linear system (which is efficient, as we explained at the beginning of this overview) makes $p_t(M)$ independent of the original right-hand side. Consequently, for all but a $\beta$ fraction of
matrices, the reduction succeeds with probability at least $1-\delta$
on \emph{every} right-hand side. Checking each candidate solution takes
$O(n^2)$ operations, so we can stop at the first success. The total
running time is $O((T+n^2)\alpha^{-1}\log(1/(\beta\delta)))$.

\paragraph{Larger updates and small fields.}
\label{par:overview-rank-k}
The rank-one walk gives a useful first guarantee, but the running time above grows with $\log(1/\beta)$, where $\beta$ is the exceptional fraction. To make this fraction exponentially small with nearly
quadratic overhead, we use larger updates. Sample
$U,V\in\F^{n\times k}$ and set $R=I_n+UV^\T$. We test invertibility using the Sylvester determinant identity, $\det(I_n+UV^\T)=\det(I_k+V^\T U)$. Since the additive shift $UV^\T$ is a sum of $k$ independent rank-one shifts, each nonconstant Fourier component shrinks by a factor at most $q^{-k}$. Conditioning on $R$ being invertible loses only a constant factor: the probability of being invertible is greater than $1/4$ (this loose bound suffices) hence every accepted step contracts every mean-zero function on $\GL$ by at most $4q^{-k}$ (\cref{lem:rank-k-update}). Using larger ranks gives us a stronger conclusion from a single update: let
$g(M)=\E[h(RM)\mid R\text{ invertible}]$ be the success probability
averaged over an accepted update from $M$. The stationary mean remains
$\mu$, like before, and the contraction gives
\[
 \E_M[(g(M)-\mu)^2]\le16q^{-2k}\mu,
 \qquad
 \Pr_M[g(M)<\mu/2]\le\frac{64q^{-2k}}{\mu}
 \le\frac{64q^{-2k}}{\alpha},
\]
where the second inequality follows because every matrix with
$g(M)<\mu/2$ contributes at least $(\mu/2)^2$ to the squared deviation. This gives a much faster rerandomization procedure: we can simply restart \emph{from the original $M$} at every attempt. Provided that $M$ is outside the exceptional set, each independent attempt has success probability at least $\alpha/8$, including the probability of accepting
the update. Hence $O(\alpha^{-1}\log(1/\delta))$ attempts suffice,
\emph{independently of $k$}. Using fast rectangular matrix multiplication, we can set $k=\lceil n^{0.32}\rceil$ with only $n^{2+o(1)}$ additional operations per attempt. For polynomially small $\alpha$, this gives the exceptional
fraction $2^{-n^{0.32}}$ in \cref{sec:hardness-self-amp} for sufficiently
large $n$, over every finite field. For the simpler
walk with rejected steps replaced by self-loops, rank two already
gives a constant spectral gap over $\mathbb F_2$ (\cref{rem: field-two}).

\subsection{A geometric viewpoint}
\label{sec:overview-geometric}

We give a geometric interpretation of our strategy. Consider
the graph whose vertices are matrices, with an edge between two matrices
when their difference has rank one. The additive walk follows these
edges with the probabilities determined by sampling $u,v$, and stays
in place when $uv^\T=0$. This is a \emph{Cayley graph} of the additive
group of matrices: the allowed increments are the same at every
vertex, so translating all vertices preserves the walk. The standard
observation for abelian Cayley graphs, as explained by Hoory, Linial, and
Wigderson~\cite{HLW06}, is that Fourier characters are eigenfunctions
of the walk; our linear-test calculation simply identifies the eigenvalues.
Our invertible walk is also a Cayley walk, on the multiplicative group $\GL$, with moves given by left multiplication by $I_n+uv^\T$. Alon and Milman~\cite{AM85} developed a connection between spectral gap and expansion which gives a geometric interpretation: let $\gamma\ge1/3$ be the spectral gap of our invertible walk.
For any set $S\subseteq\GL$ of density $s$, a uniform starting matrix
$M_0$ and one step $M_1$ satisfy $\Pr[M_0\in S,\ M_1\notin S]\ge\gamma s(1-s)$. Thus, from a uniform vertex in $S$, the probability of leaving $S$ is
at least $\gamma(1-s)$: every sufficiently dense region has many transitions crossing the boundary, with edges weighted by the transition probabilities. A region with few such transitions would have an indicator that changes little under averaging, contradicting the
spectral bound.

\begin{figure}[t]
\centering
\begin{tikzpicture}[font=\small,>=Stealth]
 \foreach \shift/\step/\near/\far/\total in {
     0/0/0.125/0.125/0.750,
     4.5/2/0.090277777778/0.118055555556/0.597,
     9/8/0.046926202180/0.066302059518/0.320} {
  \begin{scope}[xshift=\shift cm]
   \node[font=\small\bfseries] at (0,1.9) {After $\step$ steps};
   \foreach \i/\x/\y in {
       0/-1.2/1.2,1/1.2/1.2,2/1.2/-1.2,3/-1.2/-1.2,
       4/-0.5/0.5,5/0.5/0.5,6/0.5/-0.5,7/-0.5/-0.5} {
    \coordinate (v\i) at (\x,\y);
   }
   \foreach \a/\b in {0/1,1/2,2/3,3/0,4/5,5/6,6/7,7/4,
                        0/4,1/5,2/6,3/7} {
    \draw[black!22,line width=0.6pt] (v\a) -- (v\b);
   }
   \foreach \a/\b in {1/0,3/0,5/4,7/4} {
    \draw[->,ForestGreen!80!black,line width=0.7pt,
          shorten <=7pt,shorten >=7pt] (v\a) -- (v\b);
   }
   \foreach \i/\mass in {1/\near,2/\far,3/\near,
                          5/\near,6/\far,7/\near} {
    \pgfmathsetmacro{\massshade}{100*\mass/0.125}
    \filldraw[fill=black!\massshade,draw=black!35]
      (v\i) circle (0.23);
   }
   \foreach \i in {0,4} {
    \filldraw[fill=white,draw=ForestGreen!80!black,line width=1pt]
      (v\i) circle (0.23);
    \draw[ForestGreen!80!black,line width=0.6pt] (v\i) circle (0.16);
   }
   \node at (0,-1.85) {Remaining mass: $\total$};
  \end{scope}
 }
 \draw[->,black!45,line width=0.8pt] (1.75,0) -- (2.75,0);
 \draw[->,black!45,line width=0.8pt] (6.25,0) -- (7.25,0);
 \foreach \x/\shade in {0/0,0.35/50,0.7/100} {
  \filldraw[fill=black!\shade,draw=black!35]
    (\x,-2.65) circle (0.12);
 }
 \node[anchor=west,font=\scriptsize] at (0.95,-2.65)
   {darker $=$ more surviving mass};
 \draw[ForestGreen!80!black,line width=0.9pt] (6.2,-2.65) circle (0.12);
 \draw[ForestGreen!80!black,line width=0.5pt] (6.2,-2.65) circle (0.075);
 \node[anchor=west,font=\scriptsize] at (6.45,-2.65)
   {good vertex: incoming mass is removed};
\end{tikzpicture}
\caption{The hitting argument on an illustrative eight-vertex cube.
The walk stays put with probability $1/2$ and otherwise chooses one of
the three neighbors uniformly (self-loops are omitted). Initially,
each vertex has mass $1/8$, and the mass on the two green vertices is
removed. Each panel shows the result of repeatedly walking and removing
mass at green vertices. Darkness is proportional to surviving mass,
with white representing $0$ and black representing $1/8$ on the same
scale in all three panels.}
\label{fig:overview-absorption}
\end{figure}
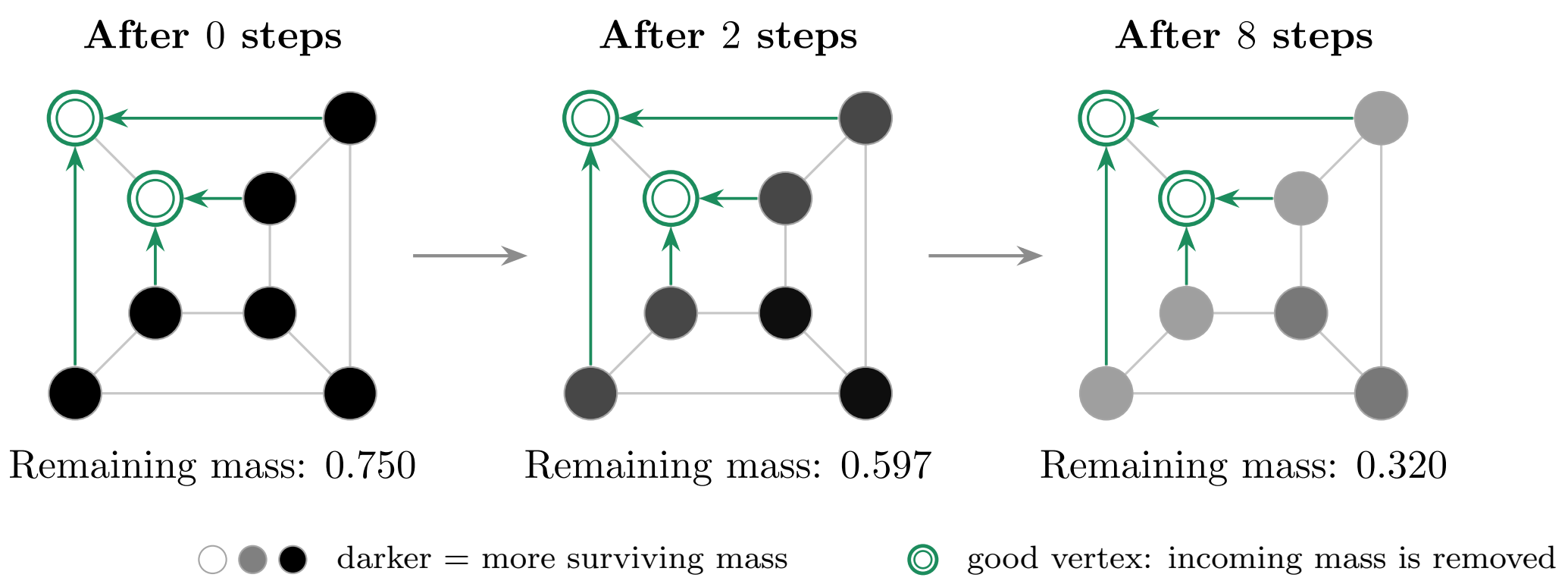

\paragraph{The good set as an absorbing boundary.}
We can extend the viewpoint to illustrate hitting.
Suppose that the success function $h$ is the indicator of a set $G$
of density $\alpha$, and mark these vertices with a green outline.
Give every vertex the same initial shade of gray, representing equal
probability mass, then erase the mass on $G$.
At each subsequent step, average the shades according to the walk
and again erase the mass at the good vertices
(\cref{fig:overview-absorption}).
The total mass after $t$ queries is exactly the probability that a
walk from a uniform starting matrix has avoided $G$ at every query. Now, a uniform vertex in the bad set enters $G$ with probability at least
$\gamma\alpha$, but the surviving mass becomes unevenly
distributed as the walk proceeds. The spectral hitting argument
accounts for this changing distribution~\cite{AKS87,HLW06} and gives $\Pr[M_1,\ldots,M_t\notin G] \le e^{-\gamma\alpha(t-1)}$. Hence, the total surviving mass decreases exponentially, even though walks from some vertices may take much longer to reach $G$. This is the geometric interpretation of the
hitting bound used in \cref{sec:overview-walk}: little mass survives on average, hence few starting vertices can have a large probability of failure.

\subsection{Hardness self-amplification for matrix multiplication}
\label{sec:overview-mm}

We turn to matrix multiplication: we consider a time-$T$ oracle $\Ocal$ that computes $AB$ with probability
$\alpha$ for independent uniform matrices $A,B\in\Mat$. Since an
instance is now a pair of matrices, we extend the additive walk by
updating both matrices independently at each step:
\[
 (A,B)\longmapsto(A',B')=(A+uv^\T,\;B+u'v'^\T),
 \qquad u,v,u',v'\gets\F^n.
\]
Compared with linear systems, this introduces two difficulties. First,
the desired answer changes at every step. Second, directly verifying
a candidate answer by recomputing the product takes
$n^{\omega+o(1)}$ operations. Both difficulties have simple solutions.

\paragraph{Correcting the answers.}
Expanding the updated product gives
$A'B'-AB=(Au')v'^\T+u(v^\T B)+(v^\T u')\cdot uv'^\T$.
Every term can be computed in $O(n^2)$ operations, using matrix-vector
and outer products. Along a walk $(A_j,B_j)$, we accumulate these
differences in $E_j=A_jB_j-A_0B_0$, starting from $E_0=0$. Thus, each
oracle answer gives the candidate $D_j=\Ocal(A_j,B_j)-E_j$ for the original product $A_0B_0$, with only $O(n^2)$ additional operations per step.

\paragraph{Verifying the answers.}
We use the test of Freivalds~\cite{Fre77}: sample $w\gets\F^n$ and check $D_jw=A_0(B_0w)$ (this takes $O(n^2)$ operations). A correct candidate always passes,
while a wrong candidate passes with probability at most $1/q$.
For at most $t\ge1$ candidates, we require
$s=\lceil\log_q(100t)\rceil$ successful tests before accepting, so the
probability of ever accepting a wrong answer is at most $1/100$.
Crucially, we discard a candidate as soon as a test fails: each wrong
candidate uses at most two tests in expectation. We stop at the first
accepted candidate, so the expected total is $O(t+s)=O(t)$ tests.
Stopping after a sufficiently large constant multiple of $t$ tests
introduces only a small constant failure probability and gives a
worst-case running-time bound.

Our spectral argument extends directly to this setting: the Fourier components
are now $\chi_C(A)\cdot\chi_D(B)$, and a step multiplies each component
by $q^{-\rank C-\rank D}\le1/q$ unless $C=D=0$. Thus, the walk has a
constant spectral gap over every finite field, including $\Ftwo$.
Starting from a uniform pair, the hitting bound ensures that
$t=O(1/\alpha)$ queries suffice to obtain a correct answer with high
constant probability. The total running time is $O((T+n^2)/\alpha)$.

For matrix multiplication, this amplification further gives a
worst-case to average-case reduction: for independent uniform masks
$X,Y\in\Mat$, the standard self-correction identity~\cite{BLR93,Lip91}
\[
 AB=(A+X)(B+Y)-(A+X)Y-X(B+Y)+XY
\]
expresses any product $AB$ as a sum and difference of four products,
each with a uniformly random input pair. Applying the amplified solver
to all four pairs and taking a union bound gives the correct product
with high constant probability, with only constant overhead
(see \cref{sec:mm}).

\subsection{Hardness self-amplification for OMv}

We now turn to the online matrix-vector multiplication (OMv) problem. In OMv, an algorithm first preprocesses a matrix $A_0$, then receives
query vectors $v$ one at a time and must output $A_0v$ before seeing
the next query. Here, our key idea is simply to \emph{preprocess the walk}: during preprocessing, we generate $t=O(1/\alpha)$ states
$A_0,\ldots,A_{t-1}$ of the walk $A_{j+1}=A_j+u_jw_j^\T$, and build
an oracle data structure for each $A_j$. For each query $v$, the correction can then
be maintained with only $O(n)$ operations per step:
\[
 e_0=0,\qquad e_{j+1}=e_j+u_j\cdot(w_j^\T v),
 \qquad e_j=(A_j-A_0)v.
\]
To make the oracle queries uniform for every fixed $v$, we also sample
independent masks $s_j\gets\F^n$ and store $A_js_j$ during
preprocessing. To obtain a candidate for $A_0v$, we query the data structure for $A_j$ on $v+s_j$ and
subtract $A_js_j+e_j$ from the answer. Lastly, verification can also be prepared in advance by preprocessing runs of the Freivalds test: sample $s$ independent
test vectors $r_i\gets\F^n$, and store $r_i^\T A_0$ during preprocessing. To check a candidate $c$, check whether $r_i^\T c=(r_i^\T A_0)v$, using only $O(n)$ operations per test and the same early rejection rule as above. As for matrix multiplication, hardness self-amplification easily translates to worst-case hardness for OMv: the reduction uses the identity $Av=(A+X)v-Xv$ for a uniform
matrix $X$ and preprocesses both $A+X$
and $X$. If the original data structure has preprocessing time $P$
and query time $T$, the resulting bounds are $O((P+n^2)/\alpha)$ and
$O((T+n)/\alpha)$, respectively, for constant error probability on
every fixed input $(A,v)$ (see \cref{sec:omv}).

\subsection{Exact computation from approximate solvers}
\label{sec:overview-partial}

The preceding reductions use an oracle that sometimes returns an
entirely correct answer. Suppose instead that its output agrees with the solution vector, or with the matrix product, on at least a
$\gamma=1/q+\eps$ fraction of the coordinates in expectation over
uniform inputs: the oracle could satisfy this guarantee while returning a wrong vector or matrix on every query. We must therefore collect enough correct votes at every coordinate, without knowing which individual values are correct. Below, we overview what can possibly go wrong when interacting with such an oracle, and outline how we use randomization to address these issues.

An adversarial oracle could always compute the same coordinates correctly and fail on the others: repeating the query would then always leave the remaining coordinates wrong. Furthermore, even if every coordinate were correct equally often, a coordinate that is correct less than half the time could be outvoted by a single repeated wrong value. Our randomization strategy therefore plays two roles: we use \emph{permutations} to spread correctness across coordinates, and \emph{random scaling} to spread nonzero errors across the $q-1$ wrong values (note that both transformations preserve zero errors). If a coordinate is correct with probability $\gamma_0\ge 1/q+\eps$, and each wrong value occurs with probability $(1-\gamma_0)/(q-1)$, then $\gamma_0-(1-\gamma_0)/(q-1)
 =(q\gamma_0-1)/(q-1)
 \ge q\eps/(q-1)>\eps$. Thus, the correct value is more frequent than every individual wrong value, even when most answers are wrong. This lets us recover each coordinate by plurality vote. However, these probabilities hold over uniform inputs. Our key observation is that the same random walk strategy that we used in the exact setting suffices to achieve (almost) the same guarantees as for random inputs for almost all inputs. Concretely, we must ensure the following: to decode successfully, the correct value must appear more often than every wrong value among the answers we collect. We must therefore show that the observed frequencies are close to the above probabilities, even though successive queries along the walk are highly correlated. This follows once more from our spectral gap argument: we used it previously to find one successful answer, and now use it again in a similar fashion to collect enough correct votes at every coordinate and limit the repetitions of each wrong value.

\paragraph{Linear systems.}
Consider a system $Mx=y$, where $M\gets\GL$ and $y$ is arbitrary.
An ideal rerandomization would simply replace $M$ by $RMP$, with uniform $R\in\GL$ and a uniform permutation matrix $P$: for every fixed
invertible $M$, the result is uniform and independent of $P$.
Together with a random mask on the right-hand side, this hides the
correspondence between the original and queried coordinates from the
oracle. However, computing $RM$ requires a matrix multiplication, which is too expensive\footnote{The permute-and-scale decoder also appears in~\cite{VZ25}. They use a trapdoor matrix for $R$, which allows for fast multiplication but looks uniform without the trapdoor. We avoid computational assumptions via low-rank updates.}. As in
\cref{sec:overview-walk}, we replace this expensive step with a sequence of inexpensive walk steps to change the
matrix, using fresh permutation and scaling at every query. Our invertible walk gives a sequence of systems $(M_j,y_j)$ with the same solution $x$. We use rank-one steps for $q>2$ and rank-two steps for $q=2$. During the walk, we use the permute-and-scale procedure which we outlined above (together with the same uniform rerandomization of the solution as we used for exact solvers): at each step, we choose a uniform permutation matrix $P$, set $B=M_jP$, and query the oracle on $(B,c)$ with $c=gy_j+Bw$, where $g\gets\F^*$ is the rescaling factor and $w\gets\F^n$ rerandomizes the solution $x$. The solution of the queried system is $gP^{-1}x+w$. Hence, from the oracle's answer $u$ we obtain the candidate $\widehat x=g^{-1}P(u-w)$. If $e=u-B^{-1}c$ is the error of this answer,
then $\widehat x-x=g^{-1}Pe$: the transformation permutes the positions of the correct coordinates and rescales the nonzero errors. Note that the rerandomizer $Bw$ is required to hide $g$ from the oracle: if we queried only on $(B,gy_j)$, the oracle could scale its error along with its input, leaving the error unchanged when we divide by $g$. With the mask, $c$ is uniform for every fixed $g$, since $B$ is invertible, and the query reveals no information about $g$ (even after fixing the walk, all permutations, all queried systems, and all oracle coins, the scalars remain independent and uniform in $\F^*$). At each fixed coordinate, every nonzero error is therefore uniform among the wrong values, independently across the queries where that coordinate is wrong.

We must still ensure that every coordinate is correct often enough. When $M_j$ is uniform, $B=M_jP$ is uniform and independent of $P$. Consequently, each coordinate of $x$ corresponds to a uniform coordinate of the oracle's answer, and has mean correctness probability at least $\gamma$. At a fixed matrix, however, the oracle may remain unreliable under every column permutation. As in \cref{sec:overview-walk}, we analyze a success function, now for each coordinate: $f_i(M)=\Pr[\widehat x_i=x_i\mid M_j=M]$. We have $\E_{M\gets\GL}[f_i(M)]\ge\gamma$. Here, the probability is over the fresh query randomization and oracle coins. The same spectral gap argument then applies: since the walk starts from a uniform matrix, it is stationary, and its constant spectral gap gives a Chernoff bound on the number of correct answers at each coordinate. Thus, we can bound the frequencies along the walk trajectory despite the fact that successive queries are correlated. The hidden scaling lets us also bound the number of repetitions of each wrong value. Together, these bounds make plurality decoding succeed after $O(\eps^{-2}\log(n/\eps))$ queries, with constant probability. Each query requires $O(n^2)$ additional operations, and we check the decoded vector against the original system. Applying the exact amplification of \cref{thm:lineq} gives a solver that succeeds with high probability on every right-hand side for almost all matrices.

We further note that over sufficiently large fields, we can decode with fewer queries: after the conditioning above, the wrong values at each fixed coordinate are independent and uniform among $q-1$ possibilities. If we choose the number of queries so that every coordinate receives at least a logarithmic number of correct answers, plurality decoding will work: when $q$ is large, no wrong value can repeat that often. This takes $O(\gamma^{-1}\log(n/\gamma))$ queries.

\paragraph{Matrix multiplication.}
For matrix multiplication, our additive walk works on pairs $(A_j,B_j)$ and allows us to maintain $E_j=A_jB_j-A_0B_0$ with $O(n^2)$ operations per step. A key observation is that subtracting $E_j$ from an answer for $A_jB_j$ gives a candidate for $A_0B_0$ with the same error. This is crucial in the approximate setting because every correct entry of an oracle answer becomes a correct entry of a candidate for the original product, regardless of errors elsewhere. The permutation and scaling strategy generalizes naturally to this setting: for each query, we independently permute and scale the rows of $A_j$ and the columns of $B_j$. Afterwards, the same spectral gap and Chernoff bound arguments bound the occurrences of each possible error. Then, a union bound over the $n^2$ entries and $q$ values implies that $O(\eps^{-2}\log(nq))$ queries suffice to decode the entire product, with probability at least $0.99$ on uniform inputs. Lastly, to handle arbitrary inputs, we apply the four-product identity
from \cref{sec:overview-mm} to this exact solver. Note that the order is important: without decoding first, their four errors could combine and destroy the small advantage over $1/q$.

\subsection{Additional results for fine-grained cryptography}
\label{sec:overview-crypto}

We conclude by returning to our original cryptographic motivation. Our hardness self-amplification immediately implies fine-grained one-way functions from ``almost'' worst-case hardness. We omit the details of ``almost-FGOWF'' via Levin's universal search as it is fairly straightforward, referring the reader to \cref{sec:worstcase}, and cover below our connection of key agreement to matrix rank.

\paragraph{Key agreement from hardness of matrix rank.}
Recall the protocol: Alice sends $(M,r=a^\T M)$ for uniform $M,a$, Bob sends $y=Mb$ for uniform $b$, and both compute $K=a^\T y=rb$. When $M$ is invertible, the transcript consists of uniform $M,r,y$, with $K=rM^{-1}y$. Given $q=2^\ell$, we also consider a single-bit-key variant where the parties output the binary inner product of a public random seed with the $\ell$-bit encoding of $K$. Then, Goldreich--Levin converts a distinguisher for this bit into a key-recovery algorithm, with a polynomial loss in $\ell$ and the inverse distinguishing advantage (\cref{thm:dense-bit}). While we do not know how to turn a key recovery algorithm into a sufficiently fast linear system solver (the straightforward coordinate-by-coordinate reconstruction introduces an $n$-fold overhead), we prove that it can be used to test matrix rank: fix a secret rank $\rho$ and let $A$ be uniform among matrices of rank $\rho$, and let $W$ be the leading $c\times c$ block of $A$. If $\rho<c$, then $W$ is singular. If $\rho\ge c$, then, over a large field, $W$ is invertible with probability $1-O(1/q)$ and  conditioned on this event, it is uniform among invertible matrices. To test which case holds, we choose independent uniform $r,b$, query the key-recovery algorithm on $(W,r,Wb)$, and check whether the answer equals $rb$. When $W$ is invertible, this is a normal random protocol transcript. In contrast, when $W$ is singular, the distribution of $b$ conditioned on the observation $Wb$ is uniform over a coset $b_0+\ker W$, and the target key becomes $rb=rb_0+r\kappa$ with $\kappa\gets\ker W$. Unless $r$ vanishes on the kernel, the hidden term $r\kappa$ is uniform in the field, hence even an unbounded algorithm passes the check with probability at most $2/q$.

We now use once more our invertible walk from \cref{sec:overview-walk} to amplify this
check. The key-recovery algorithm succeeds with probability $\alpha$ only on average over
invertible matrices, so repeated queries at a fixed $W$ are not necessarily helpful. The usual random walk argument lets us conclude: each step replaces $W$ by $(I_c+uv^\T)W$ with an invertible multiplier using $O(c^2)$ operations. Starting from a uniform invertible $W$, the
hitting bound makes a block of $O(1/\alpha)$ checks likely to contain a
correct answer. If $W$ is singular, every visited matrix remains singular,
so a block rarely contains a passing check when $q$ is sufficiently large.
We get a reliable threshold test via a majority vote over blocks, and use a binary search over $c$ to recover $\rho$.
Given time-$T$ key-recovery algorithms with success probability $\alpha$ for every dimension $c\le n$, this reduction takes $\Otilde((T+n^2)/\alpha)$ time over a sufficiently large field, for any distribution on $\rho$ with $A$ uniform conditional on $\rho$. From there, we deduce that the almost-everywhere average-case hardness of rank-finding implies infinitely-often fine-grained key agreement (see \cref{sec:rank}).

\clearpage
\section{Preliminaries and tools}
\label{sec:prelim}

\subsection{Notation and computational model}

For an integer $n\ge1$, we write $[n]=\{1,\ldots,n\}$. Throughout, $q$ is
a prime power and $\F$ is the finite field with $q$ elements. Vectors are
column vectors. A row vector is written as the transpose $a^\T$ of a column
vector, or introduced explicitly as an element of $\mathbb F_q^{1\times n}$.
We write $e_i$ for the $i$-th unit vector, $I_n$ for the $n\times n$
identity matrix, and $\Id$ for the identity operator. For a finite set $S$,
$s\gets S$ denotes a uniformly random element of $S$. For $x\in\F^n$, the
weight $\wt(x)$ is the number of nonzero coordinates of $x$, and the
Hamming distance is $d_H(x,x')=\wt(x-x')$. We write $\log$ for the natural
logarithm and $\log_2$ for the logarithm in base two. The notation
$\Otilde$ hides factors that are polylogarithmic in the parameters.

\paragraph{Model of computation.}
We measure running time in field operations. An algorithm can add,
multiply, invert and compare field elements, sample a uniform field element,
and read or write a field element, each in one step. When $q=n^{O(1)}$, this
model and a standard word-RAM model differ by at most polylogarithmic
factors. Algorithms are uniform and randomized unless we state otherwise.
A time-$T$ algorithm halts within $T$ steps on every input. When we use an algorithm as an oracle, each call uses fresh random coins. Average-case success probabilities are taken over both the input distribution and the coins of the algorithm. For a fixed input, the probability is over the coins of the algorithm alone.

\subsection{Linear-algebra problems and random matrices}
We denote by $\omega$ the exponent of matrix multiplication over the field
under consideration. The published bound $\omega<2.372$~\cite{ADVXXZ25}
suffices for our results. Recent unpublished preprints claim
$\omega\le9/4$, first over $\mathbb C$~\cite{OpenAIMM26} and then over
every field~\cite{ABZ26}. A linear system of dimension $n$ can be solved,
and the rank of an $n\times n$ matrix can be computed, with
$n^{\omega+o(1)}$ field operations~\cite{BH74,IMH82}. No
algorithm with $O(n^{\omega-\varepsilon})$ operations is known for solving
linear systems, even on random inputs.

We write $\GL$ for the group of invertible $n\times n$ matrices over $\F$. In this paper, we consider several natural linear algebra tasks:

\begin{definition}[Solving random linear systems]
\label{def:lineq}
The distributional problem $\LINEQ_{n,q}$ is defined as follows. The input
is a pair $(M,y)$, where $M\gets\GL$ and $y\gets\F^n$ are independent.
The goal is to output $M^{-1}y$. An algorithm $\Acal$ solves
$\LINEQ_{n,q}$ with probability $\alpha$ if
$\Pr_{M,y,\Acal}[\Acal(M,y)=M^{-1}y]\ge\alpha$.
\end{definition}

A candidate solution can be checked with $O(n^2)$ operations, by computing
$M\widehat x$ and comparing it with $y$. All our reductions for linear systems use this check,
and therefore never output a wrong solution. They output either the
correct solution or a failure symbol $\bot$.

\begin{samepage}
\begin{definition}[Matrix multiplication]
\label{def:matmult}
The distributional problem $\MM_{n,q}$ is defined as follows. The input
is a pair $(A,B)$, where $A,B\gets\Mat$ are independent. The goal is to
output $AB$. An algorithm $\Acal$ solves $\MM_{n,q}$ with probability
$\alpha$ if
$\Pr_{A,B,\Acal}[\Acal(A,B)=AB]\ge\alpha$.

The worst-case problem $\MM^{\mathrm{wc}}_{n,q}$ has the same input and
output, with arbitrary matrices $A,B\in\Mat$. An algorithm $\Acal$ solves
$\MM^{\mathrm{wc}}_{n,q}$ with probability $\alpha$ if
$\Pr_{\Acal}[\Acal(A,B)=AB]\ge\alpha$ for every pair $(A,B)$.
\end{definition}
\end{samepage}

\paragraph{Standard facts about random matrices.} A uniformly random matrix of $\Mat$ is singular with probability
\[
 \sigma_q:=1-\frac{|\GL|}{q^{n^2}}=1-\prod_{i=1}^n\bigl(1-q^{-i}\bigr).
\]
The value $\sigma_q$ depends on $n$, but we only use the following bound,
which is uniform in $n$:
\begin{equation}\label{eq:sigma}
 \sigma_q\le\sum_{i=1}^n q^{-i}<\frac1{q-1}.
\end{equation}
The first inequality is the union bound $1-\prod_i(1-x_i)\le\sum_ix_i$,
valid for $x_i\in[0,1]$. Similarly, for $m\ge n$, a uniformly random
matrix of $\F^{m\times n}$ has rank smaller than $n$ with probability at
most $\sum_{i=0}^{n-1}q^{i-m}<q^{n-m}/(q-1)$, by a union bound over the
$n$ columns, each of which must avoid the span of the previous ones.
We will often use the following two facts. If $M\gets\GL$ and
$R\in\GL$ is fixed, then $RM$ and $MR$ are uniform in $\GL$. If
$M\in\GL$ is fixed and $x\gets\F^n$, then $Mx$ is uniform in $\F^n$.

\subsection{Markov chains and concentration bounds}

This section introduces standard technical lemmas used in our hardness self-amplification results.

\paragraph{Markov operators.}
Let $\Omega$ be a finite set and $\pi$ a distribution on $\Omega$. We
work in the space $L^2(\pi)$ of functions $\Omega\to\mathbb R$, with the
inner product $\langle f,g\rangle=\E_{x\gets\pi}[f(x)g(x)]$ and the norm
$\|f\|=\langle f,f\rangle^{1/2}$. We write $\mathbf 1$ for the constant
function equal to one, and $\Pi f=\langle f,\mathbf 1\rangle\mathbf 1$ for
the orthogonal projection onto constants. A Markov operator $P$ with
transition probabilities $P(x,x')$ acts on functions by
$(Pf)(x)=\sum_{x'}P(x,x')f(x')$. It is reversible with respect to $\pi$
if it is self-adjoint in $L^2(\pi)$, and positive semidefinite if
$\langle f,Pf\rangle\ge0$ for all $f$. A reversible operator has $\pi$ as a
stationary distribution, and $P\mathbf 1=\mathbf 1$. We say that $P$ has
second eigenvalue at most $\lambda$ if $\langle f,Pf\rangle\le\lambda\|f\|^2$
for every $f$ orthogonal to $\mathbf 1$, and we call $1-\lambda$ the
spectral gap of $P$.

The following lemma bounds the probability that a stationary random walk
never hits a ``good'' state. It is a standard variant of the hitting
property of random walks on expanders~\cite{AKS87,HLW06}. We include the
short proof, since we use it in several settings, with success
probabilities that depend on the state.

\begin{lemma}[Hitting lemma]
\label{lem:hitting}
Let $P$ be a reversible, positive semidefinite Markov operator on $\Omega$
with stationary distribution $\pi$ and second eigenvalue at most
$\lambda<1$. Let $h:\Omega\to[0,1]$ with $\E_\pi[h]=\mu$. Let
$x_1,\ldots,x_t$ be a walk with $x_1\gets\pi$ and transitions given by
$P$, and suppose that at each visited state $x_i$, an attempt succeeds with
probability $h(x_i)$, independently of everything else given the walk.
Then the probability that all $t$ attempts fail is at most
\[
 (1-\mu)\bigl(1-(1-\lambda)\mu\bigr)^{t-1}
 \le\exp\bigl(-(1-\lambda)\mu(t-1)\bigr).
\]
\end{lemma}

\begin{proof}
Let $D$ be the operator of multiplication by $\sqrt{1-h}$. The probability
that all attempts fail is
\[
 \E\Bigl[\prod_{i=1}^t(1-h(x_i))\Bigr]
 =\bigl\langle D\mathbf 1,(DPD)^{t-1}D\mathbf 1\bigr\rangle.
\]
Indeed, expanding the right-hand side gives the sum, over all paths
$(x_1,\ldots,x_t)$, of the quantity
$\pi(x_1)\prod_{i\le t}(1-h(x_i))\prod_{i<t}P(x_i,x_{i+1})$. Therefore
\begin{equation}\label{eq:hitting-norm}
 \E\Bigl[\prod_{i=1}^t(1-h(x_i))\Bigr]
 \le\|D\mathbf 1\|^2\cdot\|DPD\|^{t-1}
 =(1-\mu)\cdot\|DPD\|^{t-1}.
\end{equation}
It remains to show that $\|DPD\|\le1-(1-\lambda)\mu$. The operator $DPD$
is self-adjoint and positive semidefinite, so the norm of $DPD$ is the
maximum of $\langle f,DPDf\rangle$ over unit vectors $f$.
Since $P$ is positive semidefinite with second eigenvalue at most
$\lambda$, we have $P\preceq\lambda\Id+(1-\lambda)\Pi$. Therefore, for every
unit vector $f$,
\[
 \langle f,DPDf\rangle
 =\langle Df,P\,Df\rangle
 \le\lambda\|Df\|^2+(1-\lambda)\langle Df,\mathbf 1\rangle^2
 \le\lambda+(1-\lambda)(1-\mu).
\]
The last inequality uses $\|Df\|^2=\E_\pi[(1-h)f^2]\le1$, and the
Cauchy--Schwarz inequality
$\langle Df,\mathbf 1\rangle^2=\E_\pi[\sqrt{1-h}\,f]^2\le\E_\pi[1-h]\cdot
\E_\pi[f^2]=1-\mu$. We obtain $\|DPD\|\le\lambda+(1-\lambda)(1-\mu)
=1-(1-\lambda)\mu$. Plugging this bound into~\eqref{eq:hitting-norm}
gives the first bound of the lemma, and the second bound follows from
$1-x\le e^{-x}$.
\end{proof}

Some of our reductions need more than one successful attempt, or need the
number of successful attempts to be close to the expected number $\mu t$.
For this, we use the following Chernoff bounds for random walks.
Chernoff bounds for random walks on expanders go back to
Gillman~\cite{Gil98}. We use the version of Le{\'o}n and
Perron~\cite{LP04}, which has explicit constants.

\begin{lemma}[Chernoff bounds for random walks]
\label{lem:chernoff}
Consider the setting of \cref{lem:hitting}, and let $Z$ be the number of
successful attempts among the $t$ attempts. Then the following holds.
\begin{enumerate}
 \item For every $\eta>0$,
 \[
  \Pr\bigl[Z\ge(\mu+\eta)t\bigr]\le\exp\Bigl(-\frac{2(1-\lambda)}{1+\lambda}\,\eta^2t\Bigr)
  \quad\text{and}\quad
  \Pr\bigl[Z\le(\mu-\eta)t\bigr]\le\exp\Bigl(-\frac{2(1-\lambda)}{1+\lambda}\,\eta^2t\Bigr).
 \]
 \item For every integer $k\ge1$,
 $\Pr[Z<k]\le\exp\bigl(k-(1-\lambda)\mu(t-1)/2\bigr)$.
\end{enumerate}
\end{lemma}

\begin{proof}
The proof is fairly standard. For the first item, we use the following
result of Le{\'o}n and Perron~\cite[Theorem~1]{LP04}. Let $(s_1,\ldots,s_t)$
be a walk of a finite, ergodic and reversible Markov chain, which starts
from the stationary distribution, and whose second eigenvalue is
$\lambda'\ge0$. Let $F$ be a function with values in $[0,1]$, whose mean
under the stationary distribution is $\nu$. Then, for every $\eta>0$ such
that $\nu+\eta<1$,
\begin{equation}\label{eq:leon-perron}
 \Pr\Bigl[\sum_{i=1}^tF(s_i)\ge(\nu+\eta)t\Bigr]
 \le\exp\Bigl(-\frac{2(1-\lambda')}{1+\lambda'}\,\eta^2t\Bigr).
\end{equation}
This bound also holds when $\nu+\eta\ge1$. Indeed, the probability is zero
when $\nu+\eta>1$, and the case $\nu+\eta=1$ follows by applying the bound
with $\eta'<\eta$ and letting $\eta'$ tend to $\eta$.

We apply this result to the chain whose states are the pairs $(x,b)$,
where $x\in\Omega$ is a state of the walk and $b\in\{0,1\}$ indicates
whether the attempt at $x$ succeeds. For $x\in\Omega$, let
$\rho(x,1)=h(x)$ and $\rho(x,0)=1-h(x)$. The chain moves from $(x,b)$ to
$(x',b')$ with probability
\[
 K\bigl((x,b),(x',b')\bigr)=P(x,x')\,\rho(x',b').
\]
The pairs $(x_1,b_1),\ldots,(x_t,b_t)$ formed by the states of the walk
and the outcomes of the attempts are a walk of this chain, which starts
from the distribution $\widetilde\pi(x,b)=\pi(x)\rho(x,b)$. The chain is
reversible with respect to $\widetilde\pi$, since
$\widetilde\pi(x,b)K((x,b),(x',b'))=\pi(x)P(x,x')\rho(x,b)\rho(x',b')$ and
$\pi(x)P(x,x')=\pi(x')P(x',x)$. In particular, $\widetilde\pi$ is
stationary.

We now bound the second eigenvalue of $K$. For a function $F$ of the pairs
$(x,b)$, let $\overline F(x)=\rho(x,0)F(x,0)+\rho(x,1)F(x,1)$. Then
$(KF)(x,b)=(P\overline F)(x)$ for every $(x,b)$, and therefore
\[
 \langle F,KF\rangle
 =\E_{(x,b)\gets\widetilde\pi}\bigl[F(x,b)\,(P\overline F)(x)\bigr]
 =\E_{x\gets\pi}\bigl[\overline F(x)\,(P\overline F)(x)\bigr]
 =\langle\overline F,P\overline F\rangle.
\]
If $F$ is orthogonal to the constant functions, then so is
$\overline F$, since $\E_\pi[\overline F]=\E_{\widetilde\pi}[F]$. Since
$P$ is positive semidefinite with second eigenvalue at most $\lambda$, we
obtain
\[
 0\le\langle F,KF\rangle\le\lambda\|\overline F\|^2\le\lambda\|F\|^2,
\]
where the last inequality uses
$\overline F(x)^2\le\rho(x,0)F(x,0)^2+\rho(x,1)F(x,1)^2$, by convexity of
the square function. Hence $K$ is positive semidefinite with second
eigenvalue $\lambda'\le\lambda$. In particular, $K$ has no eigenvalue
$-1$, and the eigenvalue $1$ is simple, so the chain is ergodic on the
support of $\widetilde\pi$. We apply~\eqref{eq:leon-perron} to the function
$F(x,b)=b$, whose mean is $\nu=\E_\pi[h]=\mu$, and whose sum over the walk
is $Z$. Since $(1-\lambda')/(1+\lambda')\ge(1-\lambda)/(1+\lambda)$, this
gives the first bound. The second bound follows by applying the same
argument to the function $1-F$, whose sum over the walk is $t-Z$.

For the second item, note that given the walk, the attempts are
independent, and the expectation of $e^{-b_i}$ is
$1-(1-e^{-1})h(x_i)$. Therefore
\[
 \E\bigl[e^{-Z}\bigr]=\E\Bigl[\prod_{i=1}^t\bigl(1-h'(x_i)\bigr)\Bigr],
 \qquad\text{where}\qquad h'=(1-e^{-1})h.
\]
The right-hand side is the probability that all attempts fail, when the
attempt at $x_i$ succeeds with probability $h'(x_i)$. The mean of $h'$ is
$(1-e^{-1})\mu\ge\mu/2$, so \cref{lem:hitting} gives
$\E[e^{-Z}]\le\exp(-(1-\lambda)\mu(t-1)/2)$. Finally, if $Z<k$, then
$e^{-Z}>e^{-k}$, so the Markov inequality gives
$\Pr[Z<k]\le e^k\,\E[e^{-Z}]$.
\end{proof}

\section{Hardness amplification for linear systems}
\label{sec:amplification}

In this section, we show that a solver for $\LINEQ_{n,q}$ with a small
success probability $\alpha$ yields a solver which, for almost all
matrices, succeeds with probability close to one on every right-hand side.
The reduction keeps the dimension $n$, makes $O(\alpha^{-1}\log(1/(\beta\delta)))$
oracle calls, and performs $O(n^2)$ additional operations per call. In the
terminology of~\cite{HS23}, this is a hardness self-amplification result:
weak average-case hardness of $\LINEQ_{n,q}$ implies strong average-case
hardness.

The idea is to re-randomize the matrix with a random walk whose steps take
$O(n^2)$ operations and preserve the solution of the system. A step multiplies the
system $(M,y)$ on the left by a random invertible matrix $R=I_n+uv^\T$,
with $u, v\in\F^n$. The new system $(RM,Ry)$ has the same solution,
and computing it takes $O(n^2)$ operations, since $R$ is a rank-one
update of the identity. The successive matrices are not independent, but
the walk has spectral gap at least $1/3$ (over every finite field of size at least 3). By the
hitting lemma (\cref{lem:hitting}), $O(\alpha^{-1})$ steps then suffice to
reach a matrix on which the oracle succeeds, with constant probability. Simultaneously, the right-hand side is
randomized using a standard self-reduction.

\subsection{Low-rank walks on matrices}
\label{sec:walks}

Let $P$ be the Markov operator on $\Mat$ which moves from $A$ to
$A+uv^\T$, for independent $u,v\gets\F^n$. We call it the additive walk,
and we compute the spectrum of $P$ with Fourier analysis. The transition
probabilities satisfy $P(A,A')=P(A',A)$, since $uv^\T$ and
$-uv^\T=(-u)v^\T$ have the same distribution. Hence $P$ is reversible with
respect to the uniform distribution on $\Mat$.

Fix a nontrivial additive character $\psi:\F\to\mathbb C^*$ (for prime
$q$, one can take $\psi(z)=e^{2i\pi z/q}$). For $C\in\Mat$, define
$\chi_C(A)=\psi(\operatorname{tr}(C^\T A))$. The $q^{n^2}$ functions
$\chi_C$ form an orthonormal basis of the space of complex-valued
functions on $\Mat$, for the inner product
$\langle f,g\rangle=\E_A[\overline{f(A)}g(A)]$. As we see below, the
functions $\chi_C$ form an eigenbasis of $P$.

\begin{lemma}[Spectrum of the additive walk]
\label{lem:additive-spectrum}
For every $C\in\Mat$, we have $P\chi_C=q^{-\rank C}\chi_C$. Consequently,
$P$ is positive semidefinite and has second eigenvalue at most
$1/q$.
\end{lemma}

\begin{proof}
The proof is a standard Fourier computation. Since
$\operatorname{tr}(C^\T uv^\T)=v^\T C^\T u$, we have
\[
 (P\chi_C)(A)=\E_{u,v}\bigl[\psi\bigl(\operatorname{tr}(C^\T(A+uv^\T))\bigr)\bigr]
 =\chi_C(A)\cdot\E_{u,v}\bigl[\psi(v^\T C^\T u)\bigr].
\]
Condition on $u$, and consider the cases:

\begin{enumerate}[label=\roman*., nosep]
\item $C^\T u\ne0$. Then, $v\mapsto v^\T C^\T u$ is a nonzero linear
form, so $v^\T C^\T u$ is uniform in $\F$ when $v$ is uniform. Since
$\sum_{z\in\F}\psi(z)=0$ for a nontrivial character, the inner expectation
over $v$ is then zero. 
\item $C^\T u=0$. Then, the expectation over $v$ is always one.
\end{enumerate}

Therefore
\[
 \E_{u,v}\bigl[\psi(v^\T C^\T u)\bigr]=\Pr_u[C^\T u=0]=q^{-\rank C},
\]
because the null space of $C^\T$ has dimension $n-\rank C$. The eigenvalues of
$P$ are nonnegative, so $P$ is positive semidefinite. 
\end{proof}

We now define the walk used by our reductions. It moves inside $\GL$ by
rank-one updates, and rejects the updates that would leave $\GL$.

\begin{definition}[Invertible walk]
\label{def:invertible-walk}
The invertible walk on $\GL$ is the following Markov chain. At a state
$M\in\GL$, sample $u, v\gets\F^n$. If $1+v^\T u\ne0$, move
to $(I_n+uv^\T)M$. Otherwise, we say the step is not accepted and stay at $M$. We denote by $K$ the
Markov operator of this walk.
\end{definition}

Applied to a linear system $(M,y)$, a step of the walk with an accepted
matrix $R=I_n+uv^\T$ produces the system $(RM,Ry)$. Both systems have the
same solution, since $(RM)^{-1}Ry=M^{-1}y$.

\begin{lemma}[Properties of the invertible walk]
\label{lem:gl-gap}
The invertible walk (over a finite field of size at least 3) satisfies the following properties.
\begin{enumerate}
 \item A step takes $O(n^2)$ operations, and the walk never leaves $\GL$.
 \item A step is accepted with probability at least $2/3$, which does not
 depend on the current state.
 \item The operator $K$ is reversible with respect to the uniform
 distribution on $\GL$, positive semidefinite, and has second eigenvalue
 at most $2/3$.
\end{enumerate}
\end{lemma}

\begin{proof}
We repeatedly use the well-known identity
$\det(I_a+XY)=\det(I_b+YX)$, valid for all $X\in\F^{a\times b}$ and
$Y\in\F^{b\times a}$.

For the first statement, the identity gives $\det(I_n+uv^\T) = 1 + v^\T u$.
So the update $R=I_n+uv^\T$ is invertible exactly when the step is
accepted, and in this case $RM$ is invertible. Computing $v^\T u$ takes
$O(n)$ operations. Computing $RM=M+u(v^\T M)$ takes $O(n^2)$ operations,
by first computing the row vector $v^\T M$.

For the second statement, the acceptance depends only on $u, v$ and not on $M$. Consider the probability $e$ that the step is not accepted:
\begin{align*}
  e &= \Pr_{u,v}[1+v^\T u = 0] \\ 
    &= \Pr_{u,v}[1+v^\T u = 0 \mid u = 0] \cdot \Pr[u = 0] + \Pr_{u,v}[1+v^\T u = 0 \mid u\neq0] \cdot \Pr[u\neq 0] \\
    &= (1-1/q^n) \cdot 1/q \\
    &\le 1/3.
\end{align*}

For the third statement, the main observation is that the invertible walk and the additive walk have the same distribution
in the event that the step is accepted. Indeed, for a fixed
$M\in\GL$, the vector $v'=M^\T v$ is uniform in $\F^n$ and
independent of $u$. Thus,
\[
 (I_n+uv^\T)M=M+uv^\T M=M+uv'^\T.
\]
Therefore, for $M\ne M'$ in $\GL$, the transition probability of the
invertible walk is $P(M,M')$, and the probability of staying at $M$ is
$P(M,M)+e$. In other words, $K=P_{\GL}+e\Id$, where $P_{\GL}$ is the
matrix $(P(M,M'))_{M,M'\in\GL}$ of transition probabilities of $P$,
restricted to $\GL$. This matrix is symmetric, so $K$ is reversible with respect to the uniform
distribution on $\GL$. It is also positive semidefinite, since $P_{\GL}$ is
a principal submatrix of the positive semidefinite matrix $P$.

It remains to bound the second eigenvalue. In this paragraph, we use
inner products with respect to the counting measure, which does not change
the ratio $\langle f,Kf\rangle/\|f\|^2$. Let $f:\GL\to\mathbb R$ with $\sum_{M\in\GL}f(M)=0$, and
let $\widetilde f$ be the extension of $f$ to $\Mat$ by zero outside $\GL$. Then
$\widetilde f$ is orthogonal to the constant functions on $\Mat$, and
$\langle f,P_{\GL}f\rangle=\langle\widetilde f,P\widetilde f\rangle$. By
\cref{lem:additive-spectrum},
\[
 \langle f, Kf\rangle
 =\langle\widetilde f,P\widetilde f\rangle+e\|f\|^2
 \le\bigl(1/q+e\bigr)\|f\|^2.
\]
By the second item, $e\le1/q \cdot (1-1/q^n)$, from which we obtain $1/q+e\le 2/q - 1/q^{n+1}$, which is at most $2/3$.
\end{proof}

Instead of following a walk, a natural alternate approach is to sample a 
single large-rank update.

\begin{definition}[Invertible update of rank $k$]
\label{def:rank-k-update}
Let $1\le k\le n$. An invertible update of rank $k$ is sampled as follows.
Sample $U,V\gets\F^{n\times k}$. If $\det(I_k+V^\T U)\ne0$, output
$R=I_n+UV^\T$, and otherwise output $\bot$. We denote by $e_k$ the
probability of outputting $\bot$, and by $K_k$ the Markov operator on $\GL$
which moves from $M$ to $RM$, where $R$ is an invertible update of rank $k$
conditioned on $R\ne\bot$.
\end{definition}

\begin{lemma}[Properties of updates of rank $k$]
\label{lem:rank-k-update}
Invertible updates of rank $k$ satisfy the following properties.
\begin{enumerate}
 \item If $R\ne\bot$, then $R$ is invertible. Given $M$, computing $RM$
 takes $O(n^2k^{\omega-2+o(1)})$ operations.
 \item The probability $e_k$ does not depend on $M$, and $1-e_k>1/4$.
 \item The operator $K_k$ is reversible with respect to the uniform
 distribution on $\GL$, positive semidefinite, and has second eigenvalue
 at most $\lambda_k=q^{-k}/(1-e_k)<4q^{-k}$.
\end{enumerate}
\end{lemma}

\begin{proof}
The proof is similar to that of \cref{lem:gl-gap}, with matrices
instead of row and column vectors. For the first statement, we again use the identity $\det(I_n+UV^\T)=\det(I_k+V^\T U)$.
We compute $RM=M+U(V^\T M)$ by block multiplication: both products split
into $O((n/k)^2)$ products of $k\times k$ matrices, padding the last blocks
if necessary. Each product takes $k^{\omega+o(1)}$ operations, for a total
of $O(n^2k^{\omega-2+o(1)})$ operations.

For the second statement, condition on the rank $\rho$ of the matrix $U$. When $\rho = 0$, the step is always accepted.
Otherwise, write $U = B_1 B_2$, where $B_1 \in \mathbb{F}^{n\times\rho}_q$ and $B_2 \in \mathbb{F}^{\rho\times k}_q$. 
Since $V^\T B_1$ is uniform in $\mathbb{F}^{k\times\rho}_q$ and $B_2$ has full row rank, $B_2 V^\T B_1$ is a uniform matrix in $\mathbb{F}^{\rho\times\rho}_q$.
Since $\det (I_k + V^\T B_1 B_2) = \det (I_{\rho} + B_2 V^\T B_1)$, the step is accepted with probability $\prod_{i = 1}^{\rho} (1-q^{-i})$,
which has the uniform bound $\prod_{i = 1}^{n} (1-q^{-i}) \ge \prod_{i = 1}^{n} (1-2^{-i}) > 1/4$, across every $\rho$, as required. 

For the third statement, fix $M\in\GL$. The matrix $V'=M^\T V$ is uniform and
independent of $U$, and $(I_n+UV^\T)M=M+\sum_{i=1}^ku_iv_i'^\T$, where $u_i$
and $v'_i$ are the columns of $U$ and $V'$. Hence the proposed matrix
follows the distribution $P^k$ of $k$ steps of the additive walk. A proposal is accepted exactly when it falls in $\GL$, so
\[
 K_k(M,M')=\frac{P^k(M,M')}{1-e_k}\qquad\text{for all }M,M'\in\GL.
\]
This matrix is symmetric and positive semidefinite, since $P^k$ is. By
\cref{lem:additive-spectrum}, the eigenvalues of $P^k$ on the
nonconstant characters are $q^{-k\rank C}\le q^{-k}$. As in the proof of
\cref{lem:gl-gap}, extending a function $f$ with mean zero on $\GL$ by zero
outside $\GL$ gives
$\langle f,K_kf\rangle\le q^{-k}\|f\|^2/(1-e_k)$.
\end{proof}

\subsection{Amplification}
\label{sec:lineq-amplification}

We now precisely state the main result of this section. We assume that
the success probability $\alpha$ of the oracle is known, since the
reduction uses it to set the number of steps.

\begin{theorem}[Hardness self-amplification for $\LINEQ$]
\label{thm:lineq}
Let $\Ocal$ be a time-$T$ algorithm that solves $\LINEQ_{n,q}$ with
probability at least $\alpha>0$. For all $0<\beta,\delta<1/2$, there is an
algorithm $\Bcal$ with the following properties.
\begin{itemize}
 \item $\Bcal$ makes $O(\alpha^{-1}\log(1/(\beta\delta)))$ calls to $\Ocal$,
 and runs in time
 \begin{equation}\label{eq:lineq-amplification}
  O\!\left(\frac{T+n^2}{\alpha}\log\frac1{\beta\delta}\right).
 \end{equation}
 \item $\Bcal$ never outputs a wrong solution.
 \item For all but a $\beta$ fraction of the matrices, $\Bcal$ succeeds with
 probability $1-\delta$ on every right-hand side:
 \[
  \Pr_{M\gets\GL}\Bigl[\forall y\in\F^n:\
  \Pr_{\Bcal}\bigl[\Bcal(M,y)=M^{-1}y\bigr]\ge1-\delta\Bigr]\ge1-\beta.
 \]
\end{itemize}
Moreover, for every $0<\varepsilon<1/2$, there is an algorithm which runs
in time $O((T+n^2)\alpha^{-1}\log(1/\varepsilon))$, never outputs a wrong
solution, and fails with probability at most $\varepsilon$ on input
$(M,y)$, where $M\gets\GL$ and $y$ is an arbitrary function of $M$.
The algorithm uses rank-one updates for $q>2$ and rank-two updates for
$q=2$, as explained in \cref{rem: field-two}.
\end{theorem}

\begin{figure}[ht]
\begin{mdframed}
\textbf{Algorithm $\Bcal^{\Ocal}(M,y)$.} Parameter: number of steps $t$.
\begin{enumerate}[leftmargin=*]
 \item Set $(M_1,y_1)=(M,y)$.
 \item For $j=1,\ldots,t$:
 \begin{enumerate}
  \item Sample $z\gets\F^n$ and compute $w=\Ocal(M_j,y_j+M_jz)$. If
  $M_jw=y_j+M_jz$, output $w-z$ and stop.
  \item For $q>2$, sample $u,v\gets\F^{n}$. If $1+v^\T u\ne0$, set
  $R=I_n+uv^\T$ and $(M_{j+1},y_{j+1})=(RM_j,Ry_j)$. Otherwise, set
  $(M_{j+1},y_{j+1})=(M_j,y_j)$.
  For $q=2$, instead sample $U,V\gets\F^{n\times2}$ and use
  $R=I_n+UV^\T$ when $\det(I_2+V^\T U)\ne0$; otherwise, stay put
  (\cref{rem: field-two}).
 \end{enumerate}
 \item Output $\bot$.
\end{enumerate}
\end{mdframed}
\caption{The amplified solver of \cref{thm:lineq}.}
\label{fig:amplified-solver}
\end{figure}

\begin{proof}
Set $c_q=3$ for $q>2$ and $c_2=8$.
We use the algorithm $\Bcal$ of \cref{fig:amplified-solver}, with
$t=1+\lceil c_q \cdot \alpha^{-1}\log(1/(\beta\delta))\rceil$. Each iteration makes
one oracle call and $O(n^2)$ other operations, which gives the running
time.

We first show that $\Bcal$ never outputs a wrong solution. All systems
$(M_j,y_j)$ have the same solution $x=M^{-1}y$, since each accepted update
is invertible. If the check of step~2a passes, then
$w=M_j^{-1}(y_j+M_jz)=x+z$, so the output $w-z$ equals $x$.

We now bound the failure probability. For $N\in\GL$, let
\[
 h(N)=\Pr_{c\gets\F^n,\,\Ocal}\bigl[\Ocal(N,c)=N^{-1}c\bigr]
\]
be the success probability of the oracle on the matrix $N$, for a uniform
right-hand side. By assumption, $\E_{N\gets\GL}[h(N)]\ge\alpha$. Observe that the matrix
trajectory can be sampled in advance, independently of the query
randomness. Then, conditioned on this trajectory, the right-hand sides
$y_j+M_jz$ are independent uniform vectors, since $M_j$ is invertible and
the mask $z$ is fresh at each iteration. With fresh oracle coins, the
success events are therefore conditionally independent, with probabilities
$h(M_j)$. Note that
the matrices $M_1,M_2,\ldots$ form a trajectory of the invertible walk
starting at $M$, and do not depend on $y$. Therefore, the probability
$f_t(M)$ that all $t$ iterations fail depends only on $M$, and not on $y$.

When $M\gets\GL$, the trajectory starts from the stationary distribution
of the walk. By \cref{lem:gl-gap,rem: field-two}, the spectral gap is at
least $1/c_q$. We can therefore apply \cref{lem:hitting} with
$\lambda=1-1/c_q$, which gives
\[
\E_{M\gets\GL}\bigl[f_t(M)\bigr]\le\exp\Bigl(-\frac{\alpha(t-1)}{c_q}\Bigr) \le\beta\delta.
\]
By the Markov inequality, $\Pr_M[f_t(M)>\delta]\le\beta$, which proves that the fraction of exceptional matrices is at most $\beta$.
For the final statement, we use the same algorithm with
$t=1+\lceil c_q \cdot \alpha^{-1}\log(1/\varepsilon)\rceil$, so that the above
expectation is at most $\varepsilon$.
\end{proof}

\begin{remark}[Dealing with $\mathbb{F}_2$]
\label{rem: field-two}
As stated, \cref{lem:gl-gap} applies only for $\mathbb{F}_q$ where $q > 2$.
Over $\mathbb{F}_2$, the bound we obtain for the spectral gap in \cref{lem:gl-gap} 
is exponentially small, which is not useful algorithmically. We remedy this by considering a walk consisting of rank two updates:
the linear system $(M, y)$ is sent to $(RM, Ry)$, where $U, V \gets \mathbb{F}^{n \times 2}_q$ and $R = I_n + UV^\T$, provided $R$ is invertible (otherwise, stay at $M$).
A calculation similar to that of \cref{lem:gl-gap} bounds the second
eigenvalue by $1/4+5/8=7/8$: the additive rank-two walk has second
eigenvalue at most $1/4$, and a proposal is rejected with probability at
most $5/8$. Thus this walk has spectral gap at least $1/8$.
We can thus extend \cref{thm:lineq} to matrices over $\mathbb{F}_2$,
using the rank-two walk and the appropriate spectral gap instead.
\end{remark}

We stress that the guarantee of \cref{thm:lineq} holds for every
right-hand side, but not for every matrix: some exceptional matrices are
unavoidable with this approach. Indeed, after $t$ steps, the rank-one walk only
visits matrices of the form $M'=R_t\cdots R_1M$. Since each matrix
$R_i-I_n$ has rank at most one, we can write $M'=M+XY^\T$ with
$X,Y\in\F^{n\times t}$. There are at most $q^{2tn}$ such matrices, which
is a fraction at most $q^{2tn-n^2}$ of $\Mat$. An oracle which fails on
all these matrices, and succeeds everywhere else, still has success
probability close to one when $t$ is much smaller than $n/2$. However, the
reduction fails on $M$ with this oracle. For the rank-two walk over
$\mathbb F_2$, we have $\rank(M'-M)\le2t$, and the same argument
gives at most $q^{4tn}$ reachable matrices. The obstruction therefore
remains when $t$ is much smaller than $n/4$.

\subsection{Updates of larger rank}
\label{sec:rank-k}

In \cref{thm:lineq}, the fraction of exceptional matrices is $\beta$,
and the number of oracle calls grows with $\log(1/\beta)$. This dependence
comes from first bounding the average failure probability by $\beta\delta$
using the hitting lemma, then applying the Markov inequality to obtain
failure probability at most $\delta$ outside a $\beta$ fraction of matrices.
We now give a reduction in which the number of calls does not depend on
the fraction of exceptional matrices. Instead of following a walk, it
restarts from $M$ at each attempt, and applies a single random invertible
update of rank $k$. The core observation is that a single update of rank
$k$ already makes the success probability of the oracle almost independent
of $M$, for all but a $64q^{-2k}/\alpha$ fraction of the matrices.

\begin{proposition}[Amplification with updates of rank $k$]
\label{prop:rank-k}
Let $\Ocal$ be a time-$T$ algorithm that solves $\LINEQ_{n,q}$ with
probability at least $\alpha>0$. For all $1\le k\le n$ and $0<\delta<1/2$,
there is an algorithm $\Bcal_k$ with the following properties.
\begin{itemize}
 \item $\Bcal_k$ makes at most $t=\lceil8\alpha^{-1}\log(1/\delta)\rceil$ calls to $\Ocal$,
 and runs in time
 \[
  O\!\left((T+n^2k^{\omega-2+o(1)})\alpha^{-1}\log(1/\delta)\right).
 \]
 \item $\Bcal_k$ never outputs a wrong solution.
 \item For all but a $64q^{-2k}/\alpha$ fraction of the matrices, $\Bcal_k$
 succeeds with probability $1-\delta$ on every right-hand side:
 \[
  \Pr_{M\gets\GL}\Bigl[\forall y\in\F^n:\
  \Pr_{\Bcal_k}\bigl[\Bcal_k(M,y)=M^{-1}y\bigr]\ge1-\delta\Bigr]
  \ge1-\frac{64}{\alpha q^{2k}}.
 \]
\end{itemize}
\end{proposition}

\begin{figure}[ht]
\begin{mdframed}
\textbf{Algorithm $\Bcal_k^{\Ocal}(M,y)$.} Parameters: rank $k$ and number of
attempts $t$.
\begin{enumerate}[leftmargin=*]
 \item For $j=1,\ldots,t$:
 \begin{enumerate}
  \item Sample an invertible update $R$ of rank $k$. If $R=\bot$, go to the
  next attempt.
  \item Sample $z\gets\F^n$ and compute $w=\Ocal(RM,R(y+Mz))$. If
  $M(w-z)=y$, output $w-z$ and stop.
 \end{enumerate}
 \item Output $\bot$.
\end{enumerate}
\end{mdframed}
\caption{The solver of \cref{prop:rank-k}.}
\label{fig:rank-k-solver}
\end{figure}

\begin{proof}
  Consider the algorithm described in \cref{fig:rank-k-solver} with $t=\lceil8\alpha^{-1} \log(1/\delta)\rceil$. When $R\ne\bot$, the matrix $RM$ is invertible, and the solution of the
queried system is $(RM)^{-1}R(y+Mz)=M^{-1}y+z$. Hence, if the check passes,
then $w-z=M^{-1}y$, and $\Bcal_k$ never outputs a wrong solution. By
\cref{lem:rank-k-update}, each attempt takes $O(n^2k^{\omega-2+o(1)})$
operations in addition to the call to $\Ocal$.

As in the proof of \cref{thm:lineq}, let $h(N)$ be the success probability
of $\Ocal$ on the matrix $N\in\GL$, for a uniform right-hand side, and let
$\mu=\E_{N\gets\GL}[h(N)]\ge\alpha$. For a fixed $R\ne\bot$, the
right-hand side $R(y+Mz)=Ry+RMz$ is uniform, since $RM$ is invertible.
Hence, an attempt succeeds with probability
\[
 (1-e_k)\cdot g(M),\qquad\text{where}\qquad
 g(M)=\E_R\bigl[h(RM)\bigr]=(K_kh)(M),
\]
independently of $y$ and of the other attempts.

We now show that $g(M)\ge\mu/2$ for all but a $64q^{-2k}/\alpha$ fraction of
the matrices. Since the uniform distribution on $\GL$ is stationary for
$K_k$, we have $\E_M[g(M)]=\mu$. The function $h-\mu$ is orthogonal to
constants, and $\|h-\mu\|^2=\E[h^2]-\mu^2\le\mu(1-\mu)$, since $0\le h\le1$.
By \cref{lem:rank-k-update}, the operator $K_k$ multiplies the norm of
functions orthogonal to constants by at most $\lambda_k<4q^{-k}$. Hence
\[
 \E_M\bigl[(g(M)-\mu)^2\bigr]=\|K_k(h-\mu)\|^2\le\lambda_k^2\,\mu(1-\mu)
 \le16q^{-2k}\mu.
\]
By the Chebyshev inequality, $\Pr_M[g(M)<\mu/2]\le16q^{-2k}\mu/(\mu/2)^2
=64q^{-2k}/\mu\le64q^{-2k}/\alpha$.

Finally, let $M$ be such that $g(M)\ge\mu/2\ge\alpha/2$. Since $1-e_k>1/4$,
each attempt succeeds with probability at least $\alpha/8$, so all $t$
attempts fail with probability at most
$(1-\alpha/8)^t\le\exp(-\alpha t/8)\le\delta$.
\end{proof}

\paragraph{How large can $k$ be?}
The number of oracle calls in \cref{prop:rank-k} does not depend on $k$.
Only the time to compute $RM=M+U(V^\T M)$ grows with $k$. With block
multiplication, it is $O(n^2k^{\omega-2+o(1)})$. With fast rectangular matrix
multiplication, the two products $V^\T M$ and $U(V^\T M)$ take
$n^{2+o(1)}$ operations when $k\le n^{\alpha_{\mathrm d}}$, where
$\alpha_{\mathrm d}>0.32$ is the dual exponent of matrix
multiplication~\cite{VXXZ24}. Hence, for $k\le\lceil n^{0.32}\rceil$, each call
requires only $n^{2+o(1)}$ additional operations. Taking
$k=\lceil n^{0.32}\rceil$ makes the exceptional fraction
$64q^{-2k}/\alpha$ at most $2^{-\Omega(n^{0.32})}/\alpha$, over
every finite field. More generally, the overhead per call is at most a
constant factor as long as computing $RM$ takes time $O(T)$. Even the
naive computation of $RM$, with $O(kn^2)$ operations, allows any
$k\le T/n^2$. Over exponentially large fields, $k=1$ already gives an
exponentially small exceptional fraction.

\paragraph{Limitation.}
Some exceptional matrices are unavoidable. All the queries of $\Bcal_k$
are made on matrices $RM$ with $\rank(RM-M)\le k$, and there are at most
$q^{2kn}$ such matrices, since $RM-M=XY^\T$ with $X,Y\in\F^{n\times k}$. Let
$Z\subseteq\GL$ be a set of $\lfloor(1-\alpha)q^{-2kn}|\GL|\rfloor$
matrices, and consider the oracle which fails on all matrices at rank
distance at most $k$ from $Z$, and succeeds everywhere else. This oracle
fails on at most $(1-\alpha)|\GL|$ invertible matrices, so it solves
$\LINEQ_{n,q}$ with probability at least $\alpha$. However, $\Bcal_k$ fails on
every matrix of $Z$. Hence, the exceptional fraction of \cref{prop:rank-k}
cannot be smaller than about $q^{-2kn}$ in general.

\section{Matrix multiplication and related problems}
\label{sec:mm}

The walks of \cref{sec:amplification} also yield reductions for matrix
multiplication and related problems. We first give a worst-case to
average-case reduction for matrix multiplication, with an overhead linear
in the inverse of the success probability of the average-case algorithm.
We then extend the approach to online matrix-vector multiplication in
\cref{sec:omv}, where the walk is performed during preprocessing, and to
the parity of the number of triangles in \cref{sec:triangles}.

Using the notation of \cref{def:matmult}, we show that a solver for
$\MM_{n,q}$ with success probability $\alpha$ yields a solver for
$\MM^{\mathrm{wc}}_{n,q}$ with success probability $1-\delta$, for every
$0<\delta<1/2$. The reduction makes
$O(\alpha^{-1}\log(1/\delta))$ oracle calls and performs $O(n^2)$
additional operations per call.
The idea is to extend the additive walk (\cref{lem:additive-spectrum}) to
pairs of matrices: sample independent vectors $u,v,u',v'\gets\F^n$ and
update $(A,B)$ to $(A+uv^\T,B+u'v'^\T)$. The product changes, but the
difference can be computed and stored efficiently:
\begin{equation}\label{eq:product-update}
 (A+uv^\T)(B+u'v'^\T)-AB=(Au')v'^\T+u(v^\T B)+u(v^\T u')v'^\T.
\end{equation}
As for $\LINEQ_{n,q}$, the hitting lemma (\cref{lem:hitting}) shows that
$O(\alpha^{-1})$ queries suffice to solve $\MM_{n,q}$ with probability
at least $0.97$. For linear systems, our reduction handles almost all
matrices. For matrix multiplication, the standard self-reduction
of~\cite{BLR93} extends the guarantee to every pair of matrices, using
independent uniform masks $X,Y\gets\Mat$ and the identity
\begin{equation}\label{eq:four-products}
 AB=(A+X)(B+Y)-(A+X)Y-X(B+Y)+XY.
\end{equation}

Unlike $\LINEQ_{n,q}$, a candidate product cannot be checked in $O(n^2)$ time using a direct calculation. We instead use the well-known
randomized test given by Freivalds~\cite{Fre77}.

\subsection{The reduction}
\label{app:mm-reduction}

We now state the theorem precisely, and assume, as before, that the success probability $\alpha$ is known.

\begin{theorem}[Worst-case to average-case reduction for matrix multiplication]
\label{thm:mm}
Let $\Ocal$ be a time-$T$ algorithm that solves $\MM_{n,q}$ with
probability at least $\alpha>0$. For every $0<\delta<1/2$, there is an algorithm which makes
$O(\alpha^{-1}\log(1/\delta))$ calls to $\Ocal$, runs in time
\[
 O\!\left(\frac{T+n^2}{\alpha}\log\frac1\delta\right),
\]
and solves $\MM^{\mathrm{wc}}_{n,q}$ with probability at least $1-\delta$.
All calls are made on pairs of $n\times n$
matrices over $\F$.
\end{theorem}

\begin{figure}[ht]
\begin{mdframed}
\textbf{Algorithm $\mathcal C^{\Ocal}(A_0,B_0)$.} Parameters: number of steps
$t$, number of tests $s=\lceil\log_q(100t)\rceil$, and test limit
$L=100(2t+s)$.
\begin{enumerate}[leftmargin=*]
 \item Set $E_0=0$.
 \item For $j=0,\ldots,t-1$:
 \begin{enumerate}
  \item Compute $C_j=\Ocal(A_j,B_j)$ and the candidate $D_j=C_j-E_j$.
  \item Repeat up to $s$ times: sample $w\gets\F^n$, and check that
  $A_0(B_0w)=D_jw$. If a check fails, stop testing $D_j$. If all $s$ checks
  pass, output $D_j$ and stop.
  \item Sample $u,v,u',v'\gets\F^n$. Set $A_{j+1}=A_j+uv^\T$ and
  $B_{j+1}=B_j+u'v'^\T$, and compute $E_{j+1}=E_j+(A_ju')v'^\T+u(v^\T B_j)
  +u(v^\T u')v'^\T$.
 \end{enumerate}
 \item Output $\bot$. Also output $\bot$ as soon as more than $L$ checks
 have been made in total.
\end{enumerate}
\textbf{Algorithm $\mathcal C_{\mathrm{wc}}^{\Ocal}(A,B)$.} Parameter: number of
repetitions $\ell$.
\begin{enumerate}[leftmargin=*]
 \item For $i=1,\ldots,\ell$: sample $X,Y\gets\Mat$, and compute
 $P_i=\mathcal C(A+X,B+Y)-\mathcal C(A+X,Y)-\mathcal C(X,B+Y)+\mathcal C(X,Y)$,
 with fresh coins for each call.
 \item Output the matrix which appears more than $\ell/2$ times among
 $P_1,\ldots,P_\ell$, or $\bot$ if there is none.
\end{enumerate}
\end{mdframed}
\caption{The reduction of \cref{thm:mm}. A call to $\mathcal C$ which
outputs $\bot$ is replaced by the zero matrix.}
\label{fig:mm}
\end{figure}

\begin{lemma}[Amplification on average]
\label{lem:mm-average}
Let $t=1+\lceil2\alpha^{-1}\log100\rceil$. The algorithm $\mathcal C$ of
\cref{fig:mm} makes at most $t$ calls to $\Ocal$, runs in time
$O((T+n^2)/\alpha)$, and solves $\MM_{n,q}$ with probability at least $0.97$.
\end{lemma}

\begin{proof}
We first show that the candidates are correct when the oracle is correct.
By~\eqref{eq:product-update}, we have $E_j=A_jB_j-A_0B_0$ for every $j$.
Hence, if $C_j=A_jB_j$, then $D_j=A_0B_0$. Each update of $E_j$ takes
$O(n^2)$ operations, since it only involves products of a matrix by a
vector, and outer products of two vectors.

\paragraph{Some oracle answer is correct.}
The pairs $(A_j,B_j)$ follow a random walk on $\Mat\times\Mat$, whose
operator is $P\otimes P$, where $P$ is the additive walk of
\cref{lem:additive-spectrum}. The eigenvectors of $P\otimes P$ are the functions
$(A,B)\mapsto\chi_C(A)\chi_{C'}(B)$, with eigenvalues
$q^{-\rank C-\rank C'}$. Hence, this walk is reversible with respect to the
uniform distribution, positive semidefinite, and has second eigenvalue
$1/q\le1/2$. Since $(A_0,B_0)$ is uniform, the walk starts from the
stationary distribution. Let $h(A,B)=\Pr_\Ocal[\Ocal(A,B)=AB]$, so that
$\E[h]\ge\alpha$. The oracle uses fresh coins at each call, so by
\cref{lem:hitting}, all the answers $C_0,\ldots,C_{t-1}$ are wrong with
probability at most
\[
 \exp\Bigl(-\frac{\alpha(t-1)}2\Bigr)\le\frac1{100}.
\]

\paragraph{Wrong candidates are rejected.}
If $D\ne A_0B_0$, then $(A_0B_0-D)w=0$ with probability at most $1/q$ for a
uniform vector $w$, since a nonzero row of $A_0B_0-D$ gives a nonzero linear
form in $w$. Hence a wrong candidate passes all $s$ checks with probability
at most $q^{-s}\le1/(100t)$. By a union bound over the at most $t$
candidates, the algorithm outputs a wrong candidate with probability at
most $1/100$.

\paragraph{Number of checks.}
A wrong candidate passes each check with probability at most $1/q\le1/2$,
independently, so the expected number of checks that it uses is at most
$\sum_{i\ge0}2^{-i}=2$. The algorithm stops after the first candidate which
passes all checks, and this candidate uses $s$ checks. Hence, the expected
total number of checks is at most $2t+s$, and by the Markov inequality, the
algorithm stops because of the limit $L$ with probability at most $1/100$.
Each check takes $O(n^2)$ operations, and $s=O(\log t)$, so the running time
is $O(t(T+n^2))=O((T+n^2)/\alpha)$.

Finally, assume that some answer is correct, that no wrong candidate passes
all the checks, and that the algorithm does not reach the limit $L$. Then
the algorithm does not stop before the first correct candidate, which
passes all the checks since it is correct, so the algorithm outputs
$A_0B_0$. Each of the three events above has probability at most $1/100$,
so the success probability is at least $0.97$.
\end{proof}

\begin{proof}[Proof of \cref{thm:mm}]
We use the algorithm $\mathcal C_{\mathrm{wc}}$ of \cref{fig:mm}, with
$\ell=\lceil4\log(1/\delta)\rceil$. The identity~\eqref{eq:four-products}
holds for all matrices $X$ and $Y$. For uniform $X,Y$, each of the four pairs
$(A+X,B+Y)$, $(A+X,Y)$, $(X,B+Y)$ and $(X,Y)$ is a pair of independent uniform
matrices. Hence, by \cref{lem:mm-average} and a union bound, $P_i=AB$ with
probability at least $1-4\cdot0.03=0.88$, independently for each $i$. By the
Hoeffding inequality, more than half of the matrices $P_i$ are equal to
$AB$, except with probability at most
$\exp(-2\ell(0.88-1/2)^2)\le\exp(-0.28\ell)\le\delta$. The majority element
can be found with $O(\ell n^2)$ operations, with the majority vote
algorithm of Boyer and Moore~\cite{BM91}, which makes $O(\ell)$
comparisons of matrices. The total number of
calls is $4\ell t=O(\alpha^{-1}\log(1/\delta))$, and the running time is
$O((T+n^2)\alpha^{-1}\log(1/\delta))$.
\end{proof}

\subsection{Comparison with previous reductions}
\label{app:mm-comparison}

\Cref{tab:mm} compares \cref{thm:mm} with previous worst-case to
average-case reductions for matrix multiplication, from algorithms which
compute the exact product with probability $\alpha$. We compare the
reductions from algorithms which only compute a fraction of the entries of
the product in \cref{app:mm-approx-comparison}. For a fair comparison, all
running times are given for a constant error probability, and we omit the
additive terms which do not depend on the oracle.

\begin{table}[ht]
\centering
{\small
\renewcommand{\arraystretch}{1.15}
\begin{tabular}{@{}>{\raggedright\arraybackslash}p{3.0cm}>{\raggedright\arraybackslash}p{4.3cm}>{\raggedright\arraybackslash}p{2.3cm}>{\raggedright\arraybackslash}p{5.2cm}@{}}
\toprule
Reduction & Domain & Success probability & Running time \\
\midrule
Self-correction \cite{BLR93,Lip91}, with the test of~\cite{Fre77} & finite rings & $\alpha>3/4$ & $O(T/(4\alpha-3))$ \\
\cite{AGGS22} & prime fields, $q\le2/\alpha$ & $\alpha$ & $2^{O(\log^5(1/\alpha))}\cdot T$ \\
\cite{AGGS22} & prime fields, $q\ge2/\alpha$ & $\alpha$ & $O(\alpha^{-4}\cdot T)$ \\
\cite[Thm.~1.6]{HS23} & finite rings & $\alpha$ &
 $O(\alpha^{-1}\log^2(1/\alpha)\log\log(1/\alpha)\cdot T)$ \\
\cite[Thm.~6.1]{VZ25} & prime fields, assuming trapdoored matrices & $\alpha$ & $\Otilde(T)$ \\
\Cref{thm:mm} & finite rings with identity & $\alpha$ & $O(\alpha^{-1}\cdot T)$ \\
\bottomrule
\end{tabular}\par}
\caption{Worst-case to average-case reductions for $n\times n$ matrix
multiplication over a ring with $q$ elements, from an algorithm running in
time $T\ge n^2$, which computes the product with probability $\alpha$ over
uniform inputs. The running times are for a constant error probability.}
\label{tab:mm}
\end{table}

\paragraph{Comparison with the reduction of Hirahara and Shimizu.}
The closest previous result is the reduction of~\cite{HS23}, which works
over every finite ring and runs in time
\[
 O\bigl(\alpha^{-1}T\log^2(1/\alpha)\log\log(1/\alpha)\bigr).
\]
This bound appears in Section~5.2 of the full version of~\cite{HS23}.
\Cref{thm:mm}, with the extension below, removes the factor
$\log^2(1/\alpha)\log\log(1/\alpha)$ over finite rings with identity.
We now explain where this factor comes from, and why our
approach avoids it.

The reduction of~\cite{HS23} splits $A$ into $b=\Theta(\log(1/\alpha))$
blocks of rows $A_1,\ldots,A_b$, and $B$ into $b$ blocks of columns
$B_1,\ldots,B_b$. A single product $AB$ then contains the $b$ products
$A_iB_i$, which are products of independent uniform rectangular matrices.
Hence, the average-case algorithm is also a direct-product algorithm for
$b$ independent rectangular products. A direct-product argument then gives
an algorithm which computes a single rectangular product with probability
$0.98$, with $O(\alpha^{-1})$ oracle calls. Computing the product of two
worst-case matrices requires the $b^2$ products $A_iB_j$, each of which is
repeated $O(\log b)$ times so that a union bound applies. This gives the
factor $b^2\log b=O(\log^2(1/\alpha)\log\log(1/\alpha))$.

The factor $\log\log(1/\alpha)$ could be removed by checking each block
with the test of Freivalds, instead of repeating the computation of every
block. A factor $\log^2(1/\alpha)$ remains in their analysis: it chooses
$b=\Theta(\log(1/\alpha))$, and each batch of $O(\alpha^{-1})$ oracle calls
only yields one of the $b^2$ blocks of the final product.

Our reduction never splits the matrices. Every query is a pair of
$n\times n$ matrices, which differs from the previous query by two
rank-one updates. Hence, by~\eqref{eq:product-update}, each answer gives a
candidate for the full product at the starting state, after a correction
which takes $O(n^2)$ operations. The amplification comes from the spectral
gap of the walk, which is a constant, rather than from the concentration
over $b$ independent blocks. Finally, a wrong candidate is rejected with a
constant number of tests of Freivalds in expectation, so that we do not
need to repeat the tests of all candidates for a union bound.

The key ingredient is the use of low-rank updates, whose effect on the
product can be corrected with few operations. These updates could also be
used without a walk, as in \cref{prop:rank-k}. A single random update of
rank $k=O(\log_q(1/\alpha))$ on both matrices gives a self-reduction with
one query, whose query distribution is a good sampler, by the argument of
the proof of \cref{prop:rank-k}. Plugged into the framework
of~\cite{HS23}, it would give a running time
$O(\alpha^{-1}(T+n^2\log_q(1/\alpha)))$. The walk removes the remaining
additive term $n^2\log_q(1/\alpha)$ per call, which matters when $T$ is
close to $n^2$, and gives an elementary proof.

\paragraph{Other reductions.}
The reductions of~\cite{AGGS22} work over prime fields, with a
quasi-polynomial dependence on $1/\alpha$ over small fields, and a
dependence $\alpha^{-4}$ over large fields. Vaikuntanathan
and Zamir~\cite{VZ25} obtain a reduction with an overhead
polylogarithmic in $n$, assuming the existence of trapdoored matrices,
which follows from variants of the learning parity with noise assumption.
They replace the uniform masks $X,Y$ of~\eqref{eq:four-products} by
pseudorandom matrices which can be multiplied quickly.

\paragraph{Extension to finite rings.}
Although stated over finite fields, \cref{thm:mm} extends to every finite
ring $R$ with identity, including noncommutative rings, with the same
asymptotic bounds. Both uses of the field structure admit simple
replacements. For a nontrivial additive character $\chi$ of $R^{n\times n}$,
the eigenvalue of the walk $A\mapsto A+uv^\T$ is
\[
 \lambda_\chi=\E_{u,v}[\chi(uv^\T)]
 =\Pr_u\bigl[\chi(uv^\T)=1\text{ for every }v\in R^n\bigr]\le\frac12.
\]
Indeed, the set of such $u$ is an additive subgroup of $R^n$. It is proper:
otherwise, $\chi$ would be trivial on every matrix $uv^\T$, hence on
every matrix with a single nonzero entry, and therefore on all matrices.
Thus the walk remains positive semidefinite with spectral gap at least
$1/2$. For verification, if $D\ne0$, the map $w\mapsto Dw$ is a nonzero
additive homomorphism, so $\Pr_w[Dw=0]\le1/2$. We therefore use
$s=\lceil\log_2(100t)\rceil$ Freivalds tests. The identities
\eqref{eq:product-update} and \eqref{eq:four-products} hold without
commutativity, and the rest of the proof is unchanged.

As in~\cite{HS23}, we assume that the success probability $\alpha$ is
known, in order to set the number of steps.

\subsection{Online matrix-vector multiplication}
\label{sec:omv}

In the online matrix-vector multiplication problem (OMv)~\cite{HKNS15}, an
algorithm first preprocesses a matrix $A$, and then answers queries $v$ by
computing $Av$. This problem is the basis of conditional lower bounds for
many dynamic problems~\cite{HKNS15}. Following~\cite{AGGS22,HS23,HS25a},
we consider the variant over $\F$, in which the matrix and the queries are
uniformly random.

\begin{definition}[Data structures for OMv]
\label{def:omv}
A data structure $\Dcal$ for OMv consists of two randomized algorithms:
\begin{enumerate}[label=\roman*., nosep]
  \item $\Dcal_{\mathrm{pre}}$ (the preprocessing algorithm): on input a matrix $A\in\Mat$, the algorithm outputs a state $\sigma$.
  \item $\Dcal_{\mathrm{query}}$ (the query algorithm): on input a state $\sigma$ and a vector $v \in \F^n$, the algorithm outputs a vector in $\F^n$, with fresh coins at each query.
\end{enumerate}

We write $\Dcal(A;v)$ for this output when
$\sigma\gets\Dcal_{\mathrm{pre}}(A)$.
The data structure has
preprocessing time $P$ and query time $T$ if these two algorithms run in
time $P$ and $T$. The data structure solves OMv with probability $\alpha$ if
$\Pr_{A,v,\Dcal}[\Dcal(A;v)=Av]\ge\alpha$, and it has expected agreement
$\gamma$ if $\E_{A,v,\Dcal}[\agr(\Dcal(A;v),Av)]\ge\gamma$. (Here,
$A\gets\Mat$ and $v\gets\F^n$ are independent, and for two vectors $c,c'$,
$\agr(c,c')$ is the fraction of the coordinates $i$ such that $c_i=c'_i$.)
\end{definition}

The state $\sigma$ is computed once, and it is not modified by the
queries. We therefore state our guarantees for a single query. A guarantee
for all the queries of a sequence of $n$ queries, fixed in advance, follows from a union
bound, with error probability $\delta/n$ for each query.

The matrix multiplication reduction extends to OMv, with an
additional feature: the walk is performed during preprocessing. Let
$A_0,\ldots,A_{t-1}$ be the states of the walk, where
$A_{j+1}=A_j+u_jw_j^\T$. We preprocess each state $A_j$ separately. Given a
query $v$, the answer at state $A_j$ gives a candidate for $A_0v$, since
\[
 A_0v=A_jv-\sum_{l<j}u_l\,(w_l^\T v).
\]
The corrections for all the states can be computed with $O(nt)$ operations
per query. To make the queries uniform, we add a random vector $s_j$ to
the query at state $A_j$, and we compute $A_js_j$ during preprocessing.
We present the resulting data structure in \cref{fig:omv}.
This allows us to amplify the success rate to a large constant, say $0.97$.
We then use a self-reduction, as for matrix multiplication, by choosing a
matrix $X$ at random, preprocessing $A+X$ and $X$, and computing the
difference. We extend this construction to data structures with small
expected agreement in \cref{sec:approximate-omv}.

\begin{figure}[ht]
\begin{mdframed}
\textbf{Data structure $\Ecal^{\Dcal}$.} Parameters: number of states $t$,
number of tests $m=\lceil\log_q(100t)\rceil$, and test limit
$L=100(2t+m)$.

\smallskip\noindent
\emph{Preprocessing, on input $A_0\in\Mat$.}
\begin{enumerate}[leftmargin=*]
 \item For $j=0,\ldots,t-2$, sample $u_j,w_j\gets\F^n$, and set
 $A_{j+1}=A_j+u_jw_j^\T$.
 \item For $j=0,\ldots,t-1$, sample $s_j\gets\F^n$, compute $A_js_j$, and
 run $\sigma_j\gets\Dcal_{\mathrm{pre}}(A_j)$.
 \item Sample $r_1,\ldots,r_m\gets\F^n$, and compute $r_i^\T A_0$ for all
 $i\in[m]$.
\end{enumerate}
\emph{Query, on input $v\in\F^n$.}
\begin{enumerate}[leftmargin=*]
 \item Set $e_0=0$.
 \item For $j=0,\ldots,t-1$:
 \begin{enumerate}
  \item Compute the candidate
  $c_j=\Dcal_{\mathrm{query}}(\sigma_j,v+s_j)-A_js_j-e_j$.
  \item For $i=1,\ldots,m$, check that $r_i^\T c_j=(r_i^\T A_0)v$. If a
  check fails, stop testing $c_j$. If all $m$ checks pass, output $c_j$
  and stop.
  \item If $j<t-1$, compute $e_{j+1}=e_j+u_j(w_j^\T v)$.
 \end{enumerate}
 \item Output $\bot$. Also output $\bot$ as soon as more than $L$ checks
 have been made in total.
\end{enumerate}
\end{mdframed}
\caption{Amplification on average for OMv.}
\label{fig:omv}
\end{figure}

\begin{lemma}[Amplification on average for OMv]
\label{lem:omv-average}
Let $\Dcal$ be a data structure with preprocessing time $P$ and query time
$T$, which solves OMv with probability $\alpha$. Set
$t=1+\lceil2\alpha^{-1}\log100\rceil$. Then $\Ecal^{\Dcal}$ has
preprocessing time $O(t(P+n^2))$ and query time $O(t(T+n))$. Moreover,
for every $v\in\F^n$, we have
$\Pr_{A_0,\Ecal}[\Ecal^{\Dcal}(A_0;v)=A_0v]\ge0.97$, where
$A_0\gets\Mat$.
\end{lemma}

\begin{proof}
Fix $v\in\F^n$. Since $e_j=(A_j-A_0)v$, the candidate $c_j$ is equal to
$A_0v$ whenever the answer at the state $A_j$ is correct. Set
$h(A)=\Pr_{z,\Dcal}[\Dcal(A;z)=Az]$, where $z\gets\F^n$ and the
probability includes preprocessing and query coins. Conditioned on the
matrix trajectory, the masks and all these coins are independent across
states. Thus the success events are conditionally independent, with
probabilities $h(A_j)$, because $v+s_j$ is uniform. Since $\E_A[h(A)]\ge\alpha$
and the stationary additive walk has second eigenvalue $1/q\le1/2$,
\cref{lem:hitting} bounds the probability of no correct candidate by
$\exp(-\alpha(t-1)/2)\le1/100$.

For this fixed query, all potential candidates are independent of the
verification vectors $r_i$. For each fixed wrong candidate, the probability
of passing all tests is at most $q^{-m}$, and the expected number of tests
up to the first failure (or the $m$-th test) is at most two.
The union bound and expected-test calculation of \cref{lem:mm-average}
therefore apply even though the vectors are reused: they require no
independence between candidates. The probability of accepting a wrong
candidate is at most $tq^{-m}\le1/100$, and the expected number of tests
before stopping is at most $2t+m$. The limit $L$ is reached with probability
at most $1/100$, giving success probability at least $0.97$.
Finally, $m=O(\log t)$ and $L=O(t)$. Preprocessing each state takes
$O(P+n^2)$ operations, and each query correction and verification test
takes $O(n)$ operations, giving the stated bounds.
\end{proof}

\begin{theorem}[Worst-case to average-case reduction for OMv]
\label{thm:omv}
Let $\Dcal$ be a data structure with preprocessing time $P$ and query time
$T$, which solves OMv with probability $\alpha$, and let $0<\delta<1/2$.
Then there is a data structure $\Ecal_{\mathrm{wc}}$ with preprocessing time
$O((P+n^2)\alpha^{-1}\log(1/\delta))$ and query time
$O((T+n)\alpha^{-1}\log(1/\delta))$, such that
$\Pr[\Ecal_{\mathrm{wc}}(A;v)=Av]\ge1-\delta$ for all $A\in\Mat$ and
$v\in\F^n$.
\end{theorem}

\begin{proof}
Fix $A,v$. Preprocess $A+X$ and $X$ with $\Ecal^{\Dcal}$ using independent
coins and a uniform $X\gets\Mat$, and subtract their answers to $v$, replacing
$\bot$ by zero. Both matrices are uniform, so by \cref{lem:omv-average}
and a union bound, their difference is equal to
$(A+X)v-Xv=Av$ with probability at least $0.94$.
Now, repeat this construction independently
$\ell=\lceil4\log(1/\delta)\rceil$ times and output the strict majority
vector, or $\bot$ if there is none. As in the proof of \cref{thm:mm},
the failure probability is at most
$\exp(-2\ell(0.94-1/2)^2)\le\delta$, and finding the majority takes
$O(n\ell)$ operations. Multiplying the preprocessing and query bounds of
\cref{lem:omv-average} by $2\ell$ gives the stated running times.
\end{proof}

Asadi, Golovnev, Gur and Shinkar~\cite{AGGS22} give a worst-case to
average-case reduction for OMv, whose query time is
$\poly(1/\alpha)$ times larger, and whose preprocessing time is
$2^{O(\log^4(1/\alpha))}$ times larger. The reduction of~\cite{HS23} makes
$\alpha^{-1}\polylog(1/\alpha)\cdot O(\log n)$ queries to the average-case
data structure, as restated in the full version of~\cite{HS25a}.
\Cref{thm:omv} makes $O(\alpha^{-1})$ queries, for a constant error
probability.

\subsection{Parity of triangles}
\label{sec:triangles}

We next apply the walk to the parity of the number of triangles in a
random graph, studied in~\cite{BBB19,HS22,HS23,SY26}.

We identify a graph on the vertex set $[n]$ with the adjacency matrix
$A\in\Ftwo^{n\times n}$ of this graph, which is symmetric with zero
diagonal. We denote
by $\Scal_n$ the set of such matrices. A uniformly random matrix
$A\gets\Scal_n$ corresponds to the Erd{\H{o}}s--R{\'e}nyi random graph
$G(n,1/2)$, in which every edge is present independently with probability
$1/2$. The parity of the number of triangles of $A$ is
\[
 \tau(A)=\sum_{1\le a<b<c\le n}A_{ab}A_{bc}A_{ac}\in\Ftwo.
\]
We consider the distributional problem $\TRI_n$, in which the input is
$A\gets\Scal_n$, and the goal is to output $\tau(A)$. We define the walk we use in the lemma below. 
The proof is similar to the Fourier computation in the proof of
\cref{lem:additive-spectrum}.

\begin{lemma}[A walk on graphs]
\label{lem:graph-walk}
Consider the Markov chain on $\Scal_n$ which moves from $A$ to
$A+uv^\T+vu^\T$, for independent $u,v\gets\Ftwo^n$. This chain never
leaves $\Scal_n$. It is reversible with respect to the uniform
distribution on $\Scal_n$, positive semidefinite, and has second
eigenvalue at most $1/4$.
\end{lemma}

\begin{proof}
The increment $uv^\T+vu^\T$ is symmetric with zero diagonal and is its own
negative, so the walk stays in $\Scal_n$ and is reversible under the
uniform distribution. Index the characters of $\Scal_n$ by strictly
upper-triangular matrices $C$, with
$\chi_C(A)=(-1)^{\sum_{a<b}C_{ab}A_{ab}}$. As in
\cref{lem:additive-spectrum}, these characters form an eigenbasis, with
eigenvalues
\[
 \lambda_C=\E_{u,v}\bigl[(-1)^{u^\T(C+C^\T)v}\bigr]
 =\Pr_v[(C+C^\T)v=0]=2^{-\rank(C+C^\T)}.
\]
All eigenvalues are nonnegative. If $C\ne0$, the symmetric matrix
$C+C^\T$ has zero diagonal and a nonzero off-diagonal entry, giving an
invertible $2\times2$ principal minor. Thus its rank is at least two,
and $\lambda_C\le1/4$.
\end{proof}

After a step of this walk, the change of the parity of the number of
triangles can be computed with $O(n^2)$ operations, as follows.

\begin{lemma}[Update of the parity of triangles]
\label{lem:triangle-update}
Let $A\in\Scal_n$, let $u,v\in\Ftwo^n$, and let $\Delta=uv^\T+vu^\T$. For
$a\in[n]$, let $\deg_A(a)=\sum_bA_{ab}$ be the degree of vertex $a$. For
$w\in\Ftwo^n$, let $e_A(w)\in\Ftwo$ be the parity of the number of edges
of $A$ whose two endpoints are in the support of $w$. For $s\in\Ftwo^2$,
let $n_s$ be the number of indices $a\in[n]$ such that $(u_a,v_a)=s$.
Then, over $\Ftwo$,
\begin{align*}
 \tau(A+\Delta)-\tau(A)={}&(Au)^\T(Av)+\sum_{a=1}^nu_av_a\deg_A(a)
 +(u^\T v)(u^\T Av)\\
 &+(v^\T v)\,e_A(u)+(u^\T u)\,e_A(v)+n_{01}n_{10}n_{11}.
\end{align*}
In particular, $\tau(A+\Delta)-\tau(A)$ can be computed with $O(n^2)$
operations.
\end{lemma}

\begin{proof}
We expand the product $(A_{ab}+\Delta_{ab})(A_{bc}+\Delta_{bc})(A_{ac}+\Delta_{ac})$
in the definition of $\tau(A+\Delta)$. The terms with three factors from
$A$ sum to $\tau(A)$, and the terms with three factors from $\Delta$ sum to
$\tau(\Delta)$. A term with exactly one factor from $\Delta$ looks like $\Delta_{ab}A_{ac}A_{bc}$. Grouping
these terms by the edge $\{a,b\}$, we obtain
\[
 L:=\sum_{a<b}\Delta_{ab}\sum_{c\notin\{a,b\}}A_{ac}A_{cb}
 =\sum_{a<b}\Delta_{ab}\,(A^2)_{ab},
\]
where the last equality uses that $A$ has zero diagonal, so that the
terms $c=a$ and $c=b$ of $(A^2)_{ab}$ vanish. In the same way, the terms
with exactly two factors from $\Delta$ sum to
$Q:=\sum_{a<b}A_{ab}(\Delta^2)_{ab}$. Therefore
$\tau(A+\Delta)=\tau(A)+L+Q+\tau(\Delta)$, and it remains to compute $L$,
$Q$ and $\tau(\Delta)$.

For $L$, we replace $\Delta_{ab}$ by $u_av_b+u_bv_a$. Since $A^2$ is
symmetric, we obtain
\[
 L=\sum_{a<b}u_av_b(A^2)_{ab}+\sum_{a<b}u_bv_a(A^2)_{ba}
 =\sum_{a\ne b}u_av_b(A^2)_{ab}
 =u^\T A^2v+\sum_{a=1}^nu_av_a(A^2)_{aa},
\]
where the last step uses $-1=1$ in $\Ftwo$. Moreover,
$u^\T A^2v=(Au)^\T(Av)$, and $(A^2)_{aa}=\sum_bA_{ab}^2=\deg_A(a)$.

For $Q$, expanding the square gives
\[
 \Delta^2=(v^\T u)(uv^\T+vu^\T)+(v^\T v)\,uu^\T+(u^\T u)\,vv^\T.
\]
As for $L$, we have
$\sum_{a<b}A_{ab}(uv^\T+vu^\T)_{ab}=\sum_{a\ne b}A_{ab}u_av_b=u^\T Av$,
since $A$ has zero diagonal. Moreover, $\sum_{a<b}A_{ab}u_au_b=e_A(u)$
and $\sum_{a<b}A_{ab}v_av_b=e_A(v)$. Hence
$Q=(u^\T v)(u^\T Av)+(v^\T v)\,e_A(u)+(u^\T u)\,e_A(v)$.

For $\tau(\Delta)$, let $s_a=(u_a,v_a)\in\Ftwo^2$. Then
$\Delta_{ab}=u_av_b+u_bv_a$ is the determinant of the $2\times2$ matrix
with columns $s_a$ and $s_b$. Hence $\Delta_{ab}=1$ if and only if $s_a$
and $s_b$ are linearly independent, that is, nonzero and distinct. Three
vectors of $\Ftwo^2$ are pairwise linearly independent if and only if
they are the three nonzero vectors $(0,1)$, $(1,0)$ and $(1,1)$.
Therefore, the triangles of $\Delta$ are the triples with one vertex of
each of these three types, and $\tau(\Delta)=n_{01}n_{10}n_{11}\bmod2$.

Finally, the vectors $Au$ and $Av$, the degrees, the scalar $u^\T Av$
and the numbers $n_s$ can be computed with $O(n^2)$ operations. The
integer $u^\T Au$, computed over the integers, is twice the number of
edges whose two endpoints are in the support of $u$, so $e_A(u)$ is the
parity of $(u^\T Au)/2$. The same holds for $e_A(v)$.
\end{proof}

We query the oracle along the walk and correct each answer using
\cref{lem:triangle-update}, so that it is a candidate for the parity at
the starting graph. If the oracle succeeds with probability at least
$1/2+\varepsilon$ on a uniform graph, each corrected answer has the same
mean success probability, since the walk starts from stationarity.
By \cref{lem:graph-walk}, the second eigenvalue is at most $1/4$, so
\cref{lem:chernoff} bounds the probability of at most $t/2$ correct answers
by $\exp(-6\varepsilon^2t/5)$. Thus, taking the most frequent corrected
answer succeeds with probability at least $1-\delta$ for the value of
$t$ below.

\begin{theorem}[Hardness self-amplification for the parity of triangles]
\label{thm:triangles}
Let $\Ocal$ be a time-$T$ algorithm such that
$\Pr_{A\gets\Scal_n,\Ocal}[\Ocal(A)=\tau(A)]\ge1/2+\varepsilon$, for some
$\varepsilon>0$. For every $0<\delta<1/2$, the algorithm $\Dcal$ of
\cref{fig:triangles} with $t=\lceil\varepsilon^{-2}\log(1/\delta)\rceil$
makes $t$ calls to $\Ocal$, runs in time
$O((T+n^2)\varepsilon^{-2}\log(1/\delta))$, and outputs $\tau(A)$ with
probability at least $1-\delta$ when $A\gets\Scal_n$. All calls are made
on graphs with $n$ vertices.
\end{theorem}

\begin{figure}[ht]
\begin{mdframed}
\textbf{Algorithm $\Dcal^{\Ocal}(A_0)$.} Parameter: number of steps $t$.
\begin{enumerate}[leftmargin=*]
 \item Set $s_0=0$.
 \item For $j=0,\ldots,t-1$:
 \begin{enumerate}
  \item Compute $b_j=\Ocal(A_j)+s_j$.
  \item Sample $u,v\gets\Ftwo^n$. Set $A_{j+1}=A_j+uv^\T+vu^\T$, and
  compute $s_{j+1}=s_j+\tau(A_j)+\tau(A_{j+1})$ with
  \cref{lem:triangle-update}.
 \end{enumerate}
 \item Output the most frequent value among $b_0,\ldots,b_{t-1}$, or $0$
 in case of a tie.
\end{enumerate}
\end{mdframed}
\caption{The algorithm of \cref{thm:triangles}. All operations on $s_j$
and $b_j$ are in $\Ftwo$.}
\label{fig:triangles}
\end{figure}

Hirahara and Shimizu~\cite{HS23} observe that, by a result of Shaltiel
and Viola~\cite{SV10}, an overhead of $\varepsilon^{-2}$ is necessary for
black-box hardness amplification of Boolean functions. The
dependence on $\varepsilon$ of the number of calls in \cref{thm:triangles}
matches this generic black-box lower bound.

\paragraph{Worst-case hardness.}
Boix-Adser{\`a}, Brennan and Bresler~\cite{BBB19} give a worst-case to
average-case reduction for the parity of the number of cliques in random
graphs, which tolerates a constant error probability. Combined with
\cref{thm:triangles}, it gives the following result:

\begin{corollary}[Worst-case to average-case reduction for the parity of triangles]
\label{cor:triangles-wc}
Let $\Ocal$ be an algorithm which, for every $n$, runs in time $T(n)$ and
satisfies $\Pr_{A\gets\Scal_n,\Ocal}[\Ocal(A)=\tau(A)]\ge1/2+\varepsilon(n)$,
where $\varepsilon(n)>0$. Then there is an algorithm which, for every $n$
and every $A\in\Scal_n$, runs in time
$O((T(3n)+n^2)\,\varepsilon(3n)^{-2})$ and outputs $\tau(A)$ with
probability at least $2/3$.
\end{corollary}

\begin{proof}
Theorem~2.9(2) of the full version of~\cite{BBB19}, with $k=3$ and $s=2$,
gives an absolute constant $0<\delta_0<1/2$ such that an algorithm with error
at most $\delta_0$ on $G(m,1/2)$ yields a worst-case triangle-parity
algorithm with error at most $1/3$ and running time
$O(T_{\mathrm{amp}}(3n)+n^2)$, where $T_{\mathrm{amp}}$ is the running
time of the average-case algorithm. By \cref{thm:triangles}, we can
reach this constant error in time
$T_{\mathrm{amp}}(m)=O((T(m)+m^2)\varepsilon(m)^{-2})$.
Substituting $m=3n$ gives the claimed bound.
\end{proof}

Hirahara and Shimizu~\cite[Theorem~1.4]{HS23} give a reduction from the
worst case of the parity of triangles to an algorithm with success
probability $1/2+\varepsilon$ on random graphs with $kn$ vertices, where
$k=O(\varepsilon^{-2}\log(1/\varepsilon))$, which runs in time
$O(T(kn)\,\varepsilon^{-2}\log(1/\varepsilon))$. Their reduction applies
to random graphs in which every edge is present with probability $p$, for
every constant $p$, but it is nonuniform. \Cref{cor:triangles-wc} gives a
uniform reduction for $p=1/2$, which only calls the average-case algorithm
on graphs with $3n$ vertices. For the number of triangles over the
integers, Shimizu and Yasunaga~\cite[Theorem~4.1]{SY26} give a nonuniform
reduction from algorithms with success probability $\alpha$ on random
graphs with $n(\log n)^{O(1)}$ vertices, for every constant edge
probability. Amireddy, Behera, Paraashar, Srinivasan and
Sudan~\cite{ABPSS25} give a uniform reduction for the edge probability
$1/2$, from algorithms with success probability $1-c$, for a small
constant $c$.

The same approach applies to the tripartite variant of the problem studied
in~\cite{HS22}. In a tripartite graph with three parts of size $n$ and
biadjacency matrices $A,B,C\in\Ftwo^{n\times n}$, the parity of the number
of triangles is $\Tr(ABC)$, where the trace is computed over $\Ftwo$. The
additive walk of \cref{sec:walks} applies to each of the three matrices,
and the update $\Tr((A+uv^\T)BC)-\Tr(ABC)=v^\T B(Cu)$ takes $O(n^2)$
operations.

\paragraph{Limitations.}
Our approach relies on the linear structure of $\Ftwo$. First, the walk
of \cref{lem:graph-walk} preserves the uniform distribution on $\Scal_n$,
so our results only apply to the random graph $G(n,1/2)$. Second, they
only apply to the parity of the number of triangles. Over the integers,
the change of the number of triangles involves the sum
$\sum_{a<b}A_{ab}\Delta_{ab}(A^2)_{ab}$, which counts the pairs formed by
a triangle of $A$ and an edge $\{a,b\}$ of this triangle such that
$\Delta_{ab}=1$. Over $\Ftwo$, this sum is multiplied by two and
vanishes. We do not know how to compute it
with $O(n^2)$ operations. Similarly, for cliques of size $\ell\ge4$, the
terms with one factor from $\Delta$ involve the number of cliques of size
$\ell-2$ in the common neighborhood of each pair of vertices, and we do
not know how to compute their sum with $O(n^2)$ operations.

\section{Exact computation from approximate answers}
\label{sec:approximate}
We extend the preceding reductions to algorithms whose answers are
only partially correct. We obtain exact solving for random linear
systems, and worst-case matrix multiplication and OMv, from algorithms
with expected agreement $1/q+\varepsilon$. These reductions combine the
matrix walks with additional randomizations of the oracle queries.

\subsection{Rerandomization and decoding}

An oracle may consistently recover the same coordinates and fail
on the others. We therefore permute the coordinates before each
query and undo the permutation afterwards. When the matrix is
uniform, this gives the same mean correctness probability at every
coordinate. The statement is averaged over the matrix; it need
not hold at each fixed matrix.

We must also control repeated wrong values. Simply multiplying
the right-hand side by a random nonzero scalar is insufficient:
the oracle may scale its error along with its input, so that
undoing the scaling leaves the error unchanged. For linear
systems, we hide the scalar with an additive mask. Writing $B$
for the matrix after permuting its columns, we query the oracle
on $c=gy+Bw$, where $g\gets\F^*$ and $w\gets\F^n$. Since $B$ is
invertible, $c$ is uniform for every fixed $g$. Thus, even after
fixing the queried system and the oracle's coins, $g$ remains
uniform. Undoing this hidden scaling distributes each nonzero
error uniformly among the nonzero field elements. We can then
recover the correct value by plurality: it only needs to occur
more often than each individual wrong value, rather than in a
majority of the answers.

The walk addresses the dependence of the oracle's reliability on
the matrix itself. Permutations alone need not change this
reliability, which may depend on global properties of the matrix.
Low-rank updates provide inexpensive changes throughout the
matrix: for example, $M\mapsto M+u(v^\T M)$ adds different
multiples of one linear combination of the rows to many rows.
Since the starting matrix is uniform, the walk is stationary.
Its spectral gap allows us to control the frequencies of oracle
errors along the trajectory, without requiring independent
matrices at successive queries.

\subsection{Approximate linear systems}
\label{sec:approximate-lineq}
We first consider algorithms that recover only a fraction of the
coordinates of the solution to a random linear system. A uniformly
random vector has expected agreement $1/q$ with the solution.
We show that, over every finite field, an algorithm with expected
agreement $1/q+\varepsilon$ yields an exact solver using
$O(\varepsilon^{-2}\log(n/\varepsilon))$ oracle calls. Over fields
of size $q\ge200/\gamma$, where $\gamma=1/q+\varepsilon$, the number
of calls improves to $O(\gamma^{-1}\log(n/\gamma))$.

\begin{theorem}[Exact solving from partial agreement]
\label{thm:approx-walk}
Let $\Ocal$ be a time-$T$ algorithm such that
\begin{equation}\label{eq:expected-agreement}
 \E_{M\gets\GL,\ c\gets\F^n,\ \Ocal}
 \left[\frac1n\cdot\Bigl|\bigl\{i\in[n]:\Ocal(M,c)_i=(M^{-1}c)_i\bigr\}\Bigr|\right]
 \ge\gamma,
\end{equation}
where $\gamma=1/q+\varepsilon$ for some $\varepsilon>0$. There is an
algorithm with the following properties.
\begin{itemize}
 \item It makes at most $m$ calls to $\Ocal$, where
 $m=O(\varepsilon^{-2}\log(n/\varepsilon))$, and
 $m=O(\gamma^{-1}\log(n/\gamma))$ when $q\ge200/\gamma$. It runs in
 time $O((T+n^2)m)$.
 \item It never outputs a wrong solution.
 \item It succeeds with probability at least $7/8$ on input $(M,y)$, where
 $M\gets\GL$ and $y$ is an arbitrary function of $M$.
\end{itemize}
Furthermore, for all $0<\beta,\delta<1/2$, multiplying the bounds on the number of calls
and the running time by $O(\log(1/(\beta\delta)))$ gives an algorithm
which, for all but a $\beta$ fraction of the matrices $M\in\GL$, succeeds
with probability at least $1-\delta$ on every right-hand side. Similarly, for
all $1\le k\le n$ and $0<\delta<1/2$, there is an algorithm which runs in
time $O(((T+n^2)m+n^2k^{\omega-2+o(1)})\log(1/\delta))$ and, for all but a
$74q^{-2k}$ fraction of the matrices $M\in\GL$, succeeds with probability
at least $1-\delta$ on every right-hand side. For $k=\lceil n^{0.32}\rceil$,
the term $n^2k^{\omega-2+o(1)}$ can be replaced by $n^{2+o(1)}$ (\cref{sec:rank-k}).
\end{theorem}

The condition $\gamma>1/q$ cannot be removed in general, since an
algorithm which outputs a uniformly random vector has expected agreement
exactly $1/q$.

\begin{figure}[ht]
\begin{mdframed}
\textbf{Algorithm $\Bcal_{\mathrm{approx}}^{\Ocal}(M,y)$.} Parameter: query
bound $m$.
\begin{enumerate}[leftmargin=*]
 \item If $m\ge n$, solve $Mx=y$ by Gaussian elimination, output the
 solution, and stop. Otherwise, set $(M_1,y_1)=(M,y)$.
 \item For $j=1,\ldots,m$:
 \begin{enumerate}
  \item Sample a uniform permutation matrix $P_j$, a scalar $g_j\gets\F^*$,
  and a vector $w_j\gets\F^n$.
  \item Set $B_j=M_jP_j$ and $c_j=g_jy_j+B_jw_j$. Compute
  $u_j=\Ocal(B_j,c_j)$ and the candidate
  $\widehat x_j=g_j^{-1}P_j(u_j-w_j)$.
  \item Take one step of the invertible walk on $(M_j,y_j)$ to obtain
  $(M_{j+1},y_{j+1})$, as in \cref{fig:amplified-solver}. For $q=2$, use the
  walk with updates of rank two of \cref{rem: field-two}.
 \end{enumerate}
 \item For each $i\in[n]$, let $\widehat x_i$ be the most frequent value
 among $\widehat x_{1,i},\ldots,\widehat x_{m,i}$, with ties broken
 arbitrarily.
 \item If $M\widehat x=y$, output $\widehat x$. Otherwise, output $\bot$.
\end{enumerate}
\end{mdframed}
\caption{The solver of \cref{thm:approx-walk}.}
\label{fig:approx-solver}
\end{figure}

\begin{proof}
We use the algorithm $\Bcal_{\mathrm{approx}}$ of \cref{fig:approx-solver}.
We set $m=\lceil30\varepsilon^{-2}\log(16nq)\rceil$ when $q<200/\gamma$.
When $q\ge200/\gamma$, we set
\[
 m=1+\lceil32k/\gamma\rceil,
 \qquad\text{where}\qquad
 k=1+\lceil\log_2(1500n/\gamma)\rceil.
\]
In the second case, $m=O(\gamma^{-1}\log(n/\gamma))$, which is also
$O(\varepsilon^{-2}\log(n/\varepsilon))$ since $\varepsilon\le\gamma$. In
the first case, $q<200/\gamma\le200/\varepsilon$, so that
$m=O(\varepsilon^{-2}\log(n/\varepsilon))$.

If $m\ge n$, Gaussian elimination always returns the correct solution in
$O(n^3)\subseteq O(mn^2)$ operations, without any oracle calls. Hence we
may assume $m<n$. The final check ensures that $\Bcal_{\mathrm{approx}}$
never outputs a wrong solution. Each iteration takes one oracle call and
$O(n^2)$ other operations. Sorting the $m$ candidate values at each
coordinate, for any fixed total order on $\F$, takes $O(nm\log m)$
operations overall. Since $m<n$, this is dominated by $O(mn^2)$.

Let $x=M^{-1}y$. As in the proof of \cref{thm:lineq}, all systems
$(M_j,y_j)$ have solution $x$. We first compute the error of a candidate.
The solution of the system queried at iteration $j$ is
\[
 B_j^{-1}c_j=g_jP_j^{-1}M_j^{-1}y_j+w_j=g_jP_j^{-1}x+w_j.
\]
Therefore, the candidate $\widehat x_j$ satisfies
\begin{equation}\label{eq:candidate-error}
 \widehat x_j-x=g_j^{-1}v_j,
 \qquad\text{where}\qquad
 v_j=P_j\bigl(u_j-B_j^{-1}c_j\bigr).
\end{equation}
In particular, $\widehat x_{j,i}=x_i$ if and only if $v_{j,i}=0$. We
call $\widehat x_{j,i}$ a correct value if $v_{j,i}=0$, and a wrong value
otherwise. For $i\in[n]$, let $Z_i$ be the number of correct values at
coordinate $i$, and let $W_i=m-Z_i$ be the number of wrong values. For
$c\ne x_i$, let $N_{i,c}$ be the number of iterations $j$ such that
$\widehat x_{j,i}=c$. The decoding of step~3 outputs $x$ if $Z_i>N_{i,c}$
for every coordinate $i$ and every $c\ne x_i$. We first show that, except
with probability $1/16$, the numbers $Z_i$ are large. We then show that,
except with probability $1/16$, the numbers $N_{i,c}$ are small. This
proves the first statement, with success probability at least $7/8$.

\paragraph{Correct values.}
For $N\in\GL$ and $i\in[n]$, let
\[
 h_i(N)=\Pr_{P,c,\Ocal}\Bigl[\Bigl(P\bigl(\Ocal(NP,c)-(NP)^{-1}c\bigr)\Bigr)_i=0\Bigr],
\]
where $P$ is a uniform permutation matrix and $c\gets\F^n$. We first show
that $\E_{N\gets\GL}[h_i(N)]\ge\gamma$ for every $i$. Let $\pi$ be the
permutation such that $Pe_l=e_{\pi(l)}$ for all $l$, so that
$(Pv)_i=v_{\pi^{-1}(i)}$ for every vector $v$. When $N\gets\GL$, the matrix
$B=NP$ is uniform in $\GL$ and independent of $P$. Therefore, the event in the
definition of $h_i$ is the event that the coordinate $\pi^{-1}(i)$ of
$\Ocal(B,c)$ is correct, where $\pi^{-1}(i)$ is uniform in $[n]$ and
independent of $(B,c)$ and of the coins of $\Ocal$. Therefore,
$\E_N[h_i(N)]$ is exactly the expected fraction of correct coordinates
in~\eqref{eq:expected-agreement}, which is at least $\gamma$.

Conditioned on the matrix trajectory, the permutations, masked right-hand
sides, and oracle coins are independent across iterations: each $P_j$ is
fresh, and $c_j$ is uniform for every fixed $P_j$, since $w_j$ is uniform
and $B_j$ is invertible. Thus, for each fixed coordinate $i$, the events
$\widehat x_{j,i}=x_i$ are conditionally independent, with probabilities
$h_i(M_j)$. When $M\gets\GL$, the matrices
$M_1,\ldots,M_m$ form a stationary trajectory of the invertible walk. By
\cref{lem:gl-gap} (for $q\ge3$) and \cref{rem: field-two} (for $q=2$), the
walk has second eigenvalue at most $7/8$, so we can apply \cref{lem:chernoff} with $\lambda=7/8$,
for which $2(1-\lambda)/(1+\lambda)=2/15$, and with the success
probability $h_i$, whose mean is at least $\gamma$. When $q<200/\gamma$,
the first item of \cref{lem:chernoff} with $\eta=\varepsilon/2$ gives
\[
 \Pr\bigl[Z_i\le(\gamma-\varepsilon/2)m\bigr]
 \le\exp\Bigl(-\frac{\varepsilon^2m}{30}\Bigr)\le\frac1{16n}.
\]
When $q\ge200/\gamma$, we have $\gamma(m-1)/16\ge2k$ and
$k\ge\log(16n)$, so the second item of \cref{lem:chernoff} gives
\[
 \Pr[Z_i<k]\le\exp\Bigl(k-\frac{\gamma(m-1)}{16}\Bigr)\le e^{-k}
 \le\frac1{16n}.
\]
By a union bound over the $n$ coordinates, except with probability at
most $1/16$, we have $Z_i>(\gamma-\varepsilon/2)m$ for all $i$ in the first
case, and $Z_i\ge k$ for all $i$ in the second case. Note that this bound
holds for every right-hand side $y$, since the success probabilities
$h_i(M_j)$ do not depend on $y$.

\paragraph{Wrong values.}
Fix the trajectory of the walk, all permutations $P_j$, all right-hand
sides $c_j$, and the coins of $\Ocal$. We claim that, conditioned on these
values, the scalars $g_1,\ldots,g_m$ are still independent and uniform in
$\F^*$. Indeed, the randomness of the walk is independent of the pairs
$(g_j,w_j)$, and for each value of $g_j$, the vector $c_j=g_jy_j+B_jw_j$ is
uniform in $\F^n$, since $w_j$ is uniform and $B_j$ is invertible. In other words,
$g_j$ and $c_j$ are independent, and conditioning on $c_j$ does not change
the distribution of $g_j$.

After this conditioning, the vectors $v_j$ of~\eqref{eq:candidate-error}
are fixed, since they only depend on $P_j$, $B_j$, $c_j$, and the coins of
$\Ocal$. Hence, the numbers $Z_i$ and $W_i$ are also fixed. At coordinate
$i$, the wrong values are $x_i+v_{j,i}/g_j$, for the $W_i$ iterations $j$
such that $v_{j,i}\ne0$. For each such $j$, the value $v_{j,i}/g_j$ is
uniform in $\F^*$, since $g_j$ is uniform in $\F^*$. Therefore, the $W_i$
wrong values at coordinate $i$ are independent and uniform in the set
$x_i+\F^*$, which contains the $q-1$ values different from $x_i$. In
particular, for every $c\ne x_i$, the number $N_{i,c}$ follows a binomial
distribution with $W_i$ trials and success probability $1/(q-1)$.

When $q<200/\gamma$, the Hoeffding inequality gives, for every $c\ne x_i$,
\[
 \Pr\Bigl[N_{i,c}\ge\frac{W_i}{q-1}+\frac{\varepsilon m}2\Bigr]
 \le\exp\Bigl(-\frac{2(\varepsilon m/2)^2}{W_i}\Bigr)
 \le\exp\Bigl(-\frac{\varepsilon^2m}2\Bigr),
\]
since $W_i\le m$. A union bound over the $n$ coordinates and the $q-1$
values $c$ shows that $N_{i,c}<W_i/(q-1)+\varepsilon m/2$ for all $i$
and $c$, except with probability at most
$nq\exp(-\varepsilon^2m/2)\le1/16$.

When $q\ge200/\gamma$, a value $c\ne x_i$ appears at least $k$ times at
coordinate $i$ with probability at most $\binom{W_i}k(q-1)^{-k}$, by a
union bound over the sets of $k$ iterations. A union bound over the $q-1$
values $c$, and the inequalities $\binom{W_i}k\le(eW_i/k)^k$ and
$W_i\le m$, show that some wrong value appears at least $k$ times at
coordinate $i$ with probability at most
\[
 \Bigl(\frac{em}k\Bigr)^k(q-1)^{1-k}
 \le(q-1)\Bigl(\frac{93}{\gamma(q-1)}\Bigr)^k
 \le\frac{93}\gamma\cdot2^{1-k}
 \le\frac1{16n}.
\]
For the first inequality, we used $m\le2+32k/\gamma\le34k/\gamma$, so
that $em/k\le93/\gamma$. For the second inequality, we wrote
$(q-1)r^k=(93/\gamma)\,r^{k-1}$ for $r=93/(\gamma(q-1))$, and we used
$q-1\ge199/\gamma$, so that $r\le1/2$. The last inequality follows from
our choice of $k$. A union bound over the $n$ coordinates
shows that $N_{i,c}<k$ for all $i$ and $c$, except with probability at
most $1/16$.

\paragraph{Decoding.}
Assume that the events of the two previous paragraphs hold. When
$q\ge200/\gamma$, we have $N_{i,c}<k\le Z_i$ for all $i$ and $c\ne x_i$.
When $q<200/\gamma$, we use $W_i=m-Z_i<(1-\gamma+\varepsilon/2)m$ and
$\gamma=1/q+\varepsilon$, which give
\[
 N_{i,c}<\frac{W_i}{q-1}+\frac{\varepsilon m}2
 <\Bigl(\frac1q-\frac{\varepsilon}{2(q-1)}+\frac\varepsilon2\Bigr)m
 \le\Bigl(\gamma-\frac\varepsilon2\Bigr)m<Z_i.
\]
In both cases, $x_i$ is the unique most frequent value at every
coordinate $i$, and the decoding of step~3 outputs $x$.

\paragraph{Amplification.}
The algorithm $\Bcal_{\mathrm{approx}}$ solves $\LINEQ_{n,q}$ with
probability at least $7/8$, and never outputs a wrong solution. The second
statement follows by applying \cref{thm:lineq} to $\Bcal_{\mathrm{approx}}$,
with $\alpha=7/8$. The third statement follows by applying
\cref{prop:rank-k} to $\Bcal_{\mathrm{approx}}$, with $\alpha=7/8$, since
$64/(7/8)<74$.
\end{proof}

\subsection{Approximate matrix multiplication}
\label{app:mm-approx}
We now consider approximate algorithms for $\MM_{n,q}$ (\cref{def:matmult}) which only compute a fraction of
the entries of the product correctly. For $C,C'\in\Mat$, let
$\agr(C,C')$ be the fraction of the entries $(a,b)\in[n]^2$ such that
$C_{ab}=C'_{ab}$. We say that an algorithm $\Ocal$ has expected agreement
$\gamma$ if
\[
 \E_{A,B,\Ocal}\bigl[\agr(\Ocal(A,B),AB)\bigr]\ge\gamma,
\]
where $A,B\gets\Mat$ are independent. The algorithm which outputs the
zero matrix has expected agreement slightly larger than $1/q$, so we
consider expected agreements $\gamma=1/q+\varepsilon$, for $\varepsilon>0$.
As in \cref{app:mm-reduction}, we assume that $\varepsilon$ is known.

The reduction follows the walk of \cref{fig:mm}, and adds a
re-randomization similar to the one in \cref{sec:approximate-lineq}. At each
step, we permute the rows of $A_j$ and the columns of $B_j$, and we
multiply them by random nonzero scalars. Vaikuntanathan and
Zamir~\cite[Theorem~6.2]{VZ25} use the same re-randomization, together
with trapdoored matrices. The permutations spread the
correct entries of the oracle evenly over all entries of the product, and
the scalars spread the wrong entries evenly over all the wrong values.
After the correction of~\eqref{eq:product-update}, each answer of the
oracle gives a candidate for the product at the starting state. We then
recover each entry as the most frequent value among the candidates.

\begin{figure}[ht]
\begin{mdframed}
\textbf{Algorithm $\mathcal C_{\mathrm{apx}}^{\Ocal}(A_0,B_0)$.} Parameter:
number of steps $t$.
\begin{enumerate}[leftmargin=*]
 \item Set $E_0=0$.
 \item For $j=0,\ldots,t-1$:
 \begin{enumerate}
  \item Sample two permutations $\sigma,\pi$ of $[n]$ and two vectors
  $d,d'\gets(\F^*)^n$. Compute the matrices $X,Y\in\Mat$ defined by
  $X_{il}=d_i(A_j)_{\sigma(i)l}$ and $Y_{lk}=(B_j)_{l\pi(k)}d'_k$, for all
  $i,k,l\in[n]$.
  \item Compute $C_j=\Ocal(X,Y)$, and the candidate $D_j\in\Mat$ defined by
  \[
   (D_j)_{\sigma(i)\pi(k)}=\frac{(C_j)_{ik}}{d_id'_k}-(E_j)_{\sigma(i)\pi(k)},
   \qquad\text{for all }i,k\in[n].
  \]
  \item Sample $u,v,u',v'\gets\F^n$. Set $A_{j+1}=A_j+uv^\T$ and
  $B_{j+1}=B_j+u'v'^\T$, and compute $E_{j+1}=E_j+(A_ju')v'^\T+u(v^\T B_j)
  +(v^\T u')\,uv'^\T$.
 \end{enumerate}
 \item Output the matrix $D$ such that, for all $(a,b)\in[n]^2$, the entry
 $D_{ab}$ is the most frequent value among
 $(D_0)_{ab},\ldots,(D_{t-1})_{ab}$, with ties broken arbitrarily.
\end{enumerate}
\end{mdframed}
\caption{Amplification on average for approximate matrix multiplication.
The algorithm of \cref{thm:mm-approx} is the algorithm
$\mathcal C_{\mathrm{wc}}$ of \cref{fig:mm}, in which the calls to
$\mathcal C$ are replaced by calls to $\mathcal C_{\mathrm{apx}}$.}
\label{fig:mm-approx}
\end{figure}

\begin{lemma}[Approximate amplification on average]
\label{lem:mm-approx-average}
Let $\Ocal$ be a time-$T$ algorithm with expected agreement
$1/q+\varepsilon$, for some $\varepsilon>0$, and let
$t=\lceil6\varepsilon^{-2}\log(100n^2q)\rceil$. The algorithm
$\mathcal C_{\mathrm{apx}}$ of \cref{fig:mm-approx} makes $t$ calls to
$\Ocal$, runs in time $O(t(T+n^2\log t))$, and outputs $A_0B_0$ with
probability at least $0.99$, when $A_0,B_0\gets\Mat$ are independent.
\end{lemma}

\begin{proof}
Each iteration makes one oracle call and $O(n^2)$ other operations, as in
the proof of \cref{lem:mm-average}. In step~3, we find the most frequent
value of each entry by sorting the $t$ candidate values, for any fixed
total order on $\F$. This takes $O(t\log t)$ operations for each of the
$n^2$ entries, which gives the running time.

\paragraph{Error of the candidates.}
For all $i,k\in[n]$, we have
$(XY)_{ik}=\sum_ld_i(A_j)_{\sigma(i)l}(B_j)_{l\pi(k)}d'_k
=d_id'_k\,(A_jB_j)_{\sigma(i)\pi(k)}$. Let $W=C_j-XY$ be the error of the
answer of the oracle. As in the proof of \cref{lem:mm-average}, we have
$E_j=A_jB_j-A_0B_0$. Therefore
\begin{equation}\label{eq:mm-candidate-error}
 (D_j-A_0B_0)_{\sigma(i)\pi(k)}=\frac{W_{ik}}{d_id'_k},
 \qquad\text{for all }i,k\in[n].
\end{equation}
In particular, the error $D_j-A_0B_0$ of the candidate only depends on
$(A_j,B_j)$ and on the randomness of iteration $j$, that is, on
$\sigma$, $\pi$, $d$, $d'$ and on the coins of $\Ocal$.

\paragraph{Distribution of the errors.}
The core of the proof is the following observation. For $(a,b)\in[n]^2$,
$c\in\F$ and $A,B\in\Mat$, let $h_{abc}(A,B)$ be the probability that
$(D_j-A_0B_0)_{ab}=c$ when $(A_j,B_j)=(A,B)$, over the randomness of
iteration $j$. Let $\gamma_0\ge1/q+\varepsilon$ be the expected agreement
of $\Ocal$. We claim that, when $A,B\gets\Mat$ are independent,
\begin{equation}\label{eq:mm-error-distribution}
 \E\bigl[h_{ab0}(A,B)\bigr]=\gamma_0
 \qquad\text{and}\qquad
 \E\bigl[h_{abc}(A,B)\bigr]=\frac{1-\gamma_0}{q-1}\quad\text{for }c\ne0.
\end{equation}
Indeed, for fixed $\sigma,\pi,d,d'$, the maps $A\mapsto X$ and $B\mapsto Y$
are bijections of $\Mat$. Hence, $X$ and $Y$ are uniform, independent, and
independent of $(\sigma,\pi,d,d')$. Let $(i,k)=(\sigma^{-1}(a),\pi^{-1}(b))$.
By~\eqref{eq:mm-candidate-error}, the error at entry $(a,b)$ is
$W_{ik}/(d_id'_k)$. The entry $(i,k)$ is uniform in $[n]^2$, and given
$(i,k)$, the product $d_id'_k$ is uniform in $\F^*$. Both are independent
of $(X,Y)$ and of the coins of $\Ocal$. Therefore, the error at $(a,b)$ is
zero with probability $\Pr[W_{ik}=0]=\gamma_0$, since $(i,k)$ is a uniform
entry. For $c\ne0$, the error equals $c$ if and only if $W_{ik}\ne0$ and
$d_id'_k=W_{ik}/c$, which happens with probability $(1-\gamma_0)/(q-1)$.

\paragraph{Concentration.}
As shown in the proof of \cref{lem:mm-average}, the pairs $(A_j,B_j)$
follow a walk which is reversible with respect to the uniform
distribution, positive semidefinite, and has second eigenvalue $1/q\le1/2$.
This walk starts from the stationary distribution, since $(A_0,B_0)$ is
uniform. The randomness of each iteration is fresh, so at iteration $j$,
the event $(D_j-A_0B_0)_{ab}=c$ happens with probability
$h_{abc}(A_j,B_j)$, independently of the other iterations given the walk.
Let $Z_{abc}$ be the number of iterations in which this event happens.
We apply \cref{lem:chernoff} with $\eta=\varepsilon/2$, for which
$2(1-\lambda)\eta^2/(1+\lambda)\ge\varepsilon^2/6$ since $\lambda\le1/2$.
By~\eqref{eq:mm-error-distribution}, we obtain
\[
 \Pr\Bigl[Z_{ab0}\le\Bigl(\gamma_0-\frac\varepsilon2\Bigr)t\Bigr]
 \le e^{-\varepsilon^2t/6}
 \qquad\text{and}\qquad
 \Pr\Bigl[Z_{abc}\ge\Bigl(\frac{1-\gamma_0}{q-1}+\frac\varepsilon2\Bigr)t\Bigr]
 \le e^{-\varepsilon^2t/6}\quad\text{for }c\ne0.
\]
By a union bound over the $n^2$ entries and the $q$ values $c$, none of
these events happens, except with probability at most
$n^2q\,e^{-\varepsilon^2t/6}\le1/100$.

\paragraph{Decoding.}
Assume that none of these events happens. Since $q\gamma_0-1\ge q\varepsilon$,
we have
\[
 \Bigl(\gamma_0-\frac\varepsilon2\Bigr)
 -\Bigl(\frac{1-\gamma_0}{q-1}+\frac\varepsilon2\Bigr)
 =\frac{q\gamma_0-1}{q-1}-\varepsilon
 \ge\frac{\varepsilon}{q-1}>0.
\]
Hence, for every entry $(a,b)$ and every $c\ne0$, we have
$Z_{abc}<Z_{ab0}$. In other words, at every entry, the correct value
$(A_0B_0)_{ab}$ is the unique most frequent value, and the algorithm
outputs $A_0B_0$.
\end{proof}

\begin{theorem}[Worst-case to average-case reduction from approximate algorithms]
\label{thm:mm-approx}
Let $\Ocal$ be a time-$T$ algorithm with expected agreement
$1/q+\varepsilon$, for some $\varepsilon>0$. For every $0<\delta<1/2$,
there is an algorithm which makes
$O(\varepsilon^{-2}\log(nq)\log(1/\delta))$ calls to $\Ocal$, runs in time
\[
 O\!\left(\frac{T+n^2\log t}{\varepsilon^2}\log(nq)\log\frac1\delta\right),
 \qquad\text{where}\qquad t=\lceil6\varepsilon^{-2}\log(100n^2q)\rceil,
\]
and outputs $AB$ with probability at least $1-\delta$, for every pair
$(A,B)\in\Mat\times\Mat$. All calls are made on pairs of $n\times n$
matrices over $\F$.
\end{theorem}

\begin{proof}
The proof of \cref{thm:mm} only uses the fact that the algorithm
$\mathcal C$ outputs the product of two independent uniform matrices with
probability at least $0.97$. By \cref{lem:mm-approx-average}, this also
holds for $\mathcal C_{\mathrm{apx}}$. Hence, the algorithm
$\mathcal C_{\mathrm{wc}}$ of \cref{fig:mm}, with
$\ell=\lceil4\log(1/\delta)\rceil$ and with $\mathcal C$ replaced by
$\mathcal C_{\mathrm{apx}}$, outputs $AB$ with probability at least
$1-\delta$. It makes $4\ell t=O(\varepsilon^{-2}\log(nq)\log(1/\delta))$
calls to $\Ocal$, and runs in time $O(\ell t(T+n^2\log t))$.
\end{proof}

When $T\ge n^2\log t$, the running time of \cref{thm:mm-approx} is
$O(\varepsilon^{-2}\log(nq)\log(1/\delta)\cdot T)$.

\subsection{Approximate online matrix-vector multiplication}
\label{sec:approximate-omv}

We now extend the OMv reduction of \cref{sec:omv} to data structures
with expected agreement $1/q+\varepsilon$, as defined in
\cref{def:omv}. As for approximate matrix multiplication, we permute and
scale the rows of each matrix during preprocessing. At query time, we
undo the corresponding changes to the answers and decode each coordinate
by plurality. All matrix updates and oracle preprocessing take place
before the queries arrive. The changes to
\cref{fig:omv} are given in \cref{fig:omv-approx}.

\begin{figure}[ht]
\begin{mdframed}
\textbf{Data structure $\Ecal_{\mathrm{apx}}^{\Dcal}$.} Parameter: number
of states $t$. Modify the data structure $\Ecal^{\Dcal}$ of
\cref{fig:omv} as follows:
\begin{enumerate}
\item In step~2 of the
preprocessing, also sample a permutation $\pi_j$ of $[n]$ and a vector
$d_j\gets(\F^*)^n$, and run $\sigma_j\gets\Dcal_{\mathrm{pre}}(B_j)$
instead, where $(B_j)_{il}=d_{j,i}(A_j)_{\pi_j(i)l}$ for all $i,l\in[n]$.
\item Skip step~3 of the preprocessing.
\item In step~2a of the query, compute
$y=\Dcal_{\mathrm{query}}(\sigma_j,v+s_j)$, and the candidate $c_j$
defined by $(c_j)_{\pi_j(i)}=y_i/d_{j,i}-(A_js_j)_{\pi_j(i)}-(e_j)_{\pi_j(i)}$
for all $i\in[n]$.
\item Skip step~2b of the query.
\item In step~3 of the query,
output the vector whose $a$-th coordinate is the most frequent value among
$(c_0)_a,\ldots,(c_{t-1})_a$, for all $a\in[n]$, with ties broken
arbitrarily.
\end{enumerate}
\end{mdframed}
\caption{Amplification on average from partial agreement for OMv.}
\label{fig:omv-approx}
\end{figure}

The next lemma follows as in \cref{lem:mm-approx-average}, using
permutations, scalings, and concentration for the $n$ output coordinates.

\begin{lemma}[Amplification from partial agreement for OMv]
\label{lem:omv-approx-average}
Let $\Dcal$ be a data structure with preprocessing time $P$ and query time
$T$, which has expected agreement $1/q+\varepsilon$, for some
$\varepsilon>0$. Set $t=\lceil6\varepsilon^{-2}\log(100nq)\rceil$.
Then $\Ecal_{\mathrm{apx}}^{\Dcal}$ has preprocessing time $O(t(P+n^2))$
and query time $O(t(T+n\log t))$. Moreover, for every $v\in\F^n$, we have
$\Pr_{A_0,\Ecal}[\Ecal_{\mathrm{apx}}^{\Dcal}(A_0;v)=A_0v]\ge0.99$,
where $A_0\gets\Mat$.
\end{lemma}

\begin{proof}
Fix $v$, write $z_j=v+s_j$, and let
$W_j=\Dcal_{\mathrm{query}}(\sigma_j,z_j)-B_jz_j$. Since
$e_j=(A_j-A_0)v$, the candidate error satisfies
\[
 (c_j-A_0v)_{\pi_j(i)}=(W_j)_i/d_{j,i}.
\]
For uniform $A_j$, the matrix $B_j$ is uniform and independent of
$(\pi_j,d_j)$, and $z_j$ is independently uniform. Thus, as in the
error-distribution calculation of \cref{lem:mm-approx-average}, at each
coordinate the error is zero with probability $\gamma_0$ and it is equal to each
nonzero value with probability $(1-\gamma_0)/(q-1)$, where
$\gamma_0\ge1/q+\varepsilon$ is the expected agreement of $\Dcal$.

Conditioned on the matrix trajectory, the errors are independent across
states, and their distributions depend only on the current state. The
walk is stationary with second eigenvalue $1/q\le1/2$. Hence the
concentration and decoding steps of \cref{lem:mm-approx-average} apply
with $n$ coordinates instead of $n^2$ entries: except with probability
at most $nq\exp(-\varepsilon^2t/6)\le1/100$, the correct value is the
unique most frequent value at every coordinate. Preprocessing takes
$O(t(P+n^2))$ operations. Corrections take $O(nt)$ operations per query,
and sorting the $t$ values at each coordinate takes $O(nt\log t)$,
giving the query bound.
\end{proof}

As in \cref{sec:omv}, preprocessing $A+X$ and $X$ for a uniform matrix
$X$, subtracting the answers, and repeating gives the worst-case
reduction below.

\begin{theorem}[Worst-case OMv from approximate data structures]
\label{thm:omv-approx}
Let $\Dcal$ be a data structure with preprocessing time $P$ and query time
$T$, which has expected agreement $1/q+\varepsilon$, for some
$\varepsilon>0$. For every $0<\delta<1/2$, there is a data structure
$\Ecal_{\mathrm{wc}}$ with preprocessing time
$O((P+n^2)\varepsilon^{-2}\log(nq)\log(1/\delta))$ and query time
$O((T+n\log t)\varepsilon^{-2}\log(nq)\log(1/\delta))$, where
$t=\lceil6\varepsilon^{-2}\log(100nq)\rceil$, such that
$\Pr[\Ecal_{\mathrm{wc}}(A;v)=Av]\ge1-\delta$ for all $A\in\Mat$ and
$v\in\F^n$.
\end{theorem}

\begin{proof}
Apply the construction in the proof of \cref{thm:omv}, using
$\Ecal_{\mathrm{apx}}^{\Dcal}$ instead of $\Ecal^{\Dcal}$.
By \cref{lem:omv-approx-average}, each difference is correct with
probability at least $1-2\cdot0.01=0.98$. The same
$\ell=\lceil4\log(1/\delta)\rceil$ independent repetitions therefore
suffice. Multiplying the preprocessing and query bounds of
\cref{lem:omv-approx-average} by $2\ell$ gives the claimed bounds.
\end{proof}

For data structures with expected agreement
$1/q+\varepsilon$, Hirahara and Shimizu~\cite[Theorem~1.5]{HS25a} give a
uniform reduction over prime fields, for a constant $\varepsilon$. In the
full version of~\cite{HS25a}, this reduction makes
$2^{\poly(q,1/\varepsilon)}\cdot O(\log n)$ queries, and
$O(\varepsilon^{-3})$ queries over fields of size larger
than $n/\varepsilon^2$. \Cref{thm:omv-approx} makes
$O(\varepsilon^{-2}\log(nq))$ queries, over every finite field. The walk itself only adds $O(n)$ operations to the query time for each
query to $\Dcal$, since all the steps of the walk are performed during
preprocessing.

\subsection{Comparison with error-correcting reductions}
\label{app:mm-approx-comparison}

\Cref{tab:mm-approx} compares \cref{thm:mm-approx} with previous
reductions from algorithms which only compute a fraction of the entries of
the product correctly. Following~\cite{HS25b}, we call them
error-correcting reductions. As in \cref{tab:mm}, the running times are given
for a constant error probability, and we omit the additive terms which do
not depend on the oracle.

\begin{table}[ht]
\centering
{\small
\renewcommand{\arraystretch}{1.15}
\begin{tabular}{@{}>{\raggedright\arraybackslash}p{2.8cm}>{\raggedright\arraybackslash}p{3.4cm}>{\raggedright\arraybackslash}p{1.8cm}>{\raggedright\arraybackslash}p{1.4cm}>{\raggedright\arraybackslash}p{5.2cm}@{}}
\toprule
Reduction & Fields & Agreement & Uniform & Running time \\
\midrule
\cite[Thm.~1.2]{GSS24} & finite fields & $>7/8$ & yes & $O(T\log n)$ \\
\cite[Thm.~1.1]{HS25a} & prime fields & $1/q+\varepsilon$ & no &
 $\Otilde(q^2\varepsilon^{-4})\cdot T\log n$ \\
\cite[Thm.~1.3]{HS25a} & fields with $q>n/\varepsilon^2$ & $\varepsilon$ & yes &
 $\poly(1/\varepsilon)\cdot T\log n$ \\
\cite[Thm.~1.4]{HS25a} & prime fields & $2/q+\varepsilon$ & yes &
 $2^{\poly(q,1/\varepsilon)}\cdot T\log n$ \\
\cite[Thm.~1]{HS25b} & prime fields, constant $q$ and $\varepsilon$ &
 $1/q+\varepsilon$ & yes & $\Otilde(T)$ \\
\cite[Thm.~3]{SS25} & prime fields, constant $\varepsilon$ &
 $1/q+\varepsilon$ & yes &
 $2^{O(1/\varepsilon^2)}q^{O(\log^2(1/\varepsilon))}\cdot\Otilde(T)$ \\
\cite[Thm.~6.2]{VZ25} & prime fields, assuming trapdoored matrices &
 $1/q+\varepsilon$ & yes & $\Otilde(T)$ \\
\Cref{thm:mm-approx} & finite fields & $1/q+\varepsilon$ & yes &
 $O(\varepsilon^{-2}\log(nq)\cdot T)$ \\
\bottomrule
\end{tabular}\par}
\caption{Worst-case to average-case reductions for $n\times n$ matrix
multiplication over a field with $q$ elements, from an algorithm running in
time $T\ge n^2$, whose output agrees with the product on a $\gamma$
fraction of the entries in expectation, over uniform inputs. The column
``Agreement'' gives the condition on $\gamma$. The running times are for a
constant error probability.}
\label{tab:mm-approx}
\end{table}

Among the reductions of \cref{tab:mm-approx} which do not rely on
computational assumptions, \cref{thm:mm-approx} is the only one which
reaches the optimal agreement $1/q+\varepsilon$ with a uniform reduction
over every finite field, including fields which are not prime, and for
$\varepsilon$ which may depend on $n$. The number of calls grows as
$\varepsilon^{-2}$. In comparison, the previous unconditional and uniform
reductions for an agreement close to $1/q$ are stated for a constant
$\varepsilon$, and their dependence on $1/\varepsilon$ is
exponential~\cite{HS25a,SS25} or doubly exponential~\cite{HS25b}. Shinkar
and Singh~\cite[Remark~2]{SS25} ask for a reduction with running time
$\poly(1/\varepsilon)\cdot\Otilde(T)$. \Cref{thm:mm-approx} gives such a
reduction, with a simple proof. Vaikuntanathan and
Zamir~\cite[Theorem~6.2]{VZ25} reach the same agreement over prime
fields, assuming the existence of trapdoored matrices.

Our decoding step is the same as in~\cite[Theorem~6.2]{VZ25}. The random
permutations and scalings make the errors at each entry uniform among the
wrong values, so that the correct value is the most frequent one.
In~\cite{VZ25}, this step is applied to the answers of the oracle on the
pairs $(A+R,B+Q)$, where $R$ and $Q$ are trapdoored matrices, so that the
corrections $AQ$, $RB$ and $RQ$ can be computed quickly. In our
reduction, the walk replaces the trapdoored matrices in the average-case
step. By~\eqref{eq:product-update}, each answer at a state of the walk
gives a candidate for the whole product at the starting state, and
by~\eqref{eq:mm-candidate-error}, the error of this candidate is the error
of a single answer, up to the permutations and scalings of step~2a. The
worst-case step then combines four outputs of $\mathcal C_{\mathrm{apx}}$
with the identity~\eqref{eq:four-products}, and these outputs are exact
with high probability. Applying this identity directly to the answers of
the oracle would combine four answers, and an error in any of them can lead
to an error in the result. For this reason, the self-correction
of~\cite[Theorem~1.2]{GSS24}, which uses this identity, requires an
agreement larger than $7/8$. To reach agreements close to $1/q$ without
computational assumptions, previous reductions instead rely on
list-decodable codes~\cite{HS25a,HS25b,SS25}, or on a version of the XOR
lemma for functions with several output bits~\cite{HS25a}.

\paragraph{Limitations.}
The number of calls of \cref{thm:mm-approx} grows with $\log q$, because
of the union bound over the $q$ possible values of each entry. Over large
fields, the reduction of~\cite[Theorem~1.3]{HS25a} has a number of calls
which does not depend on $q$. For linear systems, \cref{thm:approx-walk}
avoids this dependence when $q$ is large. In this case, the scalar which
multiplies the right-hand side is independent of the input of the oracle,
even when the trajectory of the walk is fixed, which makes the wrong
values independent. In $\mathcal C_{\mathrm{apx}}$, once the trajectory of
the walk is fixed, the input $(X,Y)$ of the oracle determines the scalars
$d$ and $d'$, and we do not know a re-randomization of matrix products
which plays the role of the vector $w_j$ of \cref{fig:approx-solver}.

Shimizu and Yasunaga~\cite[Theorem~5.2]{SY26} show that black-box
hardness amplification for functions with values in a field of prime order
$q$ requires $\Omega(q\log(1/\delta)/\varepsilon'^2)$ queries, to reach a
success probability $(1+\varepsilon')/q$. With $\varepsilon'=q\varepsilon$,
this lower bound is $\Omega(\log(1/\delta)/(q\varepsilon^2))$. This
suggests that the number of calls of \cref{thm:mm-approx} could be
improved by a factor of about $q$ over large fields.

\section{Fine-grained one-way functions}
\label{sec:owf}

We first give the one-way function associated with random linear systems,
then state the consequences of our exact and approximate reductions.
We conclude with the relaxed forms of one-wayness obtained from
worst-case hardness alone.

\subsection{The construction and its input distribution}

We use concrete security parameters. Honest algorithms run in time $t(n)$,
and an adversary has time $T(n)$. Our construction uses $t(n)=O(n^2)$;
the hardness assumptions below give a polynomial separation between
$T(n)$ and $t(n)$.

\begin{definition}[Fine-grained one-way function]
\label{def:owf}
Let $f=\{f_n\}_n$ be a family of functions $f_n:D_n\to R_n$, together with
a distribution on each domain $D_n$. The family $f$ is computable in time
$t(n)$ if both sampling an input $x\in D_n$ and computing $f_n(x)$ take time
$t(n)$. It is $(T,\alpha)$-one-way if, for every time-$T(n)$ algorithm
$\Acal$ and every large enough $n$,
\[
 \Pr_{x,\Acal}\bigl[f_n(\Acal(f_n(x)))=f_n(x)\bigr]\le\alpha(n).
\]
\end{definition}

We consider the function
\begin{equation}\label{eq:basic-function}
 f_n:\Mat\times\F^n\longrightarrow\Mat\times\F^n,
 \qquad f_n(M,x)=(M,Mx),
\end{equation}
where the input $(M,x)$ is uniformly random. Sampling the input and
computing $f_n$ take $O(n^2)$ operations. To invert $f_n$ on $(M,y)$, an
algorithm must output a pair $(M,\widehat x)$ with $M\widehat x=y$. When
$M$ is singular, $\widehat x$ can differ from $x$.

In the definition of $\LINEQ_{n,q}$, the matrix is uniform in $\GL$ rather
than in $\Mat$. This is convenient for reductions, since every system has a
unique solution, but complicates input sampling: rejecting singular matrices would require an invertibility test that we do not know how to perform in time $O(n^2)$. For this reason, $f_n$ samples $M$ uniformly from all of $\Mat$. The singular matrices form a fraction $\sigma_q<1/(q-1)$ of $\Mat$, and the following proposition shows that this is the only difference between inverting $f_n$ and solving $\LINEQ_{n,q}$.

\begin{proposition}[One-wayness from average-case hardness]
\label{prop:baseline-owf}
Let $\Acal$ be a time-$T$ algorithm that inverts $f_n$ with probability
$\alpha\ge\sigma_q$. Then there is a time-$T$ algorithm that solves
$\LINEQ_{n,q}$ with probability at least
\[
 \frac{\alpha-\sigma_q}{1-\sigma_q}\ge\alpha-\sigma_q.
\]
Conversely, if a time-$T$ algorithm solves $\LINEQ_{n,q}$ with probability
$p$, then there is a time-$(T+O(n^2))$ algorithm that inverts $f_n$ with
probability at least $(1-\sigma_q)p$.
\end{proposition}

\begin{proof}
For the first statement, we run $\Acal$ on the input $(M,y)$ of
$\LINEQ_{n,q}$ and output the second component of the answer. Let $E$ be
the event that the matrix of a uniform input of $f_n$ is invertible, so
that $\Pr[E]=1-\sigma_q$. Then
\[
 \Pr[\Acal\text{ succeeds}\mid E]
 \ge\frac{\Pr[\Acal\text{ succeeds}]-\Pr[\neg E]}{\Pr[E]}
 \ge\frac{\alpha-\sigma_q}{1-\sigma_q}.
\]
Conditioned on $E$, the
pair $(M,Mx)$ is distributed exactly as an input of $\LINEQ_{n,q}$, and
the unique preimage of $Mx$ under $x\mapsto Mx$ is $M^{-1}(Mx)$. Hence every
successful inversion is a correct solution.

For the converse, we run the solver on $(M,y)$, check that the output
$\widehat x$ satisfies $M\widehat x=y$, and output $(M,\widehat x)$. When
$M$ is invertible, which happens with probability $1-\sigma_q$, the input is
distributed as in $\LINEQ_{n,q}$, so the solver succeeds with probability
$p$.
\end{proof}

When $q$ is large, the loss $\sigma_q<1/(q-1)$ is small. Over small fields,
it can be constant, so the proposition alone only rules out inverters
whose success exceeds this loss. We therefore use sufficiently large
fields of polynomial size for the cryptographic consequences below.

\subsection{Consequences of hardness self-amplification}
\label{sec:amplification-owf}

The following corollaries translate the reductions of
\cref{sec:amplification,sec:approximate-lineq} into one-wayness statements.
The first shows that almost worst-case hardness suffices: only a tiny
fraction of matrices need remain hard on some right-hand side. The second
also rules out recovering a substantial fraction of the input coordinates.

\begin{corollary}[One-wayness from almost worst-case hardness]
\label{cor:almost-worst-case-owf}
Let $\Acal$ be a time-$T$ algorithm that inverts $f_n$ with probability
$\alpha\ge2\sigma_q$. For all $1\le k\le n$ and $0<\delta<1/2$, there is an
algorithm which runs in time
\[
 O\!\left(\frac{T+n^2k^{\omega-2+o(1)}}\alpha\log\frac1\delta\right),
\]
never outputs a wrong solution, and, for all but a
$128q^{-2k}/\alpha$ fraction of matrices $M\in\GL$, solves $Mx=y$
with probability at least $1-\delta$ for every right-hand side $y$.
For $k\le\lceil n^{0.32}\rceil$, rectangular matrix multiplication replaces
the additional work $n^2k^{\omega-2+o(1)}$ by $n^{2+o(1)}$.
\end{corollary}

\begin{proof}
By \cref{prop:baseline-owf}, the inverter solves $\LINEQ_{n,q}$ with
probability at least $\alpha-\sigma_q\ge\alpha/2$. Apply
\cref{prop:rank-k}, using the rectangular implementation described in
\cref{sec:rank-k} for the last statement.
\end{proof}

For a concrete choice of parameters, fix constants $\eta>0$ and
$0<a<0.32$. Suppose that, for every uniform time-$n^{2+\eta}$ algorithm
and every sufficiently large $n$, more than a $2^{-n^a}$ fraction of
matrices $M\in\GL$ have some right-hand side on which the algorithm
fails with probability greater than $1/3$. For any $0<\zeta<\eta$ and
polynomial-size fields with $q\ge3n^{\zeta/2}$, the family $f_n$ is then
$(n^{2+\eta-\zeta},n^{-\zeta/2})$-one-way. Indeed, an inverter with
success at least $\alpha=n^{-\zeta/2}$ gives, by the corollary with
$k=\lceil n^a\rceil$ and $\delta=1/3$, a solver running in time
\[
 O\!\left((n^{2+\eta-\zeta}+n^{2+o(1)})\cdot n^{\zeta/2}\right)
 =o(n^{2+\eta}).
\]
The exceptional fraction is at most
$128q^{-2k}n^{\zeta/2}\le2^{-n^a}$ for sufficiently large $n$, contradicting
the assumption. This gives the polynomial gap claimed in the introduction.

\paragraph{Approximate inversion.}
We say that an algorithm $\rho$-approximately inverts $f_n$ if, on input
$f_n(M,x)$, it outputs a vector $\widehat x$ with
$d_H(\widehat x,x)\le\rho n$. Here, agreement is measured against the
sampled input $x$. For invertible $M$, this is the unique preimage;
we account for singular matrices through $\sigma_q$ as above.

\begin{corollary}[Hardness of approximate inversion]
\label{cor:approximate-owf}
Let $0\le\rho<1$, and let $\Acal$ be a time-$T$ algorithm that
$\rho$-approximately inverts $f_n$ with probability $\alpha>\sigma_q$.
Set $\gamma=(\alpha-\sigma_q)(1-\rho)$ and suppose that
$\varepsilon=\gamma-1/q>0$. There is an exact solver for $\LINEQ_{n,q}$
which never outputs a wrong solution, succeeds with probability at least
$7/8$, and runs in time $O((T+n^2)m)$, where
\[
 m=O\!\left(\varepsilon^{-2}\log\frac n\varepsilon\right),
 \qquad
 m=O\!\left(\gamma^{-1}\log\frac n\gamma\right)
 \quad\text{if }q\ge200/\gamma.
\]
In particular, if $q\ge400/(\alpha(1-\rho))$, the running time is
\[
 O\!\left(\frac{T+n^2}{\alpha(1-\rho)}
             \log\frac n{\alpha(1-\rho)}\right).
\]
For every $1\le k\le n$ and $0<\delta<1/2$, composing with
\cref{prop:rank-k} gives a solver that succeeds with probability at least
$1-\delta$ on every right-hand side outside an $O(q^{-2k})$ fraction of
matrices, in time
$O(((T+n^2)m+n^2k^{\omega-2+o(1)})\log(1/\delta))$.
Again, for $k\le\lceil n^{0.32}\rceil$, the last additional-work term
can be replaced by $n^{2+o(1)}$.
\end{corollary}

\begin{proof}
Conditioned on $M$ being invertible, the probability of approximate
inversion is at least $\alpha-\sigma_q$, by the same conditioning argument
as in \cref{prop:baseline-owf}. Each success gives agreement at least
$1-\rho$, so the expected agreement is at least $\gamma$. Apply
\cref{thm:approx-walk}. Under the stated stronger condition on $q$,
we have $\sigma_q\le\alpha/2$ and hence
$\gamma\ge\alpha(1-\rho)/2$, which gives the simpler bound. Finally,
apply \cref{prop:rank-k} to the exact solver, whose success probability
is at least $7/8$.
\end{proof}

Consequently, under the same almost worst-case assumption as above, for
any fixed $\rho<1$ and $0<\zeta<\eta$, no time-$n^{2+\eta-\zeta}$ algorithm
can $\rho$-approximately invert $f_n$ with probability at least
$n^{-\zeta/2}$ when $q\ge400n^{\zeta/2}/(1-\rho)$ and $q=n^{O(1)}$.
Indeed, the last corollary with $k=\lceil n^a\rceil$ and $\delta=1/3$
would again yield an exact solver in time $o(n^{2+\eta})$ with an
exceptional fraction smaller than $2^{-n^a}$.

\subsection{Consequences of worst-case hardness}
\label{sec:worstcase}

We now assume hardness on only one system per algorithm and dimension.
We obtain two relaxed forms of one-wayness for $f_M(x)=Mx$, with $M$
as a public parameter: auxiliary-input one-wayness~\cite{OW93}, where the
hard matrix may depend on the adversary, and one-wayness with public
advice against all uniform adversaries, using Levin's universal
search~\cite{Lev73}. Both use the following elementary self-reduction.

\begin{lemma}[Randomizing the right-hand side]
\label{lem:rhs-mask}
Fix $M\in\GL$ and $0<\alpha\le1$. Let $\Acal$ be a time-$T$ algorithm such that
$\Pr_{y\gets\F^n,\Acal}[\Acal(M,y)=M^{-1}y]\ge\alpha$. Then for every
$0<\delta<1/2$, there is an algorithm which, on input $(M,y)$, makes
$\lceil\alpha^{-1}\log(1/\delta)\rceil$ calls to $\Acal$, runs in time
$O((T+n^2)\alpha^{-1}\log(1/\delta))$, and outputs $M^{-1}y$ with
probability at least $1-\delta$ for every $y\in\F^n$. It never outputs a
wrong solution.
\end{lemma}

\begin{proof}
In each trial, we sample $z\gets\F^n$ and run $\Acal$ on $(M,y+Mz)$. If
the answer $w$ satisfies $Mw=y+Mz$, we output $w-z$, which equals
$M^{-1}y$. Since $Mz$ is uniform, the vector $y+Mz$ is uniform for every
fixed $y$, so each trial succeeds with probability at least $\alpha$,
independently of the other trials. All trials fail with probability at
most $(1-\alpha)^{\lceil\alpha^{-1}\log(1/\delta)\rceil}\le\delta$.
\end{proof}

\begin{assumption}[Worst-case hardness of $\LINEQ$]
\label{ass:worst-case}
Let $S$ be a time-constructible function. We say that $\LINEQ$ over $\F$ is
$S$-hard in the worst case if, for every uniform randomized algorithm
$\Acal$ running in time $S(n)$ and every large enough $n$, there is an
invertible system $(M,y)$ of dimension $n$ such that
$\Pr_\Acal[\Acal(M,y)=M^{-1}y]<2/3$.
\end{assumption}

\Cref{ass:worst-case} asks for hardness at all large enough dimensions. If
we only assume hardness at infinitely many dimensions, all the results of
this section hold at infinitely many dimensions. We also assume that the
time bounds $T(n)$ and the success probabilities $\alpha(n)$ below can be
computed within the stated running times.

\begin{proposition}[Auxiliary-input one-wayness]
\label{prop:aux-owf}
Assume that $\LINEQ$ over $\F$ is $S$-hard in the worst case. There is an
absolute constant $C$ such that the following holds. Let $\Acal$ be a
uniform time-$T$ algorithm and $\alpha(n)>0$ such that
$C(T(n)+n^2)/\alpha(n)\le S(n)$. Then for every large enough $n$, there is a
matrix $M\in\GL$ such that
\[
 \Pr_{x\gets\F^n,\Acal}\bigl[\Acal(M,Mx)=x\bigr]<\alpha(n).
\]
\end{proposition}

\begin{proof}
Let $\Acal'$ be the algorithm of \cref{lem:rhs-mask} built from $\Acal$,
with $\delta=1/3$. It runs in time $O((T(n)+n^2)/\alpha(n))$, which is at
most $S(n)$ for a large enough constant $C$. By \cref{ass:worst-case}, for
every large enough $n$, there is a system $(M,y)$ such that $\Acal'$ solves
$(M,y)$ with probability smaller than $2/3$. By \cref{lem:rhs-mask}, if
$\Acal$ inverted $f_M$ with probability at least $\alpha(n)$, then $\Acal'$
would solve every system with matrix $M$ with probability at least $2/3$.
Hence $\Acal$ inverts $f_M$ with probability smaller than $\alpha(n)$.
\end{proof}

In \cref{prop:aux-owf}, the matrix depends on the adversary. The next
theorem gives a single sequence of matrices which is hard for all uniform
adversaries. Let $\Acal_1,\Acal_2,\ldots$ be an enumeration of all uniform
randomized algorithms, each stopped after $T(n)$ steps. For $i\ge1$, let
$\Acal'_i$ be the algorithm of \cref{lem:rhs-mask} built from $\Acal_i$,
with $\delta=1/3$ and success probability $\alpha(n)$.

\begin{theorem}[One-wayness with public advice]
\label{thm:levin-advice}
Assume that $\LINEQ$ over $\F$ is $S$-hard in the worst case. Let $s(n)$ be
a computable nondecreasing function with $s(n)\to\infty$, and assume that
running $\Acal'_1,\ldots,\Acal'_{s(n)}$ on an input of dimension $n$ takes
time at most $S(n)$. For instance, this holds when
$\Otilde(s(n)(T(n)+n^2)/\alpha(n))\le S(n)$, where $\Otilde$ accounts for
the overhead of simulating the algorithms. Then there is a sequence of
matrices $M_n\in\GL$ such that every uniform time-$T$ algorithm $\Acal$
satisfies, for every large enough $n$,
\[
 \Pr_{x\gets\F^n,\Acal}\bigl[\Acal(M_n,M_nx)=x\bigr]<\alpha(n).
\]
\end{theorem}

\begin{proof}
Consider the algorithm $\mathcal U$ which, on input $(M,y)$, runs
$\Acal'_1,\ldots,\Acal'_{s(n)}$ in sequence on $(M,y)$, and outputs the first
answer $w$ such that $Mw=y$. By assumption, $\mathcal U$ runs in time
$S(n)$. By \cref{ass:worst-case}, for every large enough $n$, there is a
system $(M_n,y_n)$ which $\mathcal U$ solves with probability smaller than
$2/3$. For smaller $n$, we choose $M_n$ arbitrarily.

Now, suppose that some $\Acal_i$ with $i\le s(n)$ inverts $f_{M_n}$ with
probability at least $\alpha(n)$. By \cref{lem:rhs-mask}, $\Acal'_i$ then
solves $(M_n,y_n)$ with probability at least $2/3$. Since $\mathcal U$
succeeds whenever one of the algorithms it runs succeeds, $\mathcal U$
also solves $(M_n,y_n)$ with probability at least $2/3$, which is a
contradiction. Hence, all of $\Acal_1,\ldots,\Acal_{s(n)}$ invert
$f_{M_n}$ with probability smaller than $\alpha(n)$. Finally, every fixed
uniform time-$T$ algorithm appears as some $\Acal_i$ in the enumeration,
and $i\le s(n)$ for every large enough $n$, since $s(n)\to\infty$.
\end{proof}

The resulting one-way function is evaluated in time $O(n^2)$, given the
public advice $M_n$, which consists of $n^2\lceil\log_2q\rceil$ bits. The
sequence $(M_n)$ can be computed by exhaustive search: the success
probability of each stopped algorithm $\Acal_i$ on $f_M$ can be computed
exactly by enumerating all vectors $x$ and all random choices of the time-bounded algorithms. One can therefore take $M_n$ to be the first matrix, in
lexicographic order, on which $\Acal_1,\ldots,\Acal_{s(n)}$ all have success
probability smaller than $\alpha(n)$. We do not know how to compute $M_n$
efficiently: an efficient algorithm for this task would give an efficiently
samplable distribution on which $\LINEQ$ is hard for uniform algorithms,
which is a form of worst-case to average-case reduction.

The restriction to uniform adversaries in \cref{thm:levin-advice} cannot
be removed: for every sequence of matrices $M_n\in\GL$, there is a nonuniform algorithm which inverts $f_{M_n}$ on every input in time $O(n^2)$, with $n^2\lceil\log_2q\rceil$ bits of advice.
The advice is $M_n^{-1}$, and on input $y$, the algorithm outputs $M_n^{-1}y$.

\section{Fine-grained key agreement}
\label{sec:protocol}

We study the two-message key agreement protocol discussed in the introduction. Our
security argument has three steps: first, we relate the full recovery of the secret key to the problem of finding \emph{one} coordinate of a random linear system. Second, we reduce full-key recovery to standard indistinguishability for a shared secret bit. Lastly, we use an efficient attacker to compute matrix ranks for a large class of distributions. Together, these results yield an infinitely-often key agreement under the almost-everywhere average-case hardness of the rank computation problem.

\subsection{Protocol and security notions}
\label{sec:protocol-definition}

\begin{definition}[Fine-grained key agreement]
\label{def:ka}
A key agreement protocol is an interactive protocol between two parties,
Alice and Bob, at the end of which both parties output a key. We denote by
$\tau$ the transcript of the protocol and by $K$ the key. We consider only perfectly correct protocols: both parties always output the same key $K$. The protocol
is computable in time $t(n)$ if each party runs in time $t(n)$. We consider the following security notions:
\begin{itemize}
 \item The protocol is $(T,\alpha)$-secure against key recovery if every
 time-$T(n)$ algorithm $\Ecal$ satisfies $\Pr[\Ecal(\tau)=K]\le\alpha(n)$,
 for every large enough $n$.
 \item When $K\in\{0,1\}$, the protocol is $(T,\varepsilon)$-secure if every
 time-$T(n)$ algorithm $\mathcal D$ satisfies, for every large enough $n$,
 \[
  \bigl|\Pr[\mathcal D(\tau,K)=1]-\Pr[\mathcal D(\tau,u)=1]\bigr|
  \le\varepsilon(n),
 \]
 where $u\gets\{0,1\}$ is independent of $\tau$.
\end{itemize}
\end{definition}

For infinitely-often security, the inequalities need only hold for all
sufficiently large $n$ in one infinite set $I$ (the set $I$ must be common
to all uniform adversaries, but the threshold for each adversary may differ).
All probabilities include the protocol coins and the adversary coins.

\paragraph{The protocol.}
We use the protocol of \cref{fig:protocol}, which is a noiseless version of
the Alekhnovich key exchange~\cite{Ale03}. A similar protocol was also studied
in the bounded memory setting~\cite{GZ19}. Both parties perform
$O(n^2)$ operations and exchange $n^2+2n$ field elements. We will relate key recovery to
one-coordinate recovery, and then to matrix rank.

\begin{figure}[ht]
\begin{mdframed}
\textbf{Protocol $\Pi_{n,q}$.} Parameters: dimension $n$ and field $\F$.
\begin{enumerate}[leftmargin=*]
 \item Alice samples $M\gets\Mat$ and $a\gets\F^n$. She sends
 $(M,r)$, where $r=a^\T M\in\mathbb F_q^{1\times n}$.
 \item Bob samples $b\gets\F^n$ and sends $y=Mb$.
 \item Alice outputs $K=a^\T y$. Bob outputs $K=rb$.
\end{enumerate}
\end{mdframed}
\caption{Key agreement. Both parties output $K=a^\T Mb$.
The transcript is $\tau=(M,r,y)$.}
\label{fig:protocol}
\end{figure}

\paragraph{The ideal distribution.}
As for our one-way functions, we sample $M$ from the set of all matrices to keep the honest computation quadratic, and analyze the protocol conditioned on $M$ being invertible, which we call the ideal distribution. Under this
conditioning, $M\gets\GL$, $r\gets\mathbb F_q^{1\times n}$, and
$y\gets\F^n$ are independent and uniform, and
\[
 K=rM^{-1}y.
\]
Indeed, for each invertible $M$, the maps $a\mapsto a^\T M$ and
$b\mapsto Mb$ are bijections. The actual and ideal distributions have
statistical distance at most $\sigma_q<1/(q-1)$, by \eqref{eq:sigma}.
The key-recovery probabilities therefore differ by at most $\sigma_q$, and
one-bit distinguishing advantages by at most $2\sigma_q$ (we will use growing fields to make this loss small).

\paragraph{Extracting a bit.}
For $q=2^\ell$, fix an $\mathbb F_2$-linear bijection
$\enc:\F\to\{0,1\}^\ell$, given by coordinates in a fixed basis.
Alice samples $s\gets\{0,1\}^\ell\setminus\{0\}$ and appends $s$ to her
first message. Both parties output $\langle s,\enc(K)\rangle$, with the
inner product taken over $\mathbb F_2$; the full transcript is $(\tau,s)$.
This preserves perfect correctness and adds $O(\log q)$ bit operations.
For a uniform field key, the extracted bit is uniform for every $s\ne0$.
The next subsection shows that a distinguisher for this bit gives a
key-recovery algorithm.

\subsection{Key recovery and one-bit security}
\label{sec:dense}

We first work with the ideal distribution. Conditioning on invertibility
makes $M,r,y$ independent, and the shared scalar is $K=rM^{-1}y$.
This lets us relate key recovery to a familiar linear-algebra task.

\subsubsection{Key recovery and one coordinate}
\label{sec:dense-coord}

\begin{definition}[One coordinate of a random system]
\label{def:coord}
The distributional problem $\COORD_{n,q}$ is defined as follows. The input
is a pair $(M,y)$, where $M\gets\GL$ and $y\gets\F^n$ are independent.
The goal is to output $(M^{-1}y)_1$, the first coordinate of the solution.
\end{definition}

The reduction relies on a simple change of basis. For a nonzero row
vector $r\in\mathbb F_q^{1\times n}$, let $i$ be the first index such that
$r_i\ne0$. Let $R_r$ be the matrix obtained from $I_n$ by replacing row $i$
with $r$, and then exchanging rows $1$ and $i$. Then $R_r$ is invertible,
since $\det R_r=\pm r_i$, and the first row of $R_r$ is $r$, that is,
$e_1^\T R_r=r$. Multiplying a matrix by $R_r$ or by $R_r^{-1}$ amounts to a
row replacement and a row exchange (or the corresponding column
operations), which takes $O(n^2)$ operations.

\begin{proposition}[Key recovery and $\COORD$]
\label{prop:dense-coordinate}
Let $\Pcal$ be a time-$T$ algorithm which recovers the key with probability
$\alpha$ on the ideal distribution of the protocol. Then there is a
time-$(T+O(n^2))$ algorithm which solves $\COORD_{n,q}$ with probability
at least $\alpha-q^{-n}$. Conversely, a time-$T$ algorithm which solves
$\COORD_{n,q}$ with probability $\alpha$ yields a time-$(T+O(n^2))$
algorithm which recovers the key with probability at least $\alpha$.
\end{proposition}

\begin{proof}
For the first statement, on input $(M,y)$, we sample
$r\gets\mathbb F_q^{1\times n}$. If $r=0$, we fail, which happens with
probability $q^{-n}$. Otherwise, we run $\Pcal$ on the transcript
$(MR_r,r,y)$. The key of this transcript is
\[
 r(MR_r)^{-1}y=rR_r^{-1}M^{-1}y=e_1^\T M^{-1}y=(M^{-1}y)_1,
\]
which is the solution of $\COORD_{n,q}$. For every fixed $r\ne0$, the
matrix $MR_r$ is uniform in $\GL$ and independent of $y$. Hence the
transcript is distributed as the ideal distribution conditioned on $r\ne0$,
and the success probability is at least $\alpha-q^{-n}$.

For the converse, on a transcript $(M,r,y)$, we output $0$ if $r=0$, which
is the correct key. Otherwise, we run the algorithm for $\COORD_{n,q}$ on
$(MR_r^{-1},y)$. The first coordinate of the solution of this system is
\[
 e_1^\T(MR_r^{-1})^{-1}y=e_1^\T R_rM^{-1}y=rM^{-1}y=K.
\]
For every fixed $r\ne0$, the matrix $MR_r^{-1}$ is uniform in $\GL$ and
independent of $y$, so the success probability is at least $\alpha$.
\end{proof}

Hence, our protocol is secure against key recovery if and only if
$\COORD_{n,q}$ is hard. Since a solver for $\LINEQ_{n,q}$ also solves
$\COORD_{n,q}$, this is a stronger assumption than the hardness of
$\LINEQ_{n,q}$; we do not know whether the converse holds (the straightforward approach of recovering coordinates one at a time already
introduces a factor $n$ in oracle calls, which removes the gap between
honest parties and adversaries).

\subsubsection{From a distinguisher to key recovery}
\label{sec:dense-bit}

We use two standard facts to go from distinguishing the extracted bit
to recovering the field key. We first recall the equivalence between
bit distinguishing and prediction, then a list-decoding version of
Goldreich--Levin.

\begin{lemma}[From distinguishing to predicting]
\label{lem:dist-to-pred}
Let $(\tau,K)$ be a pair of random variables with $K\in\{0,1\}$, and let
$\mathcal D$ satisfy $\Pr[\mathcal D(\tau,K)=1]-\Pr[\mathcal D(\tau,u)=1]
=\varepsilon$, for $u\gets\{0,1\}$. Consider the predictor $\Pcal$ which,
on input $\tau$, samples $u\gets\{0,1\}$, and outputs $u$ if
$\mathcal D(\tau,u)=1$ and $1-u$ otherwise. Then
$\Pr[\Pcal(\tau)=K]=1/2+\varepsilon$.
\end{lemma}

\begin{proof}
With probability $1/2$, the predictor tries $u=K$ and is correct if
$\mathcal D(\tau,K)=1$. With probability $1/2$, it tries $u=1-K$ and is
correct if $\mathcal D(\tau,1-K)=0$. Therefore
\begin{align*}
 \Pr[\Pcal(\tau)=K]
 &=\frac12+\frac12\Pr[\mathcal D(\tau,K)=1]-\frac12\Pr[\mathcal D(\tau,1-K)=1]\\
 &=\frac12+\Pr[\mathcal D(\tau,K)=1]-\Pr[\mathcal D(\tau,u)=1],
\end{align*}
where the last equality uses
$\Pr[\mathcal D(\tau,u)=1]
=\frac12\Pr[\mathcal D(\tau,K)=1]+\frac12\Pr[\mathcal D(\tau,1-K)=1]$.
\end{proof}

If the advantage of $\mathcal D$ is negative, the same predictor applied
to $1-\mathcal D$ has advantage $|\varepsilon|$. A reduction does not
know in general which of $\mathcal D$ and $1-\mathcal D$ has a positive
advantage. When the reduction can check the final output, it runs with
both. Otherwise, it picks one of them at random, which divides the final
success probability by two.

\begin{lemma}[Goldreich--Levin~\cite{GL89}]
\label{lem:gl}
There is a randomized algorithm which, given $\ell$, $\varepsilon>0$, and
oracle access to a randomized predictor $\Pcal$ such that
$\Pr_{s\gets\{0,1\}^\ell,\Pcal}[\Pcal(s)=\langle s,z\rangle]\ge1/2+\varepsilon$
for some $z\in\{0,1\}^\ell$, runs in time $\poly(\ell,1/\varepsilon)$ and
outputs a list of at most $\varepsilon^{-2}$ strings which contains $z$ with
probability at least $1/2$.
\end{lemma}

\begin{proof}[Sketch]
The algorithm of~\cite{GL89}, repeated a constant number of times,
outputs a list of $O(\ell/\varepsilon^2)$ candidates which contains $z$
with probability at least $3/4$. We reduce the size of this list as
follows. For a candidate $z'$, let
$a(z')=\Pr_{s,\Pcal}[\Pcal(s)=\langle s,z'\rangle]$ be the agreement of
$z'$ with the predictor. We estimate $a(z')$ with
$O(\varepsilon^{-2}\log(\ell/\varepsilon))$ samples for each candidate,
and keep the candidates whose estimate is at least $1/2+3\varepsilon/4$.
By a Chernoff bound and a union bound over the candidates, all estimates
are within $\varepsilon/4$ of the true agreements, except with
probability $1/4$. In this case, $z$ is kept, and every kept candidate
satisfies $a(z')\ge1/2+\varepsilon/2$.

We now bound the number of strings $z'$ with $a(z')\ge1/2+\varepsilon/2$.
Let $F(s)=\E_\Pcal[(-1)^{\Pcal(s)}]$, and let
$\widehat F(z')=\E_s[F(s)(-1)^{\langle s,z'\rangle}]$. Then
$\widehat F(z')=2a(z')-1$, so every such string satisfies
$\widehat F(z')\ge\varepsilon$. By the Parseval identity,
\[
 \varepsilon^2\cdot\bigl|\{z':a(z')\ge1/2+\varepsilon/2\}\bigr|
 \le\sum_{z'\in\{0,1\}^\ell}\widehat F(z')^2=\E_s\bigl[F(s)^2\bigr]\le1.
\]
On this event the list has at most $\varepsilon^{-2}$ elements. If the
computed list is larger, the algorithm outputs the empty list. Thus the
size bound always holds, and $z$ is included with probability at least
$3/4-1/4=1/2$.
\end{proof}

We apply these lemmas to the one-bit key of \cref{sec:protocol-definition}.
The transcript is $(\tau,s)$, and the key is
$\langle s,\enc(K)\rangle$ for uniform nonzero $s$.

\begin{theorem}[Key recovery from a distinguisher]
\label{thm:dense-bit}
Let $q=2^\ell$, and let $\Dcal$ be a time-$T$ algorithm which distinguishes
$(\tau,s,\langle s,\enc(K)\rangle)$ from $(\tau,s,u)$ with advantage
$\varepsilon$, on the ideal distribution of the protocol. Then there
is an algorithm which runs in time $\poly(\ell,1/\varepsilon)\cdot(T+n^2)$
and recovers $K$ with probability at least $\varepsilon^3/128$.
\end{theorem}

\begin{proof}
By \cref{lem:dist-to-pred}, applied to $\Dcal$ or to $1-\Dcal$, there is
a predictor $\Pcal(\tau,s)$ for $\langle s,\enc(K)\rangle$ with advantage
$\varepsilon$. We cannot check a candidate key, so we choose one of the two
predictors at random, which divides the final success probability by
two. Condition for now on choosing the predictor with positive advantage.
We extend $\Pcal$ to $s=0$ by letting $\Pcal(\tau,0)=0$, which
is always correct. For a uniform seed $s\gets\{0,1\}^\ell$, the advantage
is then at least $(1-2^{-\ell})\varepsilon\ge\varepsilon/2$.

For a transcript $\tau$, let $a(\tau)$ be the advantage of $\Pcal(\tau,\cdot)$
over a uniform seed. Then $a(\tau)\le1/2$ and $\E_\tau[a(\tau)]\ge\varepsilon/2$.
Let $G$ be the event that $a(\tau)\ge\varepsilon/4$. Then
\[
 \frac\varepsilon2\le\E_\tau[a(\tau)]\le\frac12\Pr[G]+\frac\varepsilon4,
\]
so $\Pr[G]\ge\varepsilon/2$. On such a transcript, the algorithm of \cref{lem:gl} with
oracle $\Pcal(\tau,\cdot)$ outputs a list of at most $16/\varepsilon^2$
strings which contains $\enc(K)$ with probability at least $1/2$. If the list is empty, we output $0$. Otherwise, we output
$\enc^{-1}(z)$ for a uniform element $z$ of the list. The success
probability is at least
\[
 \frac12\cdot\frac\varepsilon2\cdot\frac12\cdot\frac{\varepsilon^2}{16}
 =\frac{\varepsilon^3}{128}.
\]
Each call to $\Pcal$ takes time $T+O(n^2)$, and the algorithm of
\cref{lem:gl} makes $\poly(\ell,1/\varepsilon)$ calls.
\end{proof}

Combining \cref{prop:dense-coordinate,thm:dense-bit}, if no algorithm
running in time $\poly(\log q,1/\varepsilon)\cdot(T+n^2)$ solves
$\COORD_{n,q}$ with probability $\varepsilon^3/128-q^{-n}$, then the bit
version of the protocol is $(T,\varepsilon+2\sigma_q)$-secure. We note
an important limitation of this statement: a random guess solves
$\COORD_{n,q}$ with probability $1/q$, so the assumption can only hold when
$\varepsilon^3/128>1/q$. Hence, this approach only gives security for
advantages $\varepsilon>(128/q)^{1/3}$, and requires a large field.

\subsection{Connection to rank computation}
\label{sec:rank}

The hardness of $\COORD_{n,q}$ is a new assumption. We now relate the
security of the protocol to a more standard problem: computing the rank
of a matrix. Choosing the distribution on which we define the average-case rank-finding problem requires a bit of care: we cannot use uniformly random matrices, since they have full
rank with probability at least $1-1/(q-1)$, hence predicting full
rank is almost always correct. Instead, we consider distributions which first draw the rank (a natural ``hardest-looking'' candidate is for instance the distribution that picks the rank uniformly at random between $1$ and $N$), and then sample a uniformly random matrix of this rank.

\begin{definition}[Rank of random matrices]
\label{def:findrank}
Let $\mathcal R$ be a distribution over $\{0,\ldots,N\}$. The distributional
problem $\FINDRANK^{\mathcal R}_{N,q}$ is defined as follows. Sample
$\rho\gets\mathcal R$, and let $A$ be a uniformly random matrix of rank
$\rho$ in $\F^{N\times N}$. The input is $A$, and the goal is to output
$\rho$.
\end{definition}

Our reduction requires key recovery algorithms for all dimensions $c\le N$. Our main idea is to use an adversary against the key agreement to test whether $\rank A\ge c$ by running it on the
leading $c\times c$ submatrix $W$ of $A$. If $W$ is singular, part
of Bob's secret is hidden in its kernel, and the key is almost
unpredictable. The following lemma
makes the resulting information-theoretic bound precise:

\begin{lemma}[Singular matrices hide the key]
\label{lem:rank-hiding}
Let $W\in\F^{c\times c}$ be a matrix with a kernel of dimension $d$. Let
$\Pcal$ be a randomized algorithm, and let $r\gets\mathbb F_q^{1\times c}$
and $b\gets\F^c$ be independent of the coins of $\Pcal$. Then
\[
 \Pr\bigl[\Pcal(W,r,Wb)=rb\bigr]\le q^{-d}+(1-q^{-d})q^{-1}.
\]
In particular, this probability is at most $2/q$ when $W$ is singular.
\end{lemma}

\begin{proof}
Conditioned on $Wb$, the vector $b$ is uniform in a coset $b_0+\ker W$. We
write $b=b_0+\kappa$, where $\kappa$ is uniform in $\ker W$ and independent
of $(W,r,Wb)$. Then $rb=rb_0+r\kappa$. The restriction of $r$ to $\ker W$
is a uniform linear form on a space of dimension $d$, which is zero with
probability $q^{-d}$. Otherwise, $r\kappa$ is uniform in $\F$ given
$(W,r,Wb)$, and $\Pcal$ is correct with probability $1/q$.
\end{proof}

We also need the following facts about the leading block of a random
matrix of a given rank.

\begin{lemma}[Leading block of a random matrix]
\label{lem:leading-block}
Let $q\ge4$, $1\le c\le\rho\le N$, and let $A$ be a uniformly random matrix
of rank $\rho$ in $\F^{N\times N}$. Let $W$ be the leading $c\times c$
submatrix of $A$. Then $W$ is singular with probability at most $8/q$, and
conditioned on being invertible, $W$ is uniform in $\GLc c$.
\end{lemma}

\begin{proof}
Every matrix of rank $\rho$ can be written as $XY^\T$ with
$X,Y\in\F^{N\times\rho}$ of full column rank, in exactly $|\GLc\rho|$ ways
(the factorizations of $A=XY^\T$ are the pairs $(XG,YG^{-\T})$ for
$G\in\GLc\rho$). Hence, if $X$ and $Y$ are independent and uniform among
the matrices of full column rank, then $XY^\T$ is uniform among the matrices
of rank $\rho$. Let $X_c$ and $Y_c$ be the first $c$ rows of $X$ and $Y$,
so that $W=X_cY_c^\T$.

Let us first assume that $X$ and $Y$ are uniform in $\F^{N\times\rho}$.
The matrix $X_c$ has rank smaller than $c$ with probability less than
$1/(q-1)$. When $X_c$ has rank $c$, the linear map
$Z\mapsto ZX_c^\T$ from $\F^{c\times\rho}$ to $\F^{c\times c}$ is surjective, so $W^\T=Y_cX_c^\T$ is uniform in $\F^{c\times c}$, and is singular with
probability less than $1/(q-1)$. Hence, $W$ is singular with probability
less than $2/(q-1)$. Now, $X$ and $Y$ both have full column rank with
probability at least $1-2/(q-1)$. Conditioning on this event, the
probability that $W$ is singular becomes at most
\[
 \frac{2/(q-1)}{1-2/(q-1)}=\frac2{q-3}\le\frac8q,
\]
where the last inequality holds for $q\ge4$.

For the second statement, note that the distribution of $A$ is invariant
under $A\mapsto GAH$ for all $G,H\in\GLc N$. Taking $G$ and $H$ block
diagonal, with blocks $G',H'\in\GLc c$ and $I_{N-c}$, maps $W$ to $G'WH'$.
Hence, the distribution of $W$ is invariant under $W\mapsto G'WH'$. Now,
for all $W_0,W_1\in\GLc c$, taking $G'=W_1W_0^{-1}$ and $H'=I_c$ maps $W_0$
to $W_1$, so $\Pr[W=W_0]=\Pr[W=W_1]$. Therefore, $W$ is uniform in
$\GLc c$ when conditioned on being invertible.
\end{proof}

The next theorem combines these facts with the random walk of
\cref{sec:amplification}. The attacker's success probability $\alpha$ is
averaged over invertible matrices, so repeating the check on a fixed $W$
does not necessarily help. Instead, we walk through matrices $W\mapsto RW$, with
$R=I_c+uv^\T$ invertible, using $O(c^2)$ operations per step. When $W$ is
uniform and invertible, the spectral gap and the hitting lemma imply that
$O(1/\alpha)$ checks suffice to find a correct key with constant
probability. When $W$ is singular, all visited matrices remain singular,
so \cref{lem:rank-hiding} keeps the probability of each check passing
small. Grouping the checks into blocks and taking a majority gives a
reliable test of whether $\rank A\ge c$, which we use in a binary search
to recover the rank.

\begin{theorem}[Rank computation from key recovery]
\label{thm:rank-from-ka}
Let $N\ge1$ and $0<\alpha,\delta<1/2$. For every $1\le c\le N$, let
$\Pcal_c$ be a time-$T$ algorithm which recovers the key with probability
at least $\alpha$, on the ideal distribution of the protocol of
dimension $c$ over $\F$. Let
\[
 \ell_N=\lceil\log_2(N+1)\rceil,\qquad
 L=1+\left\lceil\frac8\alpha\log64\right\rceil,\qquad
 w=\left\lceil\log_4\frac{4\ell_N}\delta\right\rceil.
\]
If $q\ge\max\{128L,16\ell_N/\delta\}$, there is an algorithm which, given a
uniformly random matrix $A\in\F^{N\times N}$ of rank $\rho$, outputs $\rho$
with probability at least $1-\delta$, for every $\rho\in\{0,\ldots,N\}$. It
makes at most $\ell_NLw$ calls to the algorithms $\Pcal_c$, and runs in time
\[
 O\bigl(\ell_NLw(T+N^2)\bigr)
 =O\!\left(\frac{T+N^2}{\alpha}\log(N+1)\log\frac{\log(N+1)}\delta\right).
\]
\end{theorem}

The condition on the field size is $q\ge C\max\{\alpha^{-1},\log(N+1)/\delta\}$
for an absolute constant $C$.

\begin{figure}[ht]
\begin{mdframed}
\textbf{Test $\mathsf{Rank}_c(A)$.} Decides whether $\rank A\ge c$.
\begin{enumerate}[leftmargin=*]
 \item Let $W_1$ be the leading $c\times c$ submatrix of $A$.
 \item For $j=1,\ldots,Lw$:
 \begin{enumerate}
  \item Sample $b\gets\F^c$ and $r\gets\mathbb F_q^{1\times c}$. The check
  of step $j$ passes if $\Pcal_c(W_j,r,W_jb)=rb$.
  \item Sample $u,v\gets\F^c$. If $1+v^\T u\ne0$, set
  $W_{j+1}=(I_c+uv^\T)W_j$; otherwise, set $W_{j+1}=W_j$.
 \end{enumerate}
 \item Split the $Lw$ steps into $w$ consecutive blocks of $L$ steps. A
 block is marked if at least one of the checks of this block passes. Output ``yes'' if
 at least $w/2$ blocks are marked, and ``no'' otherwise.
\end{enumerate}
\end{mdframed}
\caption{The threshold test of \cref{thm:rank-from-ka}.}
\label{fig:rank-test}
\end{figure}

\begin{proof}
The algorithm performs a binary search for $\rho$, using the threshold test
$\mathsf{Rank}_c$ of \cref{fig:rank-test}. Note that the test can check
the answers of $\Pcal_c$, since it knows $b$. Each step of the test takes
$O(c^2)$ operations in addition to the call to $\Pcal_c$. We first show
that each test errs with probability at most $\delta/(4\ell_N)+8/q$, and
then analyze the binary search.

\paragraph{Case $\rank A<c$.}
In this case, $W_1$ is singular, and so are all the matrices $W_j$, since
the accepted updates are invertible. The vectors $b$ and $r$ are fresh at
each step, and $W_j$ only depends on the previous steps. By
\cref{lem:rank-hiding}, each check passes with probability at most $2/q$,
conditioned on the previous steps. Hence, each block is marked with
probability at most $2L/q\le1/64$, conditioned on the previous blocks.
Let $J$ be a fixed set of blocks, and let $E_i$ be the event that the
$i$-th block of $J$ is marked. Then
\[
 \Pr\Bigl[\bigcap_{i\le|J|}E_i\Bigr]=\prod_{i\le|J|}\Pr\Bigl[E_i\Bigm|\bigcap_{i'<i}E_{i'}\Bigr]
 \le64^{-|J|}.
\] If the test outputs yes, then some set
of $\lceil w/2\rceil$ blocks are all marked. A union bound over the at
most $2^w$ such sets shows that the test outputs yes with probability at most
$2^w\cdot64^{-w/2}=4^{-w}\le\delta/(4\ell_N)$. This holds for every matrix
$A$ of rank smaller than $c$.

\paragraph{Case $\rank A\ge c$.}
By \cref{lem:leading-block}, the matrix $W_1$ is singular with probability
at most $8/q$, and is uniform in $\GLc c$ when it is invertible. We now
assume that $W_1$ is uniform in $\GLc c$, so that the trajectory of the
walk is stationary. For $W\in\GLc c$, let $h(W)$ be the probability that a
check at $W$ passes. Since $Wb$ is uniform when $W$ is invertible, the
query of a check is distributed as in the ideal distribution, conditioned
on the matrix $W$. Hence $\E_W[h(W)]\ge\alpha$. As in the proof of
\cref{lem:hitting}, let $D$ be the operator of multiplication by
$\sqrt{1-h}$, and let $P$ be the operator of the invertible walk in
dimension $c$. The probability that all blocks in a fixed set $J$ are
unmarked can be written as
\[
 \bigl\langle\mathbf 1,F_1PF_2P\cdots PF_w\mathbf 1\bigr\rangle,
 \qquad\text{where}\qquad
 F_i=\begin{cases}D(DPD)^{L-1}D&\text{if }i\in J,\\P^{L-1}&\text{otherwise.}\end{cases}
\]
By \cref{lem:gl-gap}, this walk has spectral gap at least $1/3$,
and hence at least $1/8$. The proof of \cref{lem:hitting} shows that $\|DPD\|\le1-\alpha/8$, hence
$\|D(DPD)^{L-1}D\|\le(1-\alpha/8)^{L-1}\le1/64$ by the choice of $L$. All
the other operators have norm at most one. Therefore, the probability that
all the blocks in $J$ are unmarked is at most
\[
 \|\mathbf 1\|^2\cdot\prod_{i=1}^w\|F_i\|\cdot\|P\|^{w-1}\le64^{-|J|}.
\] If the test outputs no,
then some set of $\lceil w/2\rceil$ blocks are all unmarked, and the same
union bound as above shows that this happens with probability at most
$4^{-w}\le\delta/(4\ell_N)$. Including the case where $W_1$ is singular, the
test outputs no with probability at most $8/q+\delta/(4\ell_N)$, on average over
the matrices $A$ of rank $\rho\ge c$.

\paragraph{Binary search.}
The binary search over $\{0,\ldots,N\}$ makes at most $\ell_N$ tests. Fix the
rank $\rho$, and consider the sequence $c_1,\ldots,c_{\ell_N}$ of thresholds
that the search would test if all answers were correct; this sequence is
deterministic. If the search outputs a wrong rank, then the first wrong
answer is given by one of the tests $\mathsf{Rank}_{c_1}(A),\ldots,
\mathsf{Rank}_{c_{\ell_N}}(A)$. By a union bound, which does not require the
tests to be independent, the search fails with probability at most
\[
 \ell_N\Bigl(\frac8q+\frac\delta{4\ell_N}\Bigr)
 =\frac{8\ell_N}q+\frac\delta4\le\frac\delta2+\frac\delta4<\delta,
\]
where we used $q\ge16\ell_N/\delta$.
\end{proof}

By \cref{prop:dense-coordinate}, the same conclusion holds for a family of
algorithms which solve $\COORD_{c,q}$ for every $c\le N$. In particular, for
constant $\delta$, $\alpha=N^{-o(1)}$ and $T\ge N^2$, the resulting
algorithm runs in time $T\cdot N^{o(1)}$, and the field size condition
becomes $q\ge N^{o(1)}$. Hence, if computing the rank of random matrices
with a uniformly random rank requires time $N^{2+\eta}$ over such a field,
then no family of algorithms running in time $N^{2+\eta-\zeta}$, for
$\zeta>0$, can recover the key of the protocol with probability
$N^{-o(1)}$ on the ideal distribution, at every dimension $c\le N$. When
$q=2^\ell$, \cref{thm:dense-bit} gives the same conclusion for
distinguishers with advantage $\varepsilon=N^{-o(1)}$ such that
$\varepsilon^3\ge C/q$ for a large enough constant $C$. For the protocol
itself, the advantage increases by at most $2\sigma_q<2/(q-1)$.

\begin{remark}[Choice of fields]
\label{rem:rank-field-choice}
We can apply \cref{thm:rank-from-ka} to a protocol whose field is determined
by the dimension. The idea is to keep the field fixed over long intervals.
Let $n_j=2^{2^j}$ for $j\ge0$, set $q(1)=2$, and define
\[
 q(n)=n_{j+1}\qquad\text{for }n_j\le n<n_{j+1}.
\]
Thus, $q(n)$ is a power of two and $n<q(n)\le n^2$ for $n\ge2$.
At the dimensions $N_j=n_{j+1}-1$, all dimensions
$n_j\le c\le N_j$ use the same field $\mathbb F_{N_j+1}$.
For $c<n_j=\sqrt{N_j+1}$, we can recover the field key on the ideal
distribution by Gaussian elimination in time $O(c^3)=O(N_j^{3/2})$,
over the same field. Consequently, attacks on the protocol at all
sufficiently large dimensions provide the algorithms required by the
rank reduction at infinitely many dimensions $N_j$.
\end{remark}

\begin{corollary}[A win--win for rank and key agreement]
\label{cor:rank-ka-winwin}
Use the field schedule of \cref{rem:rank-field-choice}. Let $\Pi_n$ be the
protocol over $\mathbb F_{q(n)}$, with the one-bit key of
\cref{sec:protocol-definition}. Write $\tau_n$ for the full transcript, including
the extraction seed, and $B_n$ for the common bit. For each $n$, let
$\mathcal R_n$ be any distribution on $\{0,\ldots,n\}$.
For every constant $\eta>0$, there are constants $\eta',b>0$ such that at
least one of the following holds.
\begin{enumerate}[leftmargin=*]
 \item There is a uniform randomized algorithm running in time
 $O(n^{2+\eta})$ which computes the rank with probability at least $2/3$
 on $\FINDRANK_{n,q(n)}^{\mathcal R_n}$, at infinitely many dimensions $n$.
 \item The protocol $\Pi_n$ gives infinitely-often fine-grained key
 agreement. More precisely, the parties run in time $O(n^2)$, and there
 is an infinite set $I\subseteq\mathbb N$ such that every uniform
 randomized distinguisher $\Dcal$ running in time $O(n^{2+\eta'})$
 satisfies
 \[
  \left|\Pr[\Dcal(1^n,\tau_n,B_n)=1]
       -\Pr[\Dcal(1^n,\tau_n,u)=1]\right|\le n^{-b}
 \]
 for all sufficiently large $n\in I$, where $u$ is a uniform independent
 bit. The set $I$ is common to all distinguishers. The threshold above
 which the bound holds may depend on $\Dcal$.
\end{enumerate}
In particular, almost-everywhere average-case hardness of rank computation
in time $n^{2+\eta}$ implies infinitely-often fine-grained security of
$\Pi_n$.
\end{corollary}

\begin{proof}
The main point is that failure of infinitely-often security gives a finite
collection of attacks that covers all sufficiently large dimensions.
We first choose the parameters. Fix an integer $d\ge1$ such that the
polynomial overhead in \cref{thm:dense-bit} is bounded by
$O((1+\ell)^d\varepsilon^{-d})$. Choose positive rational constants
$\eta',b$ such that
\[
 3b<1,\qquad \eta'+(d+3)b<\eta.
\]

Suppose that the second alternative fails. List the uniform
distinguishers running in time $O(n^{2+\eta'})$ as
$\Dcal_1,\Dcal_2,\ldots$, and let $a_i(n)$ denote the distinguishing
advantage of $\Dcal_i$ on the actual protocol. There is an integer $h$
such that, for every sufficiently large $n$,
\[
 \max_{1\le i\le h}a_i(n)>n^{-b}.
\]
Indeed, otherwise, for every $h$ there would be infinitely many dimensions
at which $a_i(n)\le n^{-b}$ for all $i\le h$. We could then choose an
increasing sequence $m_h$ of such dimensions. The set $I=\{m_h:h\ge1\}$
would satisfy the second alternative.

Now fix a sufficiently large dimension $N=N_j$ from
\cref{rem:rank-field-choice}, so $q(N)=N+1$. For each $n_j\le c\le N$,
some $\Dcal_i$ with $i\le h$ has advantage greater than
$c^{-b}\ge N^{-b}$. The ideal and actual advantages differ by at most
$2/(q(N)-1)=2/N$. Thus, on the ideal distribution, this distinguisher has
advantage at least
\[
 \varepsilon_N=\tfrac12N^{-b}.
\]
Choose $i$ uniformly from $\{1,\ldots,h\}$ and apply
\cref{thm:dense-bit} to $\Dcal_i$, using the threshold $\varepsilon_N$.
This gives a key recovery algorithm $\Pcal_c$ with success probability
at least
\[
 \alpha_N=\frac{\varepsilon_N^3}{128h}
          =\frac{N^{-3b}}{1024h}.
\]
The key recovery procedure is stopped after the time bound given by
\cref{thm:dense-bit}, including when the chosen distinguisher has smaller
advantage. Since $\log q(N)=O(\log N)$, all these algorithms run in time
\[
 T_N=\Otilde(N^{2+\eta'+db}).
\]
For $c<n_j$, use Gaussian elimination over $\mathbb F_{N+1}$ instead.
This recovers the key with probability one on the ideal distribution,
and takes time $O(N^{3/2})$.

Since $3b<1$, the field size $q(N)=N+1$ satisfies
\[
 q(N)\ge C\max\{\alpha_N^{-1},3\log(N+1)\}
\]
for all sufficiently large $N=N_j$.
We can therefore apply \cref{thm:rank-from-ka} with $\delta=1/3$.
The resulting algorithm computes the rank with probability at least
$2/3$ for every fixed rank, and hence also for a rank drawn from
$\mathcal R_N$. The running time is
\[
 \Otilde\!\left(\frac{T_N+N^2}{\alpha_N}\right)
 =\Otilde(N^{2+\eta'+(d+3)b})
 =O(N^{2+\eta}).
\]
The finite list $\Dcal_1,\ldots,\Dcal_h$ can be included in the description
of one uniform algorithm. This algorithm runs the reduction at dimensions
$N_j$ and outputs an arbitrary value at other dimensions. It therefore
satisfies the first alternative.
\end{proof}

\section*{AI disclosure}
\addcontentsline{toc}{section}{AI disclosure}

All results in this paper were originally established by the human
authors. AI models (Claude Fable 5.1, Claude Opus 5.5, and OpenAI Astra)
were used to explore the literature and refine some of our arguments. In particular, the hitting lemma (\cref{lem:hitting}) was identified
through interactions with Claude Opus 5.5, after the authors explicitly asked for
references to Chernoff bounds for random walks on expanders, and replaced a looser bound. The same
model was asked to provide parameters and analyses for some standard arguments
in \cref{sec:protocol}: for \cref{thm:dense-bit}, it suggested using
Goldreich--Levin with a reduced list size, and it also provided the proof of
\cref{lem:leading-block}, a standard calculation in probability and linear
algebra, which the authors subsequently verified.

Claude Fable 5.1 and OpenAI Astra were also used to check the correctness
of our claims and proofs, to generate \cref{fig:overview-absorption}, and
for general editing purposes. The authors verified the correctness and originality of all content, including references.

\bibliographystyle{alpha}
\bibliography{refs}

\end{document}